\documentclass[twocolumn, twocolappendix]{aastex63}
	
\usepackage[encapsulated]{CJK}
\usepackage{ucs}
\usepackage[utf8x]{inputenc}
\usepackage{url}
\newcommand{\cntext}[1]{\begin{CJK}{UTF8}{bsmi}#1\end{CJK}}

\usepackage[italicdiff]{physics} 
\usepackage{amsmath}

\usepackage{graphicx}

\usepackage{braket}

\newcommand{\m}[1]{\underline{\underline{#1}}}

\shorttitle{H$_2$ \& CO formation in self-regulating ISM}
\shortauthors{Hu et al.}

\begin{document}


\title{Metallicity dependence of the H/H$_2$ and C$^+$/C/CO distributions in a resolved self-regulating interstellar medium}
\author[0000-0002-9235-3529]{Chia-Yu Hu (\cntext{胡家瑜})}
\affiliation{Max-Planck-Institut f\"{u}r Extraterrestrische Physik, Giessenbachstrasse 1, D-85748 Garching, Germany}
\author{Amiel Sternberg}
\affiliation{School of Physics \& Astronomy, Tel Aviv University, Ramat Aviv 69978, Israel}
\affiliation{Center for Computational Astrophysics, Flatiron Institute, 162 5th Ave, New York, NY 10010, USA}
\affiliation{Max-Planck-Institut f\"{u}r Extraterrestrische Physik, Giessenbachstrasse 1, D-85748 Garching, Germany}
\author{Ewine F. van Dishoeck}
\affiliation{Max-Planck-Institut f\"{u}r Extraterrestrische Physik, Giessenbachstrasse 1, D-85748 Garching, Germany}
\affiliation{Leiden Observatory, Leiden University, P.O. Box 9513, NL-2300 RA Leiden, the Netherlands}
\correspondingauthor{Chia-Yu Hu}
\email{cyhu.astro@gmail.com}

\begin{abstract}

We study the metallicity dependence of the H/H$_2$ and C$^+$/C/CO distributions
in a self-regulated interstellar medium (ISM)
across a broad range of metallicities ($0.1  < Z/Z_\odot < 3$).
To this end,
we conduct high-resolution (particle mass of $1\ {\rm M_\odot}$) 
hydrodynamical simulations coupled with a time-dependent H$_2$ chemistry network.
The results are then post-processed with an accurate chemistry network to model 
the associated C$^+$/C/CO abundances,
based on the time-dependent non-steady-state (``non-equilibrium'') H$_2$ abundances.
We find that the time-averaged star formation rate and the ISM structure
are insensitive to metallicity.
The column densities relevant for molecular shielding appear correlated with the volume densities
in gravitationally unstable gas.
As metallicity decreases,
H$_2$ progressively deviates from steady state (``equilibrium'')
and shows shallow abundance profiles 
until they sharply truncate at the photodissociation fronts.
In contrast,
the CO profile is sharp and controlled by photodissociation as CO quickly reaches steady state.
We construct effective one-dimensional cloud models that successfully capture the time-averaged chemical distributions in simulations.
At low metallicities,
the steady-state model significantly overestimates the abundance of H$_2$ in the diffuse medium.
The overestimated H$_2$, however, has little impact on CO.
Consequently,
the mass fraction of CO-dark H$_2$ gas is significantly lower than 
what a fully steady-state model predicts.
The mass ratios of H$_2$/C$^+$ and H$_2$/C both show a weaker dependence on $Z^{\prime}$
than H$_2$/CO,
which potentially indicates that C$^+$ and C could be alternative tracers for H$_2$ at low $Z^{\prime}$ in terms of mass budget.
Our chemistry code for post-processing is publicly available.


\end{abstract}

\section{Introduction} \label{sec:intro}

Stars are formed in molecular hydrogen (H$_2$) gas,
as ubiquitously observed
in our Milky Way \citep{2003ARA&A..41...57L, 2009ApJS..181..321E},
in nearby galaxies \citep{2008AJ....136.2846B, 2008AJ....136.2782L},
and all the way to high-redshift galaxies \citep{2010MNRAS.407.2091G, 2020ARA&A..58..157T}.
H$_2$ does not radiate efficiently at the typical temperatures of molecular clouds 
and thus is not directly visible.
As such,
carbon monoxide (CO),
which is an efficient line emitter and whose formation relies on the presence of H$_2$,
is routinely used as an observable tracer for H$_2$ \citep{2013ARA&A..51..207B}.

However,
CO emission becomes very faint and is often non-detectable at low metallicities \citep{2005ApJ...625..763L, 2012AJ....143..138S, 2014A&A...564A.121C, 2015A&A...583A.114H}.
This is expected as 
the CO formation rates are reduced and the
low dust abundance provides less shielding against the far-ultraviolent (FUV) radiation.
CO is thus further photodissociated into the clouds,
leading to a potentially large ``CO-dark'' H$_2$ reservoir in low-metallicity galaxies \citep{1997ApJ...483..200M, 1998ApJ...498..735P, 1999ApJ...513..275B, 2010ApJ...716.1191W, 2013ARA&A..51..207B}.
Therefore,
alternative tracers for H$_2$ have been proposed such as 
ionized carbon C$^+$ 
\citep{1997ApJ...483..200M, 2011MNRAS.416.2712D, 2018MNRAS.481.1976Z, 2020A&A...643A...5D, 2020A&A...643A.141M}
and atomic carbon C 
\citep{2004MNRAS.351..147P, 2011ApJ...730...18W, 2018A&A...615A.142A, 2019ApJ...880..133J, 2019MNRAS.482.3135B, 2020ApJ...890...24V},
though their feasibility and robustness require further investigations.
Therefore,
understanding the distributions of C$^+$/C/CO in different environments 
is a crucial step to assess the pros and cons of these different tracers.

However,
such theoretical considerations adopt a static view of the interstellar medium (ISM),
which is highly dynamic in reality.
Gas radiatively cools, gravitationally collapses and forms stars,
which in turn inject energy and momentum back to the ISM via stellar feedback,
heating up the gas, driving turbulence and destroying clouds.
This cycle occurs rapidly and thus a realistic picture of the ISM must include these dynamical effects.

Substantial efforts have been made using hydrodynamical simulations
to study the chemical properties of the dynamic ISM over the past decade.
\citet{2011MNRAS.412..337G} and \citet{2012MNRAS.426..377G} 
conducted simulations of turbulent clouds
to study the relationship between H$_2$ and CO
and quantify the CO-to-H$_2$ conversion factor, $X_{\rm CO}$, as a function of visual extinction and metallicity.
Cloud simulations, however, are limited by their unrealistic initial and boundary conditions 
as well as their artificial driving force,
rendering them incapable of following the formation and destruction of clouds self-consistently.
On the other hand,
the metallicity dependence of $X_{\rm CO}$ has also been investigated by
\citet{2012MNRAS.421.3127N} in galaxy mergers and
\citet{2012ApJ...747..124F} in cosmological simulations, respectively.
As these studies focus on a much larger scale, their spatial resolution is rather limited ($\sim 70$ pc),
preventing them from resolving the multi-phase ISM structure and stellar feedback
that require at least pc-scale resolution.
Consequently,
highly uncertain sub-resolution models have to be adopted,
diminishing their predictive power.
Simulations of an isolated Milky Way-type galaxy 
\citep{2015MNRAS.447.2144D, 2016MNRAS.tmp..112R}
allow for a somewhat better resolution,
but still far from what is required and thus face similar issues.

More recently,
the SILCC simulations \citep{2015MNRAS.454..238W, 2016MNRAS.456.3432G, 2017MNRAS.472.4797S}
used the ``stratified box'' setup to study a patch of the ISM regulated by stellar feedback for solar-neighborhood conditions, 
featuring a time-dependent chemistry network from \citet{2007ApJS..169..239G} and \citet{2012MNRAS.421..116G}.
The TIGRESS simulations \citep{2017ApJ...846..133K, 2020ApJ...900...61K} 
adopted a similar setup but without the time-dependent chemistry.
Instead,
their results were later post-processed by \citet{2018ApJ...858...16G} 
with their improved chemistry network \citep{2017ApJ...843...38G}
to obtain steady-state\footnote{
The term ``steady-state'' refers to situations where the formation and destruction rates of chemical species are equal 
such that the chemical abundances are constant in time in the comoving frame,
which is contrary to the ``time-dependent'' cases.
We note that
in the field of galaxy formation,
the terms ``time-dependent'' vs. ``steady-state''
are often referred to as ``non-equilibrium'' vs. ``equilibrium'',
which we cautiously reserve for thermodynamic equilibrium
where the abundances are determined solely by temperature and elemental compositions 
(instead of two-body kinetics) as in stellar atmospheres.
}
chemical abundances.
Using the same chemistry as SILCC,
``The Cloud Factory'' \citep{2020MNRAS.492.1594S} simulated
a much larger patch of the Galaxy and with a mesh refinement technique to follow 
the formation of clouds down to a remarkable mass resolution of 0.25 ${\rm M_\odot}$
albeit only for a short amount of time (4 Myr).

Another promising approach is to simulate an isolated dwarf galaxy 
that is small enough to achieve the necessary resolution.
\citet{2016MNRAS.458.3528H} and \citet{2017MNRAS.471.2151H}
studied a dwarf galaxy with $Z^{\prime}\equiv Z/Z_\odot = 0.1$ 
using the same chemistry as SILCC and found very little H$_2$ in the ISM,
concluding that the star formation reservoir is dominated by the atomic gas.
On the other hand, 
using the {\sc Grackle} chemistry network \citep{2017MNRAS.466.2217S},
\citet{2019MNRAS.482.1304E} simulated an even smaller galaxy with the same metallicity 
but found a significantly larger H$_2$ fraction. 

All of these resolved ISM simulations, however,
have been focused only on a particular metallicity.
A systematic investigation across a broad range of metallicities in the framework of 
a self-regulated ISM is therefore urgently needed.
Very recently,
\citet{2020ApJ...903..142G} post-processed the results of \citet{2020ApJ...900...61K} 
and studied the systematic variations of several physical parameters including metallicity 
in the range of $0.5 < Z^{\prime} < 2$.
However,
their simulations were all run with $Z^{\prime} = 1$ 
which is inconsistent with the post-processed chemistry.
More importantly,
by post-processing, 
they were forced to assume steady-state chemistry which, as we will demonstrate,
significantly over-produces H$_2$ (but not CO), especially at low metallicities.

While the chemistry network for H$_2$ is relatively small and can be efficiently coupled with simulations,
the network involving CO is significantly larger \citep{1976RPPh...39..573D, 1988ApJ...334..771V, 1995ApJS...99..565S}.
Therefore,
coupling the full CO network with simulations implies a formidable computational overhead,
which motivated efforts to trim down the network to a minimum level.
In a comparison study,
\citet{2012MNRAS.421..116G} showed that 
the simplest and most widely used network of \citet{1997ApJ...482..796N} (NL97),
while computationally efficient,
tends to over-produce CO.
In contrast,
the network of \citet{1999ApJ...524..923N} (NL99) is able to reproduce accurate results 
comparable to the much more comprehensive network of \citet{2010MNRAS.404....2G}.
In practice, however,
even the NL99 network is already a nontrivial computational overhead 
and therefore is rarely applied to ISM simulations.

In this work,
we study the metallicity dependence of the H/H$_2$ and C$^+$/C/CO distributions
in a stellar feedback-regulated ISM
across a broad range of metallicities $0.1 < Z^{\prime} < 3$.
We conduct stratified-box simulations which self-consistently include
a time-dependent H$_2$ chemistry and cooling,
star formation with individual stars,
and feedback from photoionization and supernovae (SNe).
To obtain the abundances of C$^+$, C and CO,
we post-process our simulations with a detailed chemistry network that includes 31 species and 286 reactions,
taking the time-dependent H$_2$ abundance into account.
With this hybrid technique,
we are able to achieve a very high resolution 
(particle mass of $1\ {\rm M_\odot}$, spatial resolution $\sim$ 0.2 pc),
capture the time-dependent nature of H$_2$ formation in a dynamical ISM,
and accurately model CO with all its major formation and destruction channels simultaneously.

This paper is organized as follows.
In Sec. \ref{sec:method}, we describe our numerical method and the setup of our simulations.
In Sec. \ref{sec:postpro}, we introduce the more detailed chemistry network 
that we use to post-process the chemical species.
In Sec. \ref{sec:result}, 
we present our simulation results,
focusing on the metallicity dependence of the ISM structure and the chemical compositions.
In Sec. \ref{sec:discuss}, we compare our results with previous studies and discuss the potential caveats.
In Sec. \ref{sec:sum}, we summarize our paper.

\section{Simulation method} \label{sec:method}

\subsection{Gravity and hydrodynamics}
We use the public version of {\sc Gizmo} \citep{2015MNRAS.450...53H},
a multi-solver code featuring the meshless Godunov method \citep{2011MNRAS.414..129G} 
on top of the TreeSPH code {\sc Gadget-3} \citep{2005MNRAS.364.1105S}.
The gravitational interaction is solved by the Barnes–Hut algorithm (``treecode'') while hydrodynamics is solved by 
the meshless finite-mass (MFM) method \citep{2015MNRAS.450...53H} 
with the number of neighboring particles in a kernel $N_{\rm ngb} = 32$.

\subsection{Cooling, heating and chemistry}\label{sec:coolchem}
We use a time-dependent hydrogen chemistry network developed in 
\citet{2007ApJS..169..239G} and \citet{2012MNRAS.421..116G},
which has been applied extensively in previous ISM simulations coupled with several different codes (e.g., \citealp{2014MNRAS.441.1628S, 2015MNRAS.454..238W, 2016MNRAS.456.3432G, 2016MNRAS.458.3528H, 2017MNRAS.471.2151H, 2017MNRAS.466.1903G, 2020MNRAS.492.1594S}).
The network includes 
H$_2$ formation on dust\footnote{
The gas-phase H$_2$ formation channel is not included 
which is appropriate in the metallicity range we explore in this work.
},
H$_2$ destruction by photodissociation, collisional dissociation and cosmic ray ionization,
and recombination in the gas phase and on dust grains.
Cooling processes include fine structure metal lines,
molecular lines,
Lyman alpha, 
H$_2$ collisional dissociation,
collisional ionization of H, and recombination of H$^+$ in the gas phase and on grains. 
Heating processes include photoelectric emission from dust, 
cosmic ray ionization, 
H$_2$ photodissociation, 
UV pumping of H$_2$ and the formation of H$_2$.

The cosmic ray ionization rate of H$_2$ ($\zeta_{\rm CR}$) and
the energy density of the unattenuated FUV radiation field ($u_{\rm UV}$) are both spatially constant but time-dependent,
linearly scaled with the star formation rate (SFR) surface density $\Sigma_{\rm SFR}$ 
projected face-on (see Sec. \ref{sec:setup}),
i.e.,
$\zeta_{\rm CR} =  (\Sigma_{\rm SFR} / \Sigma_{\rm SFR,0}) \zeta_{\rm CR,0}$,
where we use the local values for normalization:
$\zeta_{\rm CR,0} = 10^{-16}\ {\rm s^{-1}}$ \citep{2012ApJ...745...91I, 2015ApJ...800...40I}
and 
$\Sigma_{\rm SFR,0} = 2.4\times 10^{-3}\ {\rm M_\odot\ yr^{-1}\ kpc^{-2}}$ \citep{2009AJ....137..266F}.
Similarly,
$u_{\rm UV} = I_{\rm UV} u_{\rm UV,0}$ 
where $u_{\rm UV,0} = 8.94\times 10^{-14}\ {\rm erg\ cm^{-3}}$ 
is the Draine field \citep{1978ApJS...36..595D} and
$I_{\rm UV} \equiv \max(\Sigma_{\rm SFR} / \Sigma_{\rm SFR,0}, I_{\rm UV,min})$.
The floor $I_{\rm UV,min} = 0.002$ accounts for the cosmic UV background \citep{2012ApJ...746..125H}.
We define the SFR as the total mass of stars younger than $t_{\rm SFR}$ divided by $t_{\rm SFR}$,
and 
we adopt $t_{\rm SFR} = 30$ Myr motivated by the lifetime of an ${\rm 8 M_\odot}$-star.


Both dust and H$_2$ self-shielding against FUV radiation are accounted for by a method similar to \citet{2012MNRAS.420..745C}.
The {\sc HealPix} algorithm \citep{2011ascl.soft07018G} is used to define 12 different pixels centered at each gas particle.
Along each pixel,
we calculate the column densities of gas up to a pre-defined shielding length $L_{\rm sh}$.
The tree structure for gravity is used to speed up the calculation.
A treenode is assigned to a pixel if its center of mass is located within the pixel\footnote
{For simplicity, we do not let a treenode be partially assigned to multiple pixels as in \citet{2012MNRAS.420..745C}.
This is a reasonable simplification considering the small number of pixels we adopt.}.
We choose $L_{\rm sh} = 100$ pc motivated by the typical separation between OB associations in the solar neighborhood \citep{2003ApJ...584..797P, 2020ApJ...903...62B}.
The choice of $L_{\rm sh}$ only has a minor effect on the total molecular masses and is explored in Appendix \ref{app:Lsh}.

\subsection{Star formation}\label{sec:SFrecipe}
We adopt a stochastic star formation recipe commonly used in the field of galaxy formation.
Star formation occurs when the local velocity divergence becomes negative (i.e., converging flows) 
and the thermal Jeans mass of gas $M_{\rm J} = (\pi^{2.5} c_s^3)/(6 G^{1.5} \rho^{0.5})$ drops below the kernel mass $M_{\rm ker} = N_{\rm ngb} m_{\rm g}$,
where $c_s$ is the sound speed, 
$G$ is the gravitational constant, 
$\rho$ is the gas density and $m_{\rm g}$ is the gas particle mass.
Equivalently, 
this corresponds to the requirement
\begin{equation}
	n > 1.26\times 10^4 {\rm cm^{-3}} 
	\Big( \frac{T}{30 {\rm K}} \Big)^3 
	\Big(\frac{m_{\rm g}}{{\rm M_\odot}}\Big)^{-2}
	\Big(\frac{\mu}{2.3}\Big)^{-3},
\end{equation}
where $n = X_{\rm H}\rho / m_{\rm p}$ is the hydrogen number density,
$X_{\rm H} = 0.71$ is the hydrogen mass fraction,
$m_{\rm p}$ is the proton mass,
$T$ is the gas temperature and $\mu$ is the mean molecular weight.
Each gas particle that fulfills these requirements has a probability of 
$\epsilon_{\rm sf} \Delta t / t_{\rm ff}$ to be converted into a star particle,
where $\Delta t$ is the timestep, $t_{\rm ff} = \sqrt{ 3\pi /  (32 G \rho) }$ is the gas free-fall time and $\epsilon_{\rm sf}$ is the star formation efficiency.
A gas particle will therefore be converted into a star particle in roughly $t_{\rm ff} / \epsilon_{\rm sf}$.
We adopt $\epsilon_{\rm sf} = 0.5$ throughout this work as our resolution is capable of resolving the detailed structure of molecular clouds.
In addition,
we impose a threshold density $n_{\rm isf} = 10^5\ {\rm cm^{-3}}$ beyond which gas is converted into stars instantaneously,
irrespective of the above-mentioned requirements.
The effect of $n_{\rm isf}$ is discussed in Appendix \ref{app:conv}.
A star particle will inherit the position, velocity and mass of its gas progenitor but will no longer interact hydrodynamically.

\subsection{IMF sampling for individual stars}
Massive stars (initial mass $> 8\ {\rm M_\odot}$) inject energy and momentum into their surrounding gas which counters cooling and gravitational collapse, regulating the ISM.
In contrast to large-scale cosmological simulations where a star particle is massive enough to represent a star cluster that fully samples the stellar initial mass function (IMF),
care must be taken in our case as the mass of star particles 
($m_* = 1{\rm M_\odot}$) becomes too low to represent even a massive star. 
Here we present a simple method designed for high-resolution simulations.
In a uniformly sampled IMF,
there is on average one massive star in every $M_{\rm IMF} = $ 100 M$_\odot$ of stellar mass
assuming the Kroupa IMF \citep{2001MNRAS.322..231K}.
Therefore,
each star particle has a probability of $m_* / M_{\rm IMF}$ for being marked as a massive-star tracer.
The stellar mass of the tracer is determined stochastically following an IMF on the high-mass end,
which can be realized by the method of importance sampling:
\begin{equation}
m_{\rm star} = [(m_{\rm max}^\alpha - m_{\rm min}^\alpha) y +  m_{\rm min}^\alpha ]^{-1/\alpha}
\end{equation}
where $y$ is a random number uniformly distributed between 0 and 1,
$\alpha$ is the power-law index on the high-mass end,
and
$m_{\rm min} = 8 {\rm M_\odot}$ and $m_{\rm max} = 50 {\rm M_\odot}$ are the lower and upper bound for the sampled mass, respectively.
We note that $m_{\rm star}$ is insensitive to moderate variation of $m_{\rm max}$ as stars more massive than 50 ${\rm M_\odot}$ are very rare (less than 1\% of the total massive stars by number).
We adopt $\alpha = 1.3$ appropriate for the Kroupa IMF.
The sampled $m_{\rm star}$ is used to determine the stellar lifetime \citep{2012A&A...537A.146E} and luminosity of ionizing radiation from the BaSeL stellar library \citep{1997A&AS..125..229L, 1998A&AS..130...65L},
while the gravitational mass of the star particles ($m_*$) remains unchanged.

We note that our method of IMF sampling is similar to that presented in \citet{2017MNRAS.471.2151H} but different in two aspects: 
(i) we do not modify the gravitational mass
and 
(ii) we do not allow multiple tracers for a single star particle, 
which only happens when the resolution becomes so coarse that $m_* > M_{\rm IMF}$.
Our method is easy to implement 
and remains applicable when one wants to trace individual stellar masses below $8 {\rm M_\odot}$ 
(e.g. for the purpose of metal enrichment from the asymptotic giant branch stars) 
by lowering $m_{\rm min}$ (and adjusting $M_{\rm IMF}$ accordingly).

\subsection{Stellar feedback}
Massive stars emit hydrogen-ionizing radiation continuously throughout their lifetimes.
Our photoionization feedback is based on \citet{2017MNRAS.471.2151H}
which uses an iterative method to define the ionization front within which the gas is heated up to $10^4$ K.
The balance between recombination and photoionization is evaluated using the density of the photoionized gas which can be different from the gas density around the star.
This approach captures the dynamical evolution of a D-type photoionization front in a uniform medium 
as accurately as radiative transfer methods.
Moreover,
it can be applied to overlapping HII regions where a naive Str\"{o}mgren-sphere method would fail.

Massive stars explode as core-collapse SNe by the end of their lifetimes.
Our SN feedback is purely thermal,
i.e., an SN injects $10^{51}$ erg into its nearest 32 gas particles as thermal energy weighted by the kernel function.
As we are working at the resolution where the Sedov-Taylor phase of SN remnant is always well-resolved,
the dynamical effect of SN feedback on the ISM is insensitive to the form of energy injection (i.e., kinetic, thermal, or mixed) 
as shown in \citet{2019MNRAS.483.3363H}.


\subsection{Simulation setup}\label{sec:setup}

We use the ``stratified box'' setup which consists of an elongated box representing a patch of a disk galaxy,
that, in our case, is a model for the solar neighborhood.
The box size is 1 kpc along the $x$- and $y$-axes with periodic boundary conditions and 10 kpc along the $z$-axis with outflow boundary conditions.
The origin is defined at the box center and $z = 0$ represents the mid-plane of the galaxy
such that $z = \pm 5$ kpc are the (open) boundaries.
The gas initially follows a vertical (along the $z$-axis) density profile:
$\rho(z) = (\Sigma_{\rm g}/(2 H_{\rm g})\sech^2(z/H_{\rm g})$,
where $ \Sigma_{\rm g} = 10\ {\rm M_\odot pc^{-2}}$ is the gas surface density and $H_{\rm g} = 0.25$ kpc is the scale-height of the gaseous disk.
The resulting vertical gravitational acceleration due to self-gravity is
\begin{equation}
a_{\rm g} = -2\pi G \Sigma_{\rm g} \tanh\Big(\frac{z}{H_{\rm g}}\Big),
\end{equation}
and evolves according to the movement of gas.

We include external gravity from the stellar disk that exerts vertical gravitational acceleration:
\begin{equation}
a_* = -2\pi G \Sigma_* \tanh\Big(\frac{z}{H_*}\Big),
\end{equation}
where $\Sigma_* = 40\ {\rm M_\odot pc^{-2}}$ is the stellar surface density and $H_* = $ 0.25 kpc is the scale-height of the stellar disk.
In addition, 
we include vertical gravitational acceleration from a dark matter halo which follows an NFW profile \citep{1997ApJ...490..493N}
with a virial mass $M_{\rm vir} = 10^{12} {\rm M_\odot}$ and concentration $c = 12$:
\begin{equation}
a_{\rm DM} = - \frac{G m(r) z}{r^3}
\end{equation}
where
$r = \sqrt{z^2+R_0^2}$ is the spherical radius from the center of the halo,
$m(r) = 4\pi r_s^3 \rho_s \ln[(1 +r/r_s) - (r/r_s)(1 + r/r_s)]$ is the enclosed mass, 
$r_s = 17$ kpc, $\rho_s = 9.5\times 10^{-3} {\rm M_\odot pc^{-3}}$,
and $R_0 = 8$ kpc is the galactocentric radius of the Sun.

Our periodic boundary conditions for self-gravity are only pseudo-periodic.
Namely,
we only account for self-gravity from the nearest 3-by-3 periodic images in the $x$ and $y$ directions.
This is a good approximation as the large-scale gravitational potential is dominated by the external stellar disk\footnote{
	In a setup where the gas self-gravity dominates,
	the large-scale vertical gravitational acceleration will be underestimated if too few images are included,
	and the convergence of this summation is known to be slow.
	Numerical techniques like the two-dimensional (2D) Ewald summation can be applied to improve convergence,
	which is not yet supported in the {\sc Gizmo} code.}.
Furthermore,
mathematically robust periodic boundary conditions are not necessarily a better approximation of a galaxy, 
which is not infinitely large and has a radial distribution of gas.
We have tested and confirmed that our results are largely unchanged even when we include 9-by-9 periodic images.

We adopt a mass resolution of $m_{\rm g} = 1\ {\rm M_\odot}$ 
which safely resolves every SN event in our simulations \citep{2016MNRAS.458.3528H, 2019MNRAS.483.3363H}.
Given $X_{\rm H} = 0.71$ and $N_{\rm ngb} = 32$,
the radius of the MFM kernel (which corresponds to the spatial resolution) is
\begin{equation}
	h = 6\ {\rm pc} \Big(\frac{n}{{\rm cm^{-3}}}\Big)^{-1/3} \Big( \frac{m_{\rm g}}{{\rm M_\odot}} \Big)^{1/3}.
\end{equation}
At the typical density where the Jeans mass becomes unresolved and star formation occurs,
which turns out to be $n \approx 10^4\ {\rm cm^{-3}}$,
$h\approx 0.2$ pc.
Therefore,
the gravitational softening length is chosen to be $\epsilon_{\rm soft} = $ 0.2 pc for both gas and stars.

The gas temperature is initially set to $10^4$ K uniformly.
The metallicity and dust-to-gas mass ratio (DGR) are both constant throughout the simulations.
We define a normalized metallicity $Z^{\prime} = Z / Z_\odot$ such that  $Z^{\prime} = 1$ represents solar metallicity.
Similarly,
we define a normalized DGR that scales linearly with metallicity such that $Z_d^{\prime} = Z^{\prime}$
and $Z_d^{\prime} = 1$ corresponds a DGR of 1\%.
The initial gas-phase metal abundances are set to be 
$x_{\rm C,0} = 1.4\times 10^{-4} Z^{\prime}$, 
$x_{\rm O,0} = 3.2\times 10^{-4} Z^{\prime}$ and 
$x_{\rm Si,0} = 1.7\times 10^{-6} Z^{\prime}$ based on \citet{1994ApJ...420L..29C} and \citet{2000ApJ...528..310S}.

We run a set of four simulations at different metallicities $Z^{\prime} = $ 3, 1, 0.3 and 0.1.
Each simulation is run for 600 Myr.
In the initial 100 Myr,
we artificially set the lifetime of massive stars to zero. 
This forces SN feedback to occur intermediately after star formation and in dense environments by construction
which results in a less violent ISM \citep{2015MNRAS.449.1057G, 2015MNRAS.454..238W}.
We do so in order to circumvent the problematic initial transient phase where 
the entire disk is first violently blown off by SN feedback and then slowly falls back to the mid-plane.
This is a simple alternative method to the initial artificial turbulent driving \citep{2015MNRAS.454..238W, 2017ApJ...846..133K} 
or random SN driving \citep{2017MNRAS.471.2151H} in the literature.
As the first 100 Myr is an artificial phase,
there is no need to run it with four different metallicities.
To save computational resources,
simulations with $Z^{\prime} = $ 3, 0.3 and 0.1 are started from $t = 100$ Myr using 
the snapshot in the $Z^{\prime} = 1$ case.

\section{Post-processing Chemistry}\label{sec:postpro}

\subsection{Chemistry network}

Once the simulations are complete,
we apply another much more comprehensive chemistry network in post-processing to calculate the CO abundance for each gas particle.
Our network follows 31 species:
H, H$^-$, H$_2$, H$^+$, H$_2^+$, H$_3^+$, e$^-$, He, He$^+$, HeH$^+$,
C, C$^+$, CO, HCO$^+$,
O, O$^+$, OH, OH$^+$, H$_2$O$^+$, H$_3$O$^+$, H$_2$O,
O$_2$, CO$^+$, O$_2^+$,
CH$_2$, CH$_2^+$, CH, CH$^+$, CH$_3^+$,
Si$^+$ and Si.
We include all chemical reactions in the UMIST database 
\citep{2013A&A...550A..36M} that exclusively involve the above-mentioned species 
(including photoreactions, cosmic-ray ionization and cosmic-ray-induced photoreactions),
which leads to 286 reactions in total.
The rate coefficients are also taken from the UMIST database.
As the UMIST database assumes a constant $\zeta_{\rm CR,u} = 1.2\times 10^{-17} {\rm s^{-1}}$,
we scale the appropriate rate coefficients for all the cosmic-ray processes by a factor of $\zeta_{\rm CR} / \zeta_{\rm CR,u}$ accordingly.
As there are five chemical elements involved in our network,
we can derive six species from the mass and charge conservations without explicitly solving them.
We choose to derive the neutral atomic form of each element and the free electron (i.e., H, He, C, O, Si and e$^-$).

\begin{figure*}
	\centering
	\includegraphics[width=0.99\linewidth]{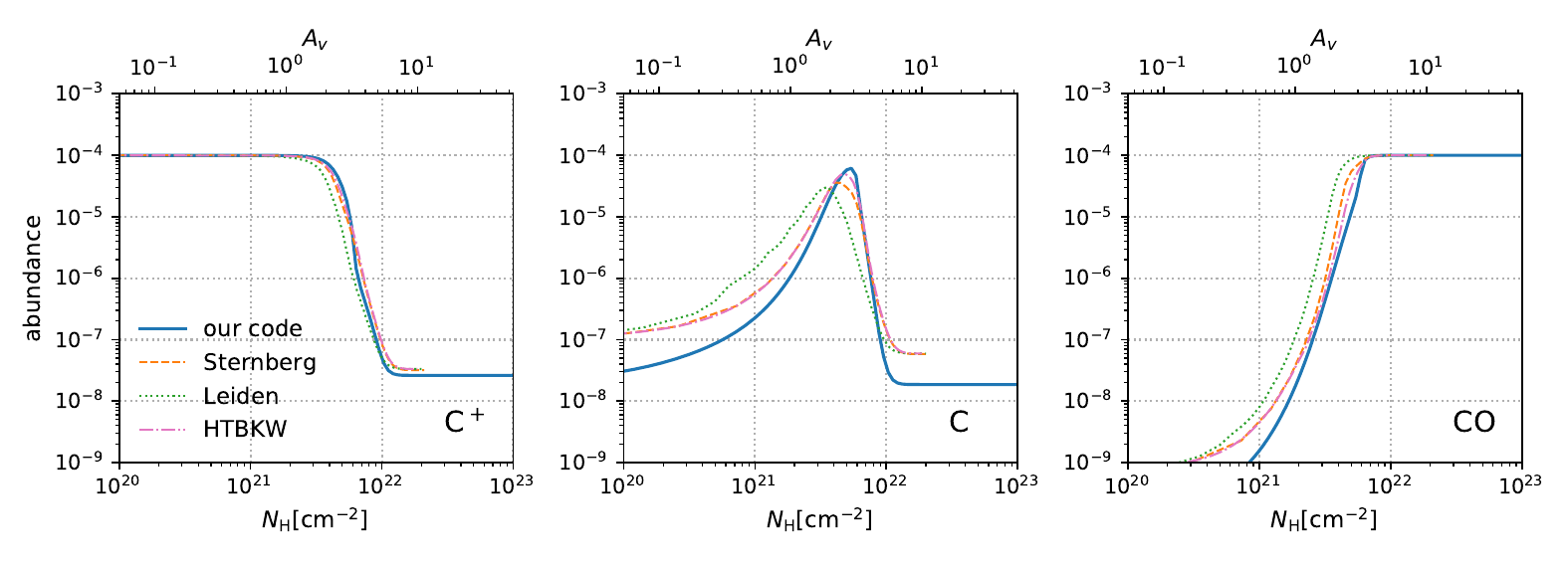}
	\caption{
		Chemical abundances as a function of $N$ (lower $x$-axis) and $A_V$ (upper $x$-axis)
		for C$^+$ (left), C (middle), and CO (right) in a 1D PDR calculation. 
		The solid blue line is from our chemistry network,
		while the others are from three different codes in the code comparison project \citep{2007A&A...467..187R}.
		The physical parameters are
		$n = 1000\ {\rm cm^{-3}}$,
		$T = 50$ K,
		$Z^{\prime} = 1$,
		$I_{\rm UV} = 10$,
		and $\zeta_{\rm CR} = 10^{-16}{\rm s^{-1}}$ (the ``F1'' test).
		Our results agree very well with the others in general,
		demonstrating that our network can accurately capture the C$^+$/C/CO transitions.		
	}
	\label{fig:pdrf1test}
\end{figure*}

In addition to the reactions from the UMIST database,
we also include H$_2$ formation on dust and photodissociation of H$_2$ and CO by the FUV radiation.
Following \citet{1979ApJS...41..555H} and 
assuming a dust temperature of 15 K,
the rate coefficient of H$_2$ formation on dust is
\begin{equation}\label{eq:rH2}
	R_{\rm H_2} =   \frac{ 3\times 10^{-17} \sqrt{T_2} Z_d^{\prime} }{1 + 0.4 \sqrt{T_2 + 0.15} + 0.2 T_2 + 0.08 T_2^2  }\ 
	{\rm cm^{3}s^{-1}},
\end{equation}
where 
$T_2\equiv T/(100\ {\rm K})$.
The H$_2$ photodissociation rate coefficient is
\begin{equation}\label{eq:kH2}
	k_{\rm H_2} = I_{\rm UV} k_{\rm 0,H_2} f_{\rm dust,H_2} f_{\rm ss,H_2},
\end{equation}
where $k_{\rm 0,H_2}$ is the unattenuated rate coefficient for a Draine field,
$f_{\rm dust,H_2}$ is the dust shielding factor
which depends on the visual extinction 
\begin{equation}\label{eq:AV}
	A_V = f_{A_V} N ,
\end{equation}
where $f_{A_V} \equiv 5.35\times 10^{-22} Z_d^{\prime}$ 
and $N$ is the total hydrogen column density.
$f_{\rm ss,H_2}$ is the H$_2$ self-shielding factor which is a function of H$_2$ column density $N_{\rm H_2}$\footnote{
We note that the rate coefficients in Eqs. \ref{eq:rH2} and \ref{eq:kH2} are 
similar to what our time-dependent H$_2$ chemistry network adopts
and are only used in our fully steady-state H$_2$ model (see Sec. \ref{sec:td_vs_ss}).}.
Similarly,
the CO photodissociation rate coefficient is
\begin{equation}
k_{\rm CO} = I_{\rm UV} k_{\rm 0,CO} f_{\rm dust,CO} f_{\rm gas,CO},
\end{equation}
where $k_{\rm 0,CO}$ is the unattenuated rate coefficient,
$f_{\rm dust,CO}$ is the dust shielding factor 
and $f_{\rm gas,CO}$ is the shielding factor from H$_2$ shielding and CO self-shielding, 
which is a function of both $N_{\rm H_2}$ and $N_{\rm CO}$ (CO column density). 
Following \citet{2017A&A...602A.105H}\footnote{\url{https://home.strw.leidenuniv.nl/~ewine/photo/index.html}},
we adopt 
the two-sided slab model where
$k_{\rm 0,H_2} = 5.68\times 10^{-11}\ {\rm s^{-1}}$ \citep{2014ApJ...790...10S},
$k_{\rm 0,CO} = 2.43\times 10^{-10}\ {\rm s^{-1}}$,
$f_{\rm dust,H_2} = \alpha_{\rm H_2}\exp(-\gamma_{\rm H_2} A_V)$ where $\alpha_{\rm H_2} = 0.52$ and $\gamma_{\rm H_2} = 3.85$,
and  
$f_{\rm dust,CO} = \alpha_{\rm CO}\exp(-\gamma_{\rm CO} A_V)$ where $\alpha_{\rm CO} = 0.48$ and $\gamma_{\rm CO} = 3.51$.
For the self-shielding factors,
we follow \citet{1996ApJ...468..269D} for $f_{\rm ss,H_2}$ 
(assuming an H$_2$ velocity dispersion of 2 ${\rm km\ s^{-1}}$)
and \citet{2009A&A...503..323V} for $f_{\rm gas,CO}$ 
(assuming a CO excitation temperature of 5 K and a CO linewidth of 0.3 km s$^{-1}$).
Recombination on grains 
(X$^+$ +  gr $\rightarrow$ X + gr$^+$, where X is a chemical element and gr stands for grains)
for H, He, C and Si are also included following \citet{2001ApJ...563..842W} 
(see the discussion in Appendix \ref{app:grRec}).
We use the solver {\sc Differentialequations.jl} \citep{rackauckas2017} to integrate the system of rate equations.

As a validation of our chemistry network,
we perform the ``F1'' test from the code comparison project in \citet{2007A&A...467..187R}.
The test consists of a one-dimensional (1D) calculation of a photon-dominated region (PDR)
with $n = 1000\ {\rm cm^{-3}}$,
$T = 50$ K,
$Z^{\prime} = 1$,
$I_{\rm UV} = 10$,
and $\zeta_{\rm CR} = 10^{-16}\ {\rm s^{-1}}$.
The initial elemental abundances relative to hydrogen are, respectively,
$x_{\rm He,0} = 0.1$,
$x_{\rm C,0} = 1.0\times 10^{-4}$, 
$x_{\rm O,0} = 3.0\times 10^{-4}$, 
and $x_{\rm Si,0} = 0$.
Recombination on grains is switched off in order to be consistent with \citet{2007A&A...467..187R}.
Fig.~\ref{fig:pdrf1test} shows the chemical abundances as a function of $N$ (lower $x$-axis) and $A_V$ (upper $x$-axis)
for C$^+$ (left), C (middle), and CO (right).
The solid blue line is from our chemistry network,
while the others are from three different codes (among many others) participated in \citet{2007A&A...467..187R}:
\textit{Sternberg}
in orange dashed \citep{1995ApJS...99..565S, 2005ApJ...632..302B}
\textit{Leiden} in green dotted \citep{1987ApJ...322..412B, 1988ApJ...334..771V},
and
\textit{HTBKW} in pink dash-dotted \citep{1985ApJ...291..722T, 1999ApJ...527..795K}.
Our results agree very well with these codes in general,
demonstrating that our network can accurately capture the C$^+$/C/CO transitions in a 1D uniform medium.

\subsection{Time-dependent vs. steady-state H$_2$ model}\label{sec:td_vs_ss}

For each simulation snapshot,
we run our network for every gas particle up to steady state (run for 1 Gyr),
taking the local gas density and temperature as input parameters.
The same metallicity and solar abundances ($x_{\rm C,0} = 1.4\times 10^{-4} Z^{\prime}$, 
	$x_{\rm O,0} = 3.2\times 10^{-4} Z^{\prime}$ and 
	$x_{\rm Si,0} = 1.7\times 10^{-6} Z^{\prime}$) are adopted as those in the simulations,
as well as the time-dependent $I_{\rm UV}$ and $\zeta_{\rm CR}$ which scales with the current SFR 
(see Sec. \ref{sec:coolchem}).
However,
since $x_{\rm H_2}$ and $x_{\rm H^+}$ are already calculated from our time-dependent hydrogen network,
these two species are used as input parameters\footnote{In the simulations,
$x_{\rm H_2}$ and $x_{\rm H^+}$ are solved explicitly while $x_{\rm H}$ is derived from mass conservation.
However, 
we cannot use all three time-dependent abundances from the simulations as given,
as otherwise there would be no room for other species involving hydrogen (e.g. CH, OH, etc.).}.
By doing so,
we allow H$_2$ to be out of steady state
which happens when 
the dynamical time controlled by turbulence motions or cloud destruction is shorter than
the H$_2$ formation time 
\begin{equation}\label{eq:tform}
t_{\rm H_2,form} = \frac{1}{2R_{\rm H_2} n}\approx 15\ {\rm Myr}\frac{1}{n_2 Z_d^{\prime}},
\end{equation}
where $n_2\equiv n / ({\rm 100\ cm^{-3}})$.
All the other species, including C$^+$, C and CO, are evolved up to steady state.
As shown in Appendix \ref{app:chemtime},
the chemical timescales for C$^+$, C and CO to reach steady state are short compared to the dynamical time in our simulations,
justifying our steady-state assumption.
We will refer to these results as our \textit{time-dependent H$_2$ model},
which is our default model.

For comparison purposes,
we run another set of post-processing 
without using $x_{\rm H_2}$ and $x_{\rm H^+}$ from the simulations 
and obtain steady-state solutions for all species.
We will refer to these results as our \textit{steady-state H$_2$ model},
which, while less realistic than the time-dependent one, can provide useful insights. 
Note that the CO abundance can be different in these models 
despite being in steady state in both cases 
because (i) CO formation is initiated by H$_2$ and 
(ii) H$_2$ provides shielding for CO.


\subsection{Radiation shielding and column densities}\label{sec:shieldcolumn}
Radiation shielding is accounted for using the same {\sc HealPix}-based method as in the time-dependent network (see Sec. \ref{sec:coolchem}).
Because of self-shielding,
local abundances of H$_2$ and CO depend on those from a distance and therefore a few iterations are required.
We find that three iterations are enough to obtain converged results (see Appendix \ref{app:chem_iter}).

We distinguish between the \textit{observed} total hydrogen column density projected along the $z$-axis $N^{\rm obs}$ 
(e.g., as shown in Fig.~\ref{fig:loadmolmapsneq48}) 
and the angle-averaged \textit{effective} total hydrogen column density relevant for dust shielding against FUV radiation.
The former is obtained by mapping the particle information onto a Cartesian mesh with a mesh size of 2 pc.
The latter is defined as 
$N^{\rm eff} = A_V^{\rm eff} / f_{A_V}$ and $A_V^{\rm eff}$ is the \textit{effective} visual extinction
such that $\exp(- \gamma A^{\rm eff}_V) = \sum_{i=1}^{N_{\rm pix}} ( \exp(- \gamma A_{V,i})  )$
where 
$A_{V,i} = f_{A_V} N_i$,
$N_{\rm pix}$ is the number of {\sc HealPix} pixels,
and
$N_i$ is the total hydrogen column density along a {\sc HealPix} pixel $i$ integrated up to $L_{\rm sh}$.
We adopt $N_{\rm pix} = 12$ in this work.
Therefore,
it follows
\begin{equation}\label{eq:Neff}
	A_V^{\rm eff} \equiv -\frac{1}{\gamma} \ln\Bigg[ \frac{1}{N_{\rm pix}}
	\sum_{i=1}^{N_{\rm pix}} 
	\Big(   \exp(- \gamma A_{V,i})  \Big) \Bigg],
\end{equation}
where the free parameter $\gamma$ is chosen to be 3.51
such that $\exp(- \gamma A^{\rm eff}_V)$ represents exactly the dust shielding factor for CO photodissociation\footnote{
This is not exactly the case for dust shielding against H$_2$ photodissociation where 
$\gamma = 3.85$ but the difference is rather small.}.
Note that $N^{\rm eff}$ is a 3D property associated with each gas particle,
while $N^{\rm obs}$ is a 2D property associated with each sightline in projection.
The latter is directly observable but
may include material which happens to be in the sightline but is not physically relevant for shielding.

Another angle-averaged column density we use in Sec. \ref{sec:nHvsNH} is the geometric mean
$\langle N \rangle \equiv (\prod_{i=1}^{N_{\rm pix}} N_i )^{1/N_{\rm pix}} $,
which is, equivalently, the arithmetic average in log-space:
\begin{equation}\label{eq:Nave}
	\log_{10} \langle N \rangle \equiv 
	\frac{1}{N_{\rm pix}} \sum_{i=1}^{N_{\rm pix}} \log_{10}N_i.
\end{equation}
It has the advantage of not being strongly biased towards high column densities as is the case for the arithmetic average in linear-space.
Note that $\langle N \rangle$ depends only on the cloud structures and is independent of chemistry.



\section{Results} \label{sec:result}

\subsection{Time evolution}

\begin{figure*}
	\centering
	\includegraphics[trim = 0cm 0cm 0cm 0cm, clip, width=0.99\linewidth]{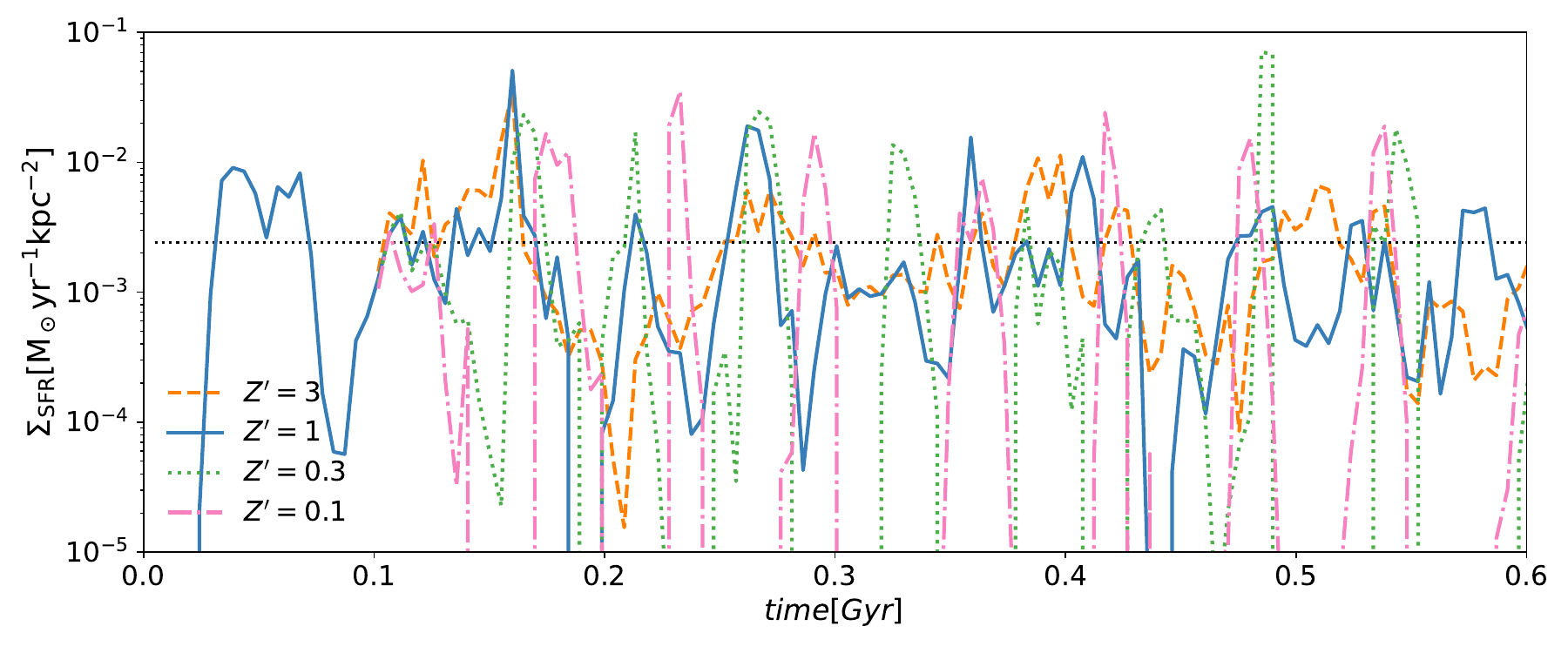}
	\caption{
		Time evolution of the star formation rate (SFR) surface density averaged over the entire simulation domain (face-on).
		Our $Z^{\prime} = 1$ run agrees very well with the observed value in the solar neighborhood 
		from \citet{2009AJ....137..266F} 
		shown as the horizontal black dotted line.
		The time-averaged SFR is insensitive to $Z^{\prime}$ (see also Table \ref{tab:sumstats}),
		but the temporal fluctuation of SFR significantly increases as $Z^{\prime}$ decreases.
	}
	\label{fig:sfrtime}
\end{figure*}

\begin{figure}
	\centering
	\includegraphics[width=0.99\linewidth]{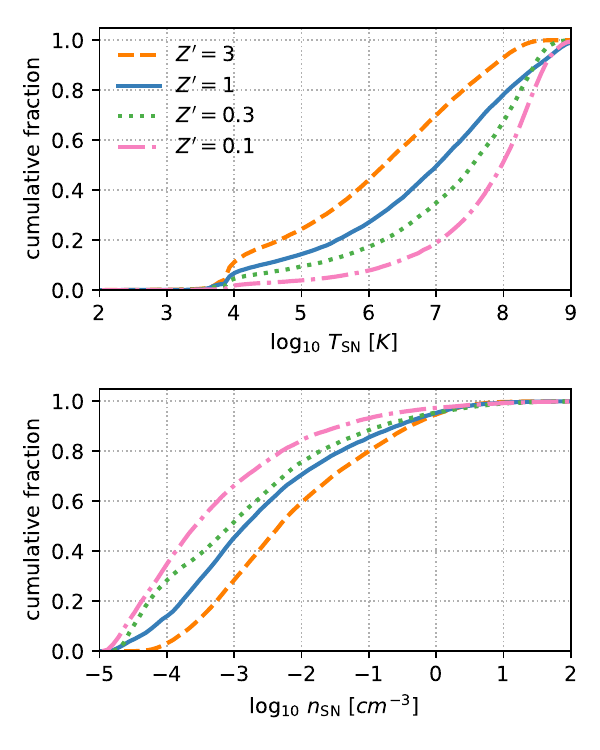}
	\caption{
		Cumulative distribution of the gas temperature ($T_{\rm SN}$, top panel) and hydrogen number density ($n_{\rm SN}$, bottom panel) where SNe occur.
		The vast majority of SNe go off in hot ($T_{\rm SN} > 10^5$ K) and under-dense ($n_{\rm SN} < 0.1 {\rm cm^{-3}}$) regions,
		indicating that SNe are clustered and frequently go off in super-bubbles.
	}
	\label{fig:cdf_TSN_nSN}
\end{figure}

\begin{figure}
	\centering
	\includegraphics[width=0.99\linewidth]{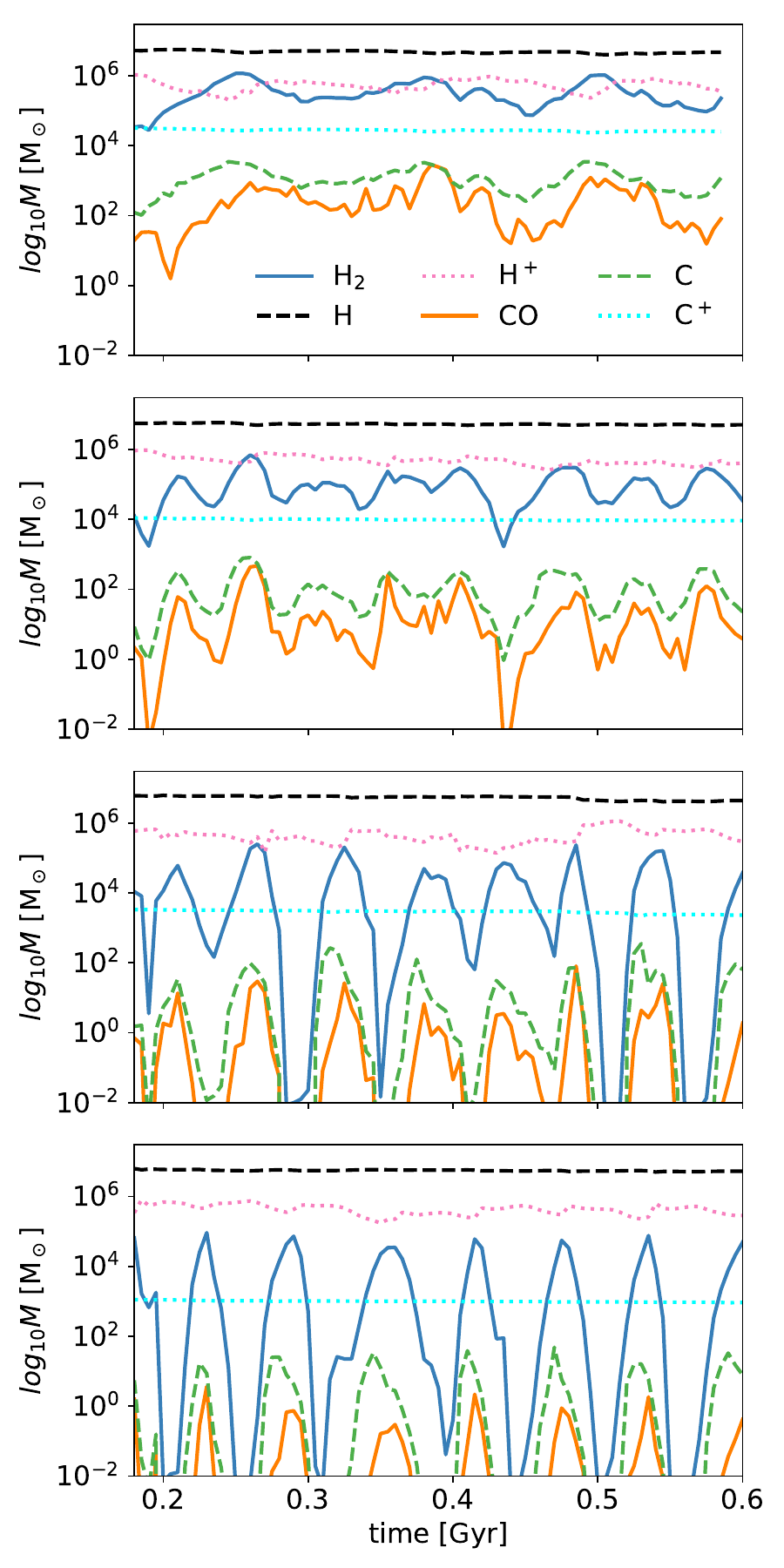}
	\caption{
		Total masses of H$^+$, H, H$_2$, C$^+$, C and CO in the time-dependent model as a function of time 
		for $Z^{\prime} = $ 3, 1, 0.3 and 0.1 from top to bottom.
	}
	\label{fig:timehphih2cpcicoall}
\end{figure}

Fig.~\ref{fig:sfrtime} shows the time evolution of the SFR surface density 
averaged over the entire simulation domain in the face-on projection
at different metallicities.
Our $Z^{\prime} = 1$ run agrees very well with the observed value in the solar neighborhood,
$\Sigma_{\rm SFR,0} = 2.4\times 10^{-3}\ {\rm M_\odot\ yr^{-1}\ kpc^{-2}}$ (\citealp{2009AJ....137..266F},
shown as the horizontal black dotted line).
As demonstrated in several previous studies,
the key to this success is the proper inclusion of stellar feedback (see, e.g., \citealp{2017MNRAS.466.1903G, 2017ApJ...846..133K, 2018MNRAS.480..800H}).
In fact,
the time-averaged SFR is rather insensitive to $Z^{\prime}$ (see also Table \ref{tab:sumstats}).
However,
the temporal fluctuation of SFR is significantly larger at lower $Z^{\prime}$,
partly because the fine-structure metal-line cooling rate is linearly proportional to $Z^{\prime}$.
Once the cold clouds are destroyed and heated up by stellar feedback,
it takes longer for the gas to cool down from $10^4$ K, 
which delays the gravitational collapse and star formation.
Another reason for the burstiness is that
star formation occurs at increasingly higher densities as $Z^{\prime}$ decreases 
because the clouds are slightly warmer 
(as will be shown below).
Consequently,
the SFR is elevated locally, which explains the higher peaks,
and the subsequent stellar feedback is more clustered and destroys clouds more efficiently.
The absolute burstiness is expected to depend on the box size,
as different patches of the ISM may be in different phases of the fluctuation in a real galaxy.
However, the systematic trend with metallicity should be robust.

	Fig.~\ref{fig:cdf_TSN_nSN} shows the cumulative distribution of the gas temperature ($T_{\rm SN}$, top panel) 
	and hydrogen number density ($n_{\rm SN}$, bottom panel) where SNe occur at different metallicities.
	The vast majority of SNe explode in hot ($T_{\rm SN} > 10^5$ K) and under-dense ($n_{\rm SN} < 0.1 {\rm cm^{-3}}$) regions,
	indicating that SNe are clustered and frequently go off in super-bubbles.
	Furthermore, 
	as metallicity decreases,
	$T_{\rm SN}$ increases while $n_{\rm SN}$ decreases.
	SNe are more clustered at lower metallicities, which leads to more efficient feedback and thus burstier SFRs.

Fig.~\ref{fig:timehphih2cpcicoall} shows the total masses of H$^+$, H, H$_2$, C$^+$, C and CO as a function of time 
for $Z^{\prime} = $ 3, 1, 0.3 and 0.1 from top to bottom.
Most hydrogen is in the form of H while most carbon is in the form of C$^+$,
both of which remain almost constant with time.
In contrast,
the masses of H$_2$, C and CO show significant time variations as they trace the cold and dense gas
and their modulation is correlated with the SFR.
The H$^+$ mass anti-correlates with the cold-gas tracers (H$_2$, C and CO) as H$^+$ is generated primarily by supernovae and photoionization.
Therefore, it peaks at the ``destruction'' phase of the ISM cycles when the cold-gas tracers are at their minima.

\begin{figure}
	\centering
	\includegraphics[trim = 3cm 0cm 3cm 0cm, clip, width=0.99\linewidth]{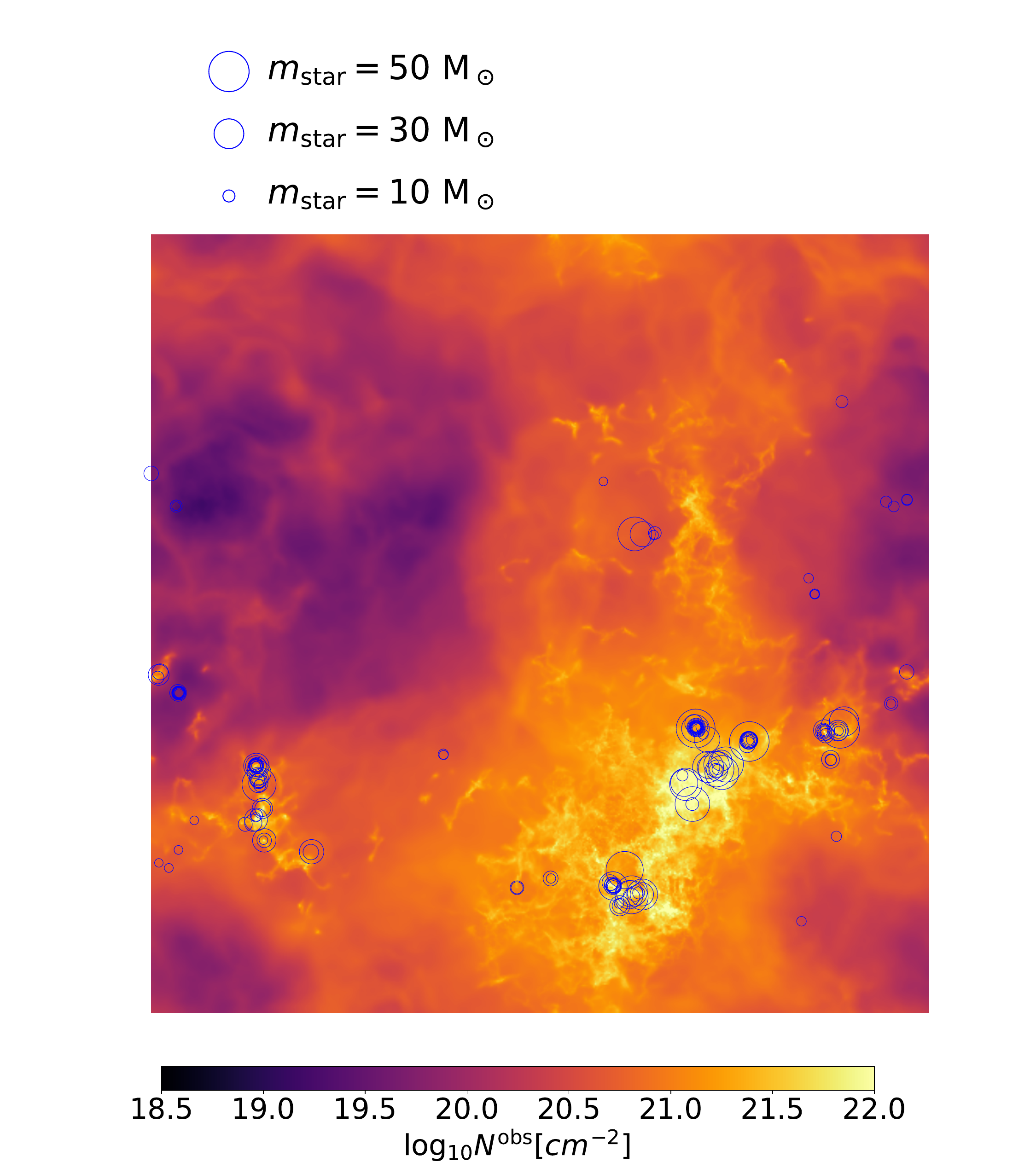}
	\caption{
		Column density map of total hydrogen at $t = 530$ Myr in the $Z^{\prime} = 1$ run (solar metallicity).
		Massive stars are overplotted as the blue circles (sizes are scaled with the stellar masses).
	}
	\label{fig:loadntotmapswithstarsneq620}
\end{figure}

\begin{figure*}
	\centering
	\includegraphics[trim = 0cm 1cm 0cm 0cm, clip, width=1.0\linewidth]{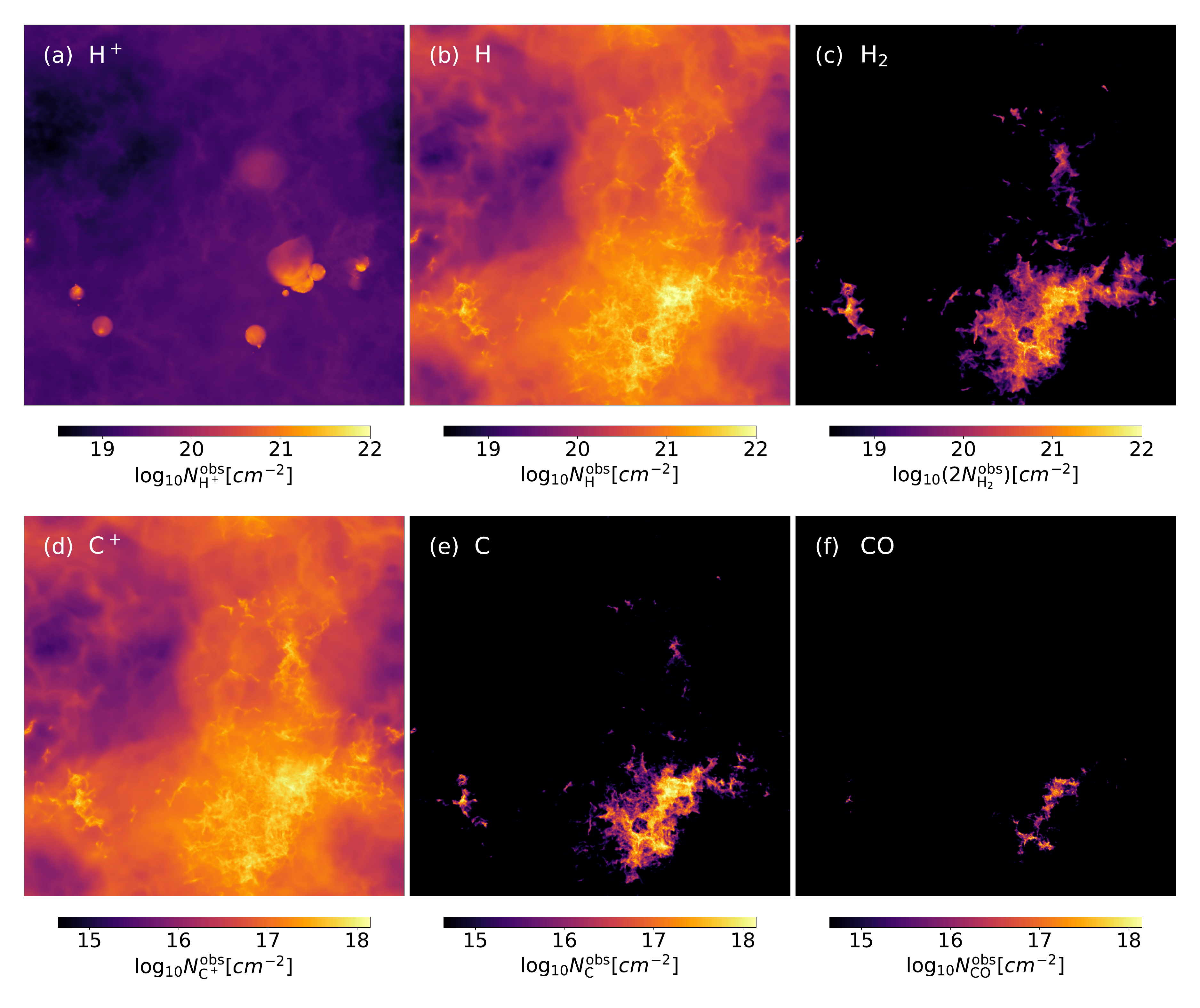}
	\caption{
		Column density maps of 
		(a) ionized hydrogen,
		(b) atomic hydrogen,
		(c) molecular hydrogen,
		(d) ionized carbon,
		(e) atomic carbon, and
		(f) carbon monoxide 
		at $t = 530$ Myr in the $Z^{\prime} = 1$ run (solar metallicity).
	}
	\label{fig:loadmolmapsneq48}
\end{figure*}

For a visual impression,
Fig.~\ref{fig:loadntotmapswithstarsneq620}
shows the 
observed column density maps of total hydrogen at $t = 530$ Myr in the $Z^{\prime} = 1$ run,
which is designed to be a model for the solar neighborhood.
The entire simulation domain of 1 kpc$^2$ is shown.
A multi-phase hierarchical structure is evident,
and consists of dense molecular clouds, a diffuse medium and SN-driven cavities.
Massive stars are overplotted as the blue circles (sizes are scaled with the stellar masses).
They are spatially clustered, especially for the more massive ones,
and mostly located within the dense clouds.
Fig.~\ref{fig:loadmolmapsneq48} shows the 
observed column density maps of H$^+$, H, H$_2$, C$^+$, C and CO from panel (a) to (f), respectively,
for the same snapshot as Fig.~\ref{fig:loadntotmapswithstarsneq620}.
Recent feedback events are traced by the small bubbles in the H$^+$ map.
Carbon is mostly in the form of C$^+$ in the volume-filling diffuse medium.
H$_2$ is a tracer for dense clouds while CO traces the even denser part of the clouds.
The atomic carbon is an intermediate phase between C$^+$ and CO.
Qualitatively,
the chemical properties of the ISM are broadly consistent with standard PDR calculations \citep{1988ApJ...334..771V, 1995ApJS...99..565S, 2007A&A...467..187R, 2017ApJ...839...90B}.

Given the strong temporal variations,
any analysis for a particular snapshot will be biased and incomplete.
As such,
in the following sections,
all results we present will be ``time-averaged''
over the snapshots between 150 and 600 Myr with a time interval of 5 Myr (91 snapshots in total) for each simulation.
For a histogram (either 1D or 2D),
time-averaging is simply done bin by bin over all snapshots.
A cumulative distribution is generated from the time-averaged histogram 
(which is different from time-averaging 91 cumulative distributions bin-wise).
A correlation plot between $x$ and $y$ that shows
the median of $y$ in each $x$-bin by a line (and sometimes also the scatter of $y$ by a shaded area)
is constructed from a time-averaged 2D histogram of $x$ and $y$.
By doing so,
the temporal and spatial scatters are treated on an equal footing and can be shown simultaneously.
Each snapshot can also be viewed equivalently as a different patch of a galaxy.

\subsection{Thermodynamical properties}

\begin{figure*}
	\centering
	\includegraphics[width=0.99\linewidth]{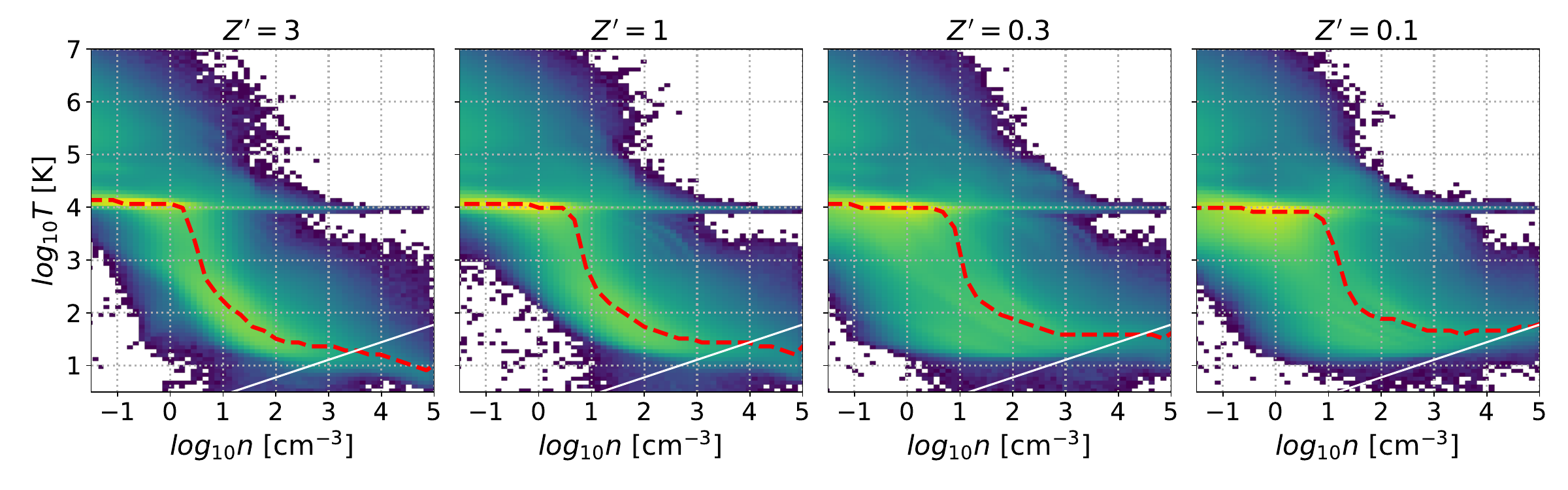}
	\caption{
		Time-averaged 2D histograms of density vs. temperature 
		for $Z^{\prime} = $ 3, 1, 0.3 and 0.1 from left to right.
		The red dashed lines represent the median temperature within each density bin.
		The white solid lines indicate the star formation threshold where $M_{\rm J} = 32\ {\rm M_\odot}$.
		The slightly higher temperature in the cold phase at lower $Z^{\prime}$ 
		shift star formation to higher densities.
	}
	\label{fig:pdmedian}
\end{figure*}

Fig.~\ref{fig:pdmedian} shows the time-averaged
2D histograms of $n$ vs. $T$ for $Z^{\prime} = $ 3, 1, 0.3 and 0.1 from left to right.
The red dashed lines represent the median temperature within each density bin.
The white solid lines indicate the star formation threshold where $M_{\rm J} = 32\ {\rm M_\odot}$.
The overall gas distribution broadly follows the classical curves for bistable warm/cold thermal equilibrium.
The scatter increases with decreasing $Z^{\prime}$ as the bursty SFR implies large temporal variations of $I_{\rm UV}$ and $\zeta_{\rm CR}$, 
which in turn alter the thermal-equilibrium curves.
In addition,
the hot gas ($T > 10^5$ K) is generated by SN feedback
while the narrow gas distribution at $T = 10^4$ K and $n_{\rm H} >10\ {\rm cm^{-3}}$ 
originates from photoionization in the HII regions.
The majority of gas is in the warm and diffuse phase ($T\sim 10^4$ K and $n \sim 1\ {\rm cm^{-3}}$) 
whose temperature is rather insensitive to $Z^{\prime}$ as it is controlled by the Lyman alpha cooling.
In contrast,
the cold and dense phase ($T< 10^2$ K and $n > 10\ {\rm cm^{-3}}$) 
has a typical temperature that is slightly higher at lower $Z^{\prime}$.
This is partly due to less efficient dust shielding that enhances photoelectric heating,
and, at the highest densities, due to heating from UV pumping and H$_2$ formation 
\citep{2019ApJ...881..160B}.
As such,
gas has to collapse to higher densities to reach the same level of gravitational instability (quantified by $M_{\rm J}$) to form stars,
making the SFR burstier at lower $Z^{\prime}$ as discussed above.

\begin{figure*}
	\centering
	\includegraphics[trim = 2cm 0cm 3cm 0cm, clip, width=0.95\linewidth]{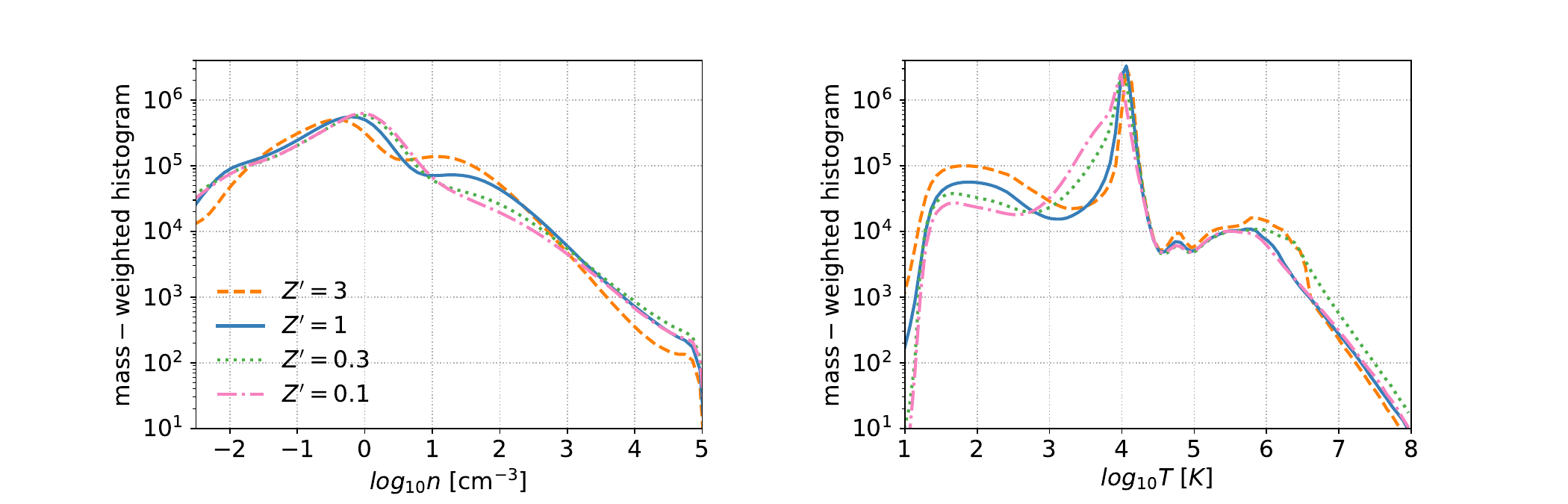}
	\caption{
		Time-averaged mass-weighted histograms of density (left) and temperature (right) at different metallicities.
		The two-phase feature due to thermal instability is more pronounced at high $Z^{\prime}$.		
	}
	\label{fig:densitypdf}
\end{figure*}

To be more quantitative,
Fig.~\ref{fig:densitypdf} shows the time-averaged mass-weighted histograms of 
$n$ (left) and $T$ (right)
at different metallicities.
The density histogram peaks at $n \sim 1\ {\rm cm^{-3}}$ which represents the typical density of the diffuse background medium.
At higher metallicities ($Z^{\prime} = $ 3 and $Z^{\prime} = $ 1),
the two-phase feature due to thermal instability \citep{1969ApJ...155L.149F, 2003ApJ...587..278W, 2019ApJ...881..160B} is clearly visible 
with two distinct peaks at $T\sim 10^4$ K and $T\sim 10^2$ K in the temperature histogram.
At lower metallicities ($Z^{\prime} = $ 0.3 and $Z^{\prime} = $ 0.1),
the cold phase is less prominent 
and more gas can be found in the unstable intermediate temperature regime.
This occurs because the cooling time becomes comparable to or longer than the dynamical time at low $Z^{\prime}$
and so the gas is constantly driven away from its thermally stable phases.

\begin{figure*}
	\centering
	\includegraphics[trim = 0cm 1cm 0cm 1cm, clip, width=0.99\linewidth]{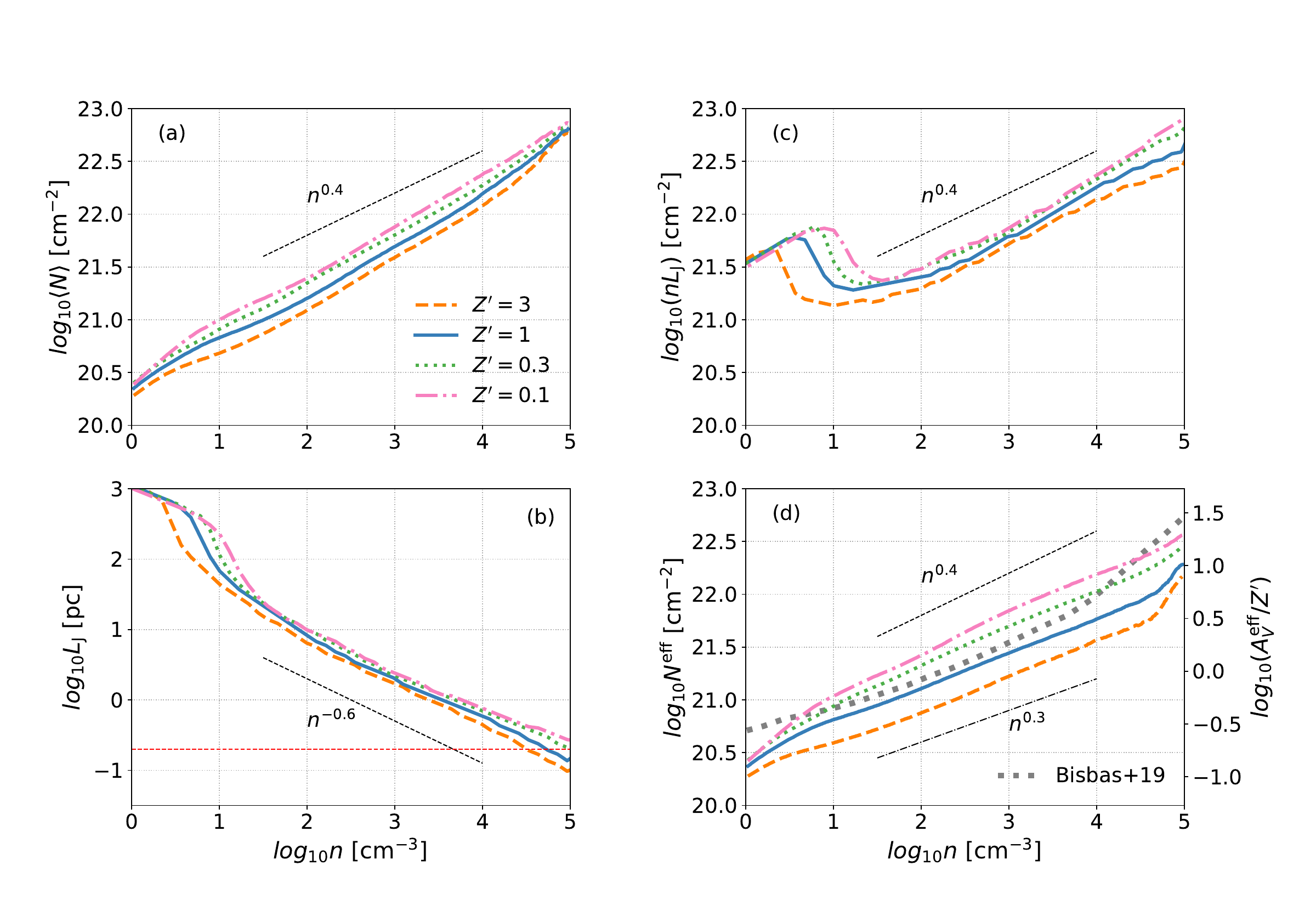}
	\caption{
		(a)
		Angle-averaged (geometric mean) hydrogen column density
		$\langle N \rangle$ as a function of $n$
		at different metallicities.
		The black dashed line indicates a power-law relation of $n^{0.4}$.
		(b)
		Same as (a), but showing the local Jeans length $L_{\rm J}$ on the $y$-axis.
		The black dashed line indicates a power-law relation of $n^{-0.6}$,
		and the horizontal red dashed line shows 
		the adopted gravitational softening length $\epsilon_{\rm soft} = $ 0.2 pc.
		(c)
		Same as (a), but showing $n L_{\rm J}$ on the $y$-axis.
		(d)
		Same as (a), but showing the effective column density $N^{\rm eff}$ 
		(i.e., angle-averaged via Eq. \ref{eq:Neff}) on the $y$-axis.		
		The corresponding $A_V/Z^{\prime}$ is shown on the right $y$-axis.		
		The grey dotted line shows the average of four independent hydrodynamical simulations with solar-metallicity 
		in the literature
		as compiled by \citet{2019MNRAS.485.3097B}.
		All quantities are time-averaged.
	}
	\label{fig:nhvsnh}
\end{figure*}

\subsection{Correlation between $n$ and $N^{\rm eff}$}\label{sec:nHvsNH}

The amount of available shielding material against the FUV radiation is an important factor for the formation of H$_2$ and CO.
Panel (a) in Fig.~\ref{fig:nhvsnh} shows 
the angle-averaged geometric mean of the hydrogen column density
$\langle N \rangle$ (see Eq. \ref{eq:Nave}) as a function of $n$.
The geometric mean 
roughly scales as $n^{0.4}$ (the black dashed line)
and is insensitive to $Z^{\prime}$ besides a slight difference in normalization.
Panel (b) in Fig.~\ref{fig:nhvsnh} shows 
the thermal Jeans length 
$L_{\rm J} = \sqrt{\pi c_s^2 / (G\rho) }$ 
as a function of $n$.
Our adopted gravitational softening length $\epsilon_{\rm soft} = $ 0.2 pc is shown by the horizontal red dashed line in panel (b),
which represents the minimum length scale at which the gravity can be resolved.
All snapshots are combined and the median values are shown.

In the cold and dense phase $n> 30\ {\rm cm^{-3}}$,
the temperature decreases very slowly as density increases (``isothermal regime''),
which leads to $L_{\rm J} \propto n^{-0.6}$.
Assuming the Jeans length is the typical cloud size at a given density,
we can estimate the hydrogen column density by
$n L_{\rm J}$,
which is shown in panel (c) of Fig.~\ref{fig:nhvsnh} as a function of $n$
and agrees well with $\langle N \rangle$ in panel (a)
in the range of 
$10 < n < 10^4\ {\rm cm^{-3}}$,
suggesting that 
the correlation between $n$ and $\langle N \rangle$ is a consequence of gravitational instability.
This Jeans-length argument has been demonstrated by \citet{2013MNRAS.430.2427R}\footnote{This is in the context of atomic hydrogen self-shielding against the meta-galactic UV radiation.} and \citet{2017MNRAS.465..885S}.
However,
we caution that \citet{2010MNRAS.404....2G} found a similar correlation in their turbulent box simulations without self-gravity,
implying that the Jeans-length argument might also be a coincidence.
The correlation is insensitive to $Z^{\prime}$ because the relationship between density and temperature (and thus the Jeans length) is insensitive to $Z^{\prime}$ (see Fig.~\ref{fig:pdmedian}). 
At a given density,
$\langle N \rangle$ is only slightly larger at lower $Z^{\prime}$ as the temperature is higher.
At $n < 10\ {\rm cm^{-3}}$,
there is a sudden increase in $n L_{\rm J}$ corresponding to the abrupt transition 
from $T\sim 100$ K to $T\sim 10^4$ K.
The same jump is not seen in $\langle N \rangle$,
as the warm and diffuse medium is gravitationally stable against its self-gravity 
(otherwise the entire disk would have been collapsing)
and thus gas does not cluster on a scale of $L_{\rm J}$.

Note that our adopted shielding length $L_{\rm sh} = 100$ pc is much larger than the Jeans length of the clouds.
Therefore,
we expect $n L_{\rm J}$ to be a lower limit of $\langle N \rangle$
even if $L_{\rm J}$ is a good proxy of the typical cloud size,
as material beyond $L_{\rm J}$ and up to $L_{\rm sh}$ to can still make a contribution.
The good agreement found between $n L_{\rm J}$ and $\langle N \rangle$ suggests that 
this contribution is not very important and the majority of shielding occurs in the vicinity of the clouds.
We explore different choices of $L_{\rm sh}$ in Appendix \ref{app:Lsh}.

On the other hand,
in terms of dust shielding of H$_2$ and CO,
the more relevant quantity is 
the effective column density $N^{\rm eff}$ (see Eq. \ref{eq:Neff}) rather than $\langle N \rangle$.
Although $N^{\rm eff}$ is also an angle average like $\langle N \rangle$,
it has a different weighting that is biased towards pixels with the lowest column densities.
Therefore, $N^{\rm eff}$ roughly corresponds to the optical depth of a cloud.
From Eq. \ref{eq:Neff},
pixels with $A_{V,i} \gg 1$ where dust shielding becomes effective are
significantly down-weighted,
which corresponds to a lower $N_i$ at higher $Z^{\prime}$.
Therefore,
in Panel (d) of Fig.~\ref{fig:nhvsnh} where we show $n$ vs. $N^{\rm eff}$,
we see the curves are down-shifted and flattened compared to panel (a),
to a larger extent at higher $Z^{\prime}$.
The grey dotted line shows the average of four independent hydrodynamical simulations at solar-metallicity in the literature, 
\citet{2010MNRAS.404....2G},
\citet{2013ApJ...764...36V},
\citet{2017MNRAS.465..885S} and
\citet{2017MNRAS.472.4797S},
as compiled by \citet{2019MNRAS.485.3097B}.
Our $Z^{\prime} = 1$ run agrees very well with those simulations\footnote{
	The agreement is further improved if we adopt $\gamma = 2.5$ as used in those studies instead of $\gamma = 3.51$.
}
in the range of $10 < n < 10^4\ {\rm cm^{-3}}$.
Above this density range,
the Jeans length drops below our resolution limit and thus gravity becomes underestimated (softened),
which presumably explains the discrepancy.
Quantitatively,
we find the correlation can be well-described by 
\begin{equation}\label{eq:nvsN}
	N^{\rm eff} = A (n/{\rm cm^{-3}})^\alpha
\end{equation}
where $A/(10^{20}{\rm cm^{-2}} ) = $ 2, 3, 4 and 4.5
while $\alpha = $ 0.3, 0.33, 0.36 and 0.39
for $Z^{\prime} = $ 3, 1, 0.3 and 0.1, respectively.

\begin{deluxetable*}{cccccccccc}
	\tablecaption{
		Densities, effective column densities and effective visual extinctions (see Eq. \ref{eq:Neff}) at which 
		H/H$_2$, C$^+$/C and C/CO conversions occur in the time-dependent H$_2$ model.
	}
	\label{tab:nNtrans}
	\tablewidth{0pt}
	\tablehead{
		\colhead{$Z^{\prime}$} & 
		\colhead{$n^*_{\rm H_2}$}  & 
		\colhead{$n^*_{\rm C}$}  & 
		\colhead{$n^*_{\rm CO}$}  & 
		\colhead{$N^*_{\rm H_2}$} &
		\colhead{$N^*_{\rm C}$} &
		\colhead{$N^*_{\rm CO}$} &
		\colhead{$A^*_{V,{\rm H_2}}$  } &
		\colhead{$A^*_{V,{\rm C}}$  } &
		\colhead{$A^*_{V,{\rm CO}}$  } \\
		\colhead{ } & 
		\colhead{$[{\rm cm^{-3}}]$}  & 
		\colhead{$[{\rm cm^{-3}}]$}  & 
		\colhead{$[{\rm cm^{-3}}]$}  & 
		\colhead{$[{\rm 10^{21} cm^{-2}}]$  } &
		\colhead{$[{\rm 10^{21} cm^{-2}}]$  } &
		\colhead{$[{\rm 10^{21} cm^{-2}}]$  } &
		\colhead{ } &
		\colhead{ } &
		\colhead{ } 
	}
	\decimalcolnumbers
	\startdata
	3    &   90     &     68    &     720   &  0.88  &  0.78  &  1.7  &  1.4  &  1.2  &  2.7\\
	1    &   470    &    270    &    3100   &  2.9   &  2.1   &  4.6  &  1.5  &  1.1  &  2.5\\
	0.3  &   3300   &   1400    &   14000   &  9.4   &  7.2   &   14  &  1.5  &  1.2  &  2.2\\
	0.1  &   16000  &   6800    &   69000   &  22    &  16    &   34  &  1.2  &  0.87 &  1.8\\
	\enddata
	\tablecomments{	
		(1) Normalized metallicity; 
		(2)-(4) Conversion densities for H/H$_2$, C$^+$/C, and C/CO, respectively.
		(5)-(7) Conversion effective column densities for H/H$_2$, C$^+$/C, and C/CO, respectively.
		(8)-(10) Conversion effective visual extinctions for H/H$_2$, C$^+$/C, and C/CO, respectively.
	}
\end{deluxetable*}

\begin{deluxetable*}{cccccccccc}
	\tablecaption{Same as Table \ref{tab:nNtrans} but in the steady-state H$_2$ model.}
	\label{tab:nNtrans_eq}
	\tablewidth{0pt}
	\tablehead{
		\colhead{$Z^{\prime}$} & 
		\colhead{$n^*_{\rm H_2}$}  & 
		\colhead{$n^*_{\rm C}$}  & 
		\colhead{$n^*_{\rm CO}$}  & 
		\colhead{$N^*_{\rm H_2}$} &
		\colhead{$N^*_{\rm C}$} &
		\colhead{$N^*_{\rm CO}$} &
		\colhead{$A^*_{V,{\rm H_2}}$  } &
		\colhead{$A^*_{V,{\rm C}}$  } &
		\colhead{$A^*_{V,{\rm CO}}$  } \\
		\colhead{ } & 
		\colhead{$[{\rm cm^{-3}}]$}  & 
		\colhead{$[{\rm cm^{-3}}]$}  & 
		\colhead{$[{\rm cm^{-3}}]$}  & 
		\colhead{$[{\rm 10^{21} cm^{-2}}]$  } &
		\colhead{$[{\rm 10^{21} cm^{-2}}]$  } &
		\colhead{$[{\rm 10^{21} cm^{-2}}]$  } &
		\colhead{ } &
		\colhead{ } &
		\colhead{ } 
	}
	\decimalcolnumbers
	\startdata
	3    &   11     &     66    &     620   &  0.43  &  0.78  &  1.6  &  0.69 &  1.2  &  2.5\\
	1    &   33     &    270    &    2400   &  0.97  &  2.1   &  4.1  &  0.52 &  1.1  &  2.2\\
	0.3  &   68     &   1400    &   11000   &  2.1   &  7.2   &   12  &  0.34 &  1.2  &  2.0\\
	0.1  &   200    &   6500    &   54000   &  5.2   &  16    &   30  &  0.28 &  0.85 &  1.6\\
	\enddata
\end{deluxetable*}

\subsection{Atomic-molecular distributions}\label{sec:transition}

\begin{figure*}
	\centering
	\includegraphics[trim=0 1cm 0 1cm, clip, width=0.99\linewidth]{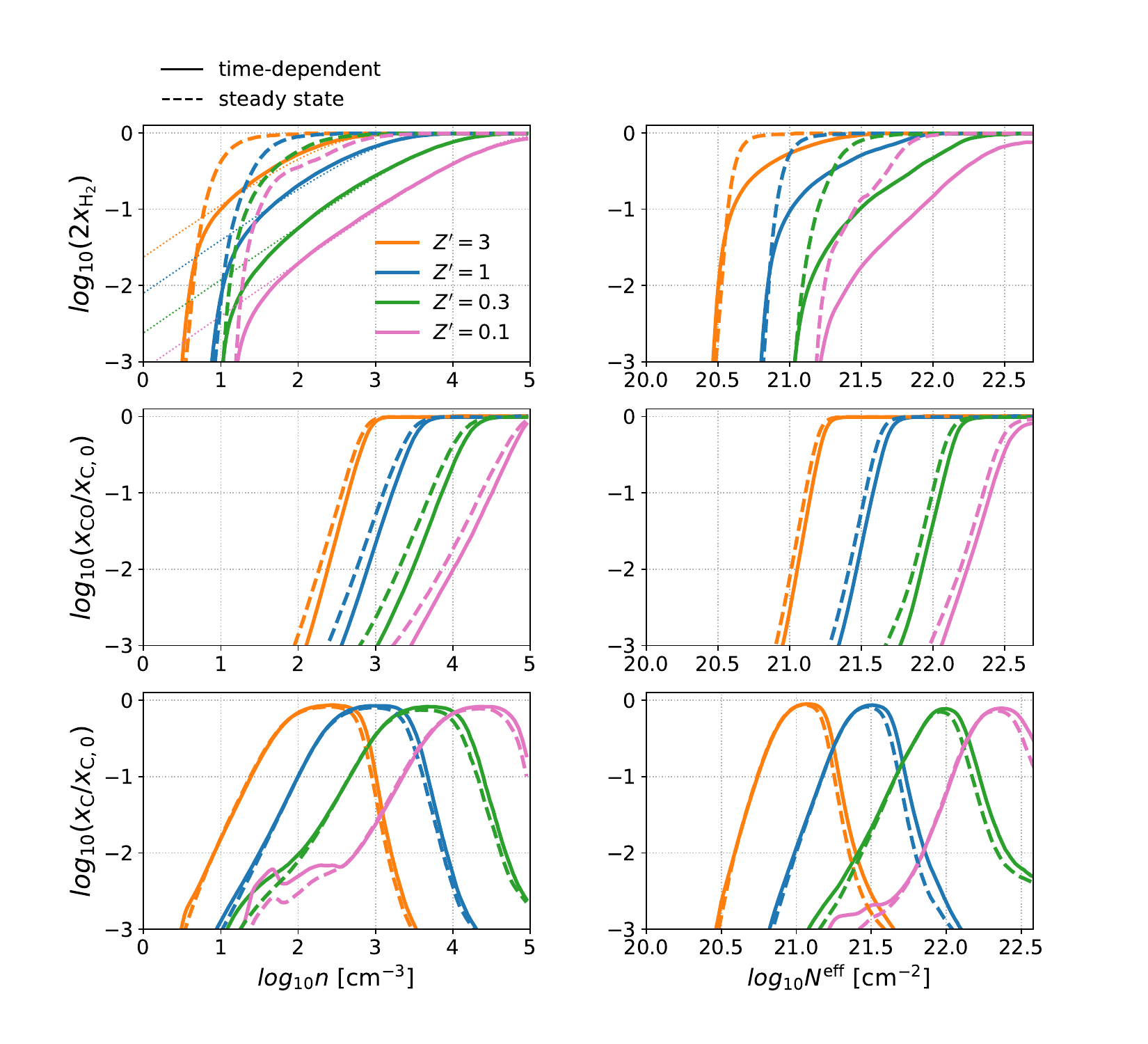}
	\caption{
		Time-averaged
		normalized fractional chemical abundances of 
		H$_2$ ($2x_{\rm H_2}$, top panel), 
		CO ($x_{\rm CO} / x_{\rm C,0}$, middle panel) and 
		C ($x_{\rm C} / x_{\rm C,0}$, bottom panel)
		as a function of $n$ (left) and $N^{\rm eff}$ (right)
		at different metallicities.
		The time-dependent H$_2$ model is shown in solid lines
		while the steady-state H$_2$ model is shown in dashed lines.
		The thin dotted lines in the top left panel show the analytic solutions (Eq. \ref{eq:xh2analytic}).
		The time-dependent effect makes the H$_2$ abundance profile much shallower than its steady-state counterpart.
	}
	\label{fig:transitioncompareeqneq}
\end{figure*}

\begin{figure*}
	\centering
	\includegraphics[width=0.99\linewidth]{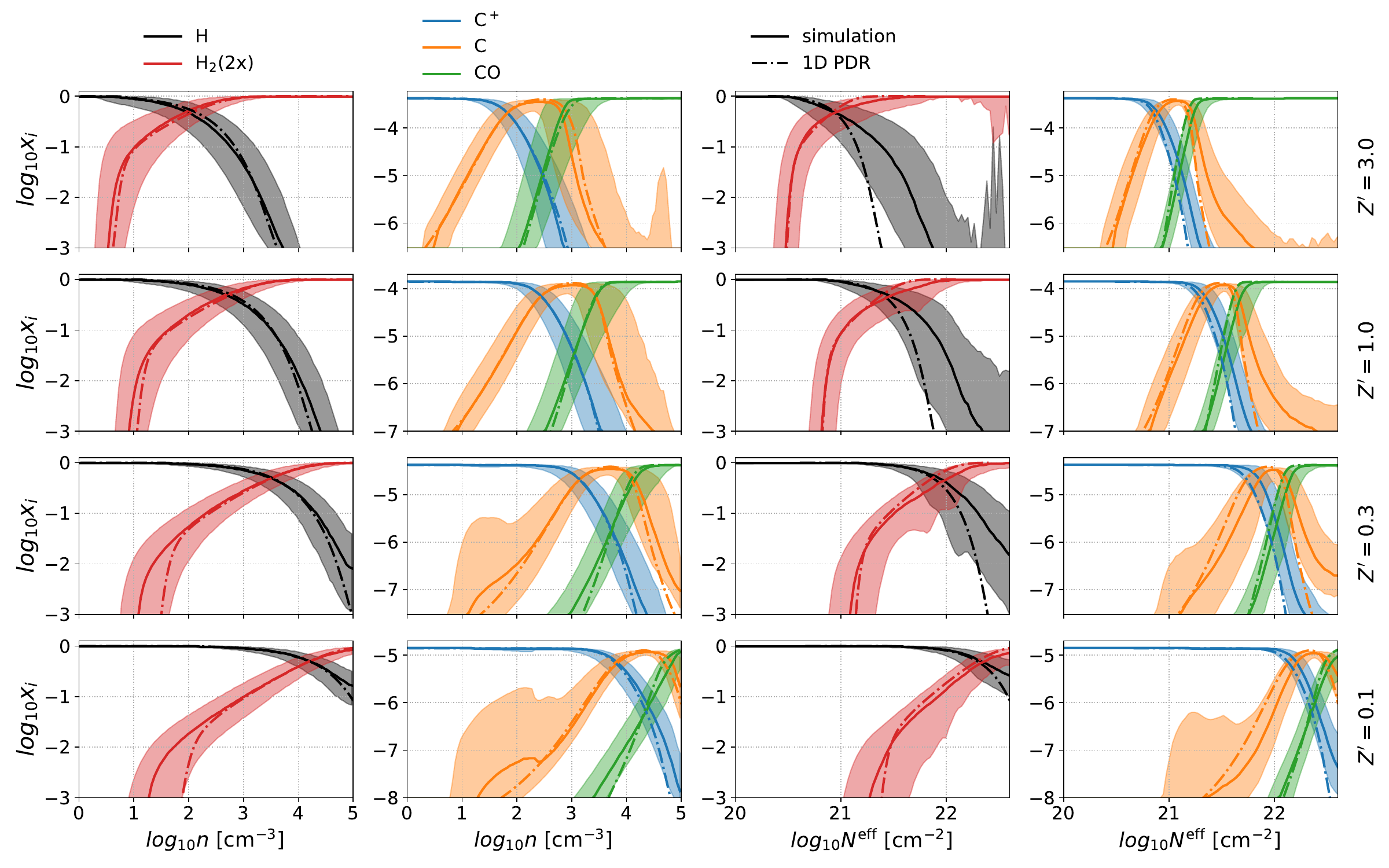}
	\caption{
		Chemical abundances of H (black), H$_2$ (red, multiplied by 2), C$^+$ (blue), C (orange) and CO (green) 
		for the time-dependent H$_2$ model
		as a function of $n$ (1st and 2nd columns) and $N^{\rm eff}$ (3rd and 4th columns) for $Z^{\prime} = $ 3, 1, 0.3 and 0.1 from top to bottom.		
		The solid lines show the medium abundances in each $n$- and $N^{\rm eff}$-bin while the shaded area brackets the 16 and 84 percentiles.
		The dash-dotted lines show the effective 1D PDR model,
		which agrees remarkably well with the median abundances in simulations.
	}
	\label{fig:transitionneq}
\end{figure*}

\begin{figure*}
	\centering
	\includegraphics[trim= 0 2cm 0 2cm, clip, width=0.9\linewidth]{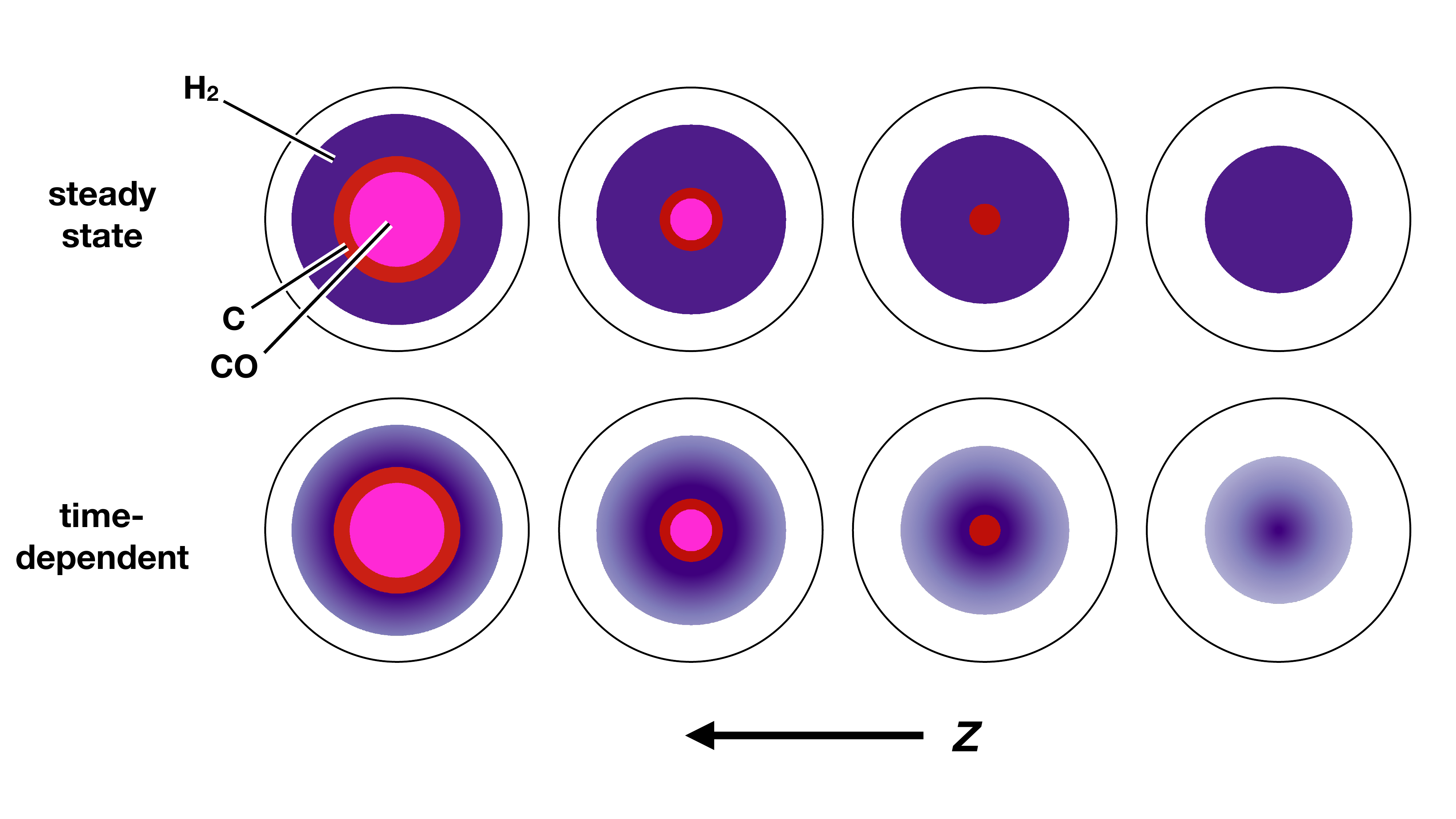}
	\caption{
		Schematic illustration of a typical cloud and its chemical composition at increasing metallicity from right to left.
		Top: the steady-state model. 		
		The transitions of H/H$_2$, C$^+$/C and C/CO 
		occur at the boundaries of 
		the white/purple, purple/red and red/magenta, respectively.
		The transition surfaces are sharp and well-defined.
		As the metallicity decreases,
		the C$^+$/C and C/CO surfaces contract faster than the H/H$_2$ surface.
		Bottom: the time-dependent model. 
		The C$^+$/C and C/CO transitions are largely unchanged.
		The H$_2$ abundance shows a shallow radial profile which declines gradually with increasing radius 
		up to the sharp H$_2$ photodissociation front.
	}
	\label{fig:cloudz}
\end{figure*}

Having discussed the gas densities and thermal states,
we now present the chemical compositions focusing on the atomic-to-molecular conversions.
Fig.~\ref{fig:transitioncompareeqneq} shows 
the normalized fractional chemical abundances of 
H$_2$ ($2x_{\rm H_2}$, top panel), 
CO ($x_{\rm CO} / x_{\rm C,0}$, middle panel) and 
C ($x_{\rm C} / x_{\rm C,0}$, bottom panel)
as functions of $n$ (left) and $N^{\rm eff}$ (right)
for different metallicities.
The time-dependent H$_2$ model is shown in solid lines
while the steady-state H$_2$ model is shown in dashed lines.
In addition,
we define conversion densities for H$_2$ ($n^*_{\rm H_2}$), C ($n^*_{\rm C}$) and CO ($n^*_{\rm CO}$) 
as the densities where 
$x_{\rm H} = 2x_{\rm H_2}$, $x_{\rm C^+} = x_{\rm C}$, and $x_{\rm C} = x_{\rm CO}$, respectively.
Likewise,
the conversion (effective) column densities of H$_2$ ($N^*_{\rm H_2}$), C ($N^*_{\rm C}$) and CO ($N^*_{\rm CO}$) are defined in a similar fashion,
with the associated conversion visual extinctions 
$A^*_{V,{\rm H_2}}$, $A^*_{V,{\rm C}}$ and $A^*_{V,{\rm CO}}$, respectively.
These quantities are summarized in Tables \ref{tab:nNtrans} and \ref{tab:nNtrans_eq}.


Our steady-state H$_2$ model is qualitatively consistent with standard PDR expectations.
The abundance profiles\footnote{
We refer to the curves in Fig.~\ref{fig:transitioncompareeqneq} as ``abundance profiles'' in analogy to the profiles in 1D PDR models (Fig.~\ref{fig:pdrf1test})}
 of H$_2$ and CO are both shifted towards higher $n$ and $N^{\rm eff}$ at decreasing $Z^{\prime}$
(recall that $N^{\rm eff}$ is roughly a monotonically increasing function of $n$).
At any given metallicity,
the H/H$_2$ conversion occurs at the lowest density and column density,
then followed by C$^+$/C and C/CO as the column density increases.
These profiles are mainly determined by a balance between the formation via two-body reactions
and the destruction via photodissociation or photoionization.
The sharp conversions occur when the accumulated column densities are such that enough FUV radiation is attenuated.
At the conversion density of H$_2$,
the corresponding dimensionless parameter $\alpha G$ from \citet{2014ApJ...790...10S} (their Eqs. 44 and 46),
which quantifies the relative importance between dust-shielding and self-shielding,
is 2.7, 0.91, 0.44 and 0.15 for $Z^{\prime} = $ 3, 1, 0.3 and 0.1, respectively.
Therefore,
self-shielding becomes increasingly important at decreasing $Z^{\prime}$.

In contrast,
in the more realistic time-dependent H$_2$ model,
the H$_2$ profile is much shallower,
spanning almost two orders of magnitude in $n$ from $2x_{\rm H_2} = 0.1$ to $2x_{\rm H_2} = 1$. 
In addition,
the H/H$_2$ conversions occur at much higher densities than their steady-state counterparts,
by a factor of 8, 14, 39 and 64 for $Z^{\prime} = $ 3, 1, 0.3 and 0.1, respectively.
In other words,
H$_2$ is further away from steady state at lower $Z^{\prime}$.
This is expected as $t_{\rm H_2,form} \propto 1 / Z^{\prime}$ 
becomes progressively longer than the dynamical time $t_{\rm dyn}$,
defined here as the available time for which H$_2$ can form unperturbed.
We expect $t_{\rm dyn}$ to be insensitive to $Z^{\prime}$
as it is mainly controlled by the SFR (via stellar feedback) which is insensitive to $Z^{\prime}$ in our simulations.
Empirically,
we find the dynamical time in the following form\footnote{
We note that Eq. \ref{eq:tdyn} is a purely empirical form and is not measured from simulations.
		Physically, the dynamical time represents the typical ``lifetime'' a gas particle spends at a fixed $n$.
		Measuring this quantity from simulations requires a very high frequency of output ($\lesssim$ 0.2 Myr) that is impractical for us.
}
\begin{equation}\label{eq:tdyn}
	t_{\rm dyn} = 3\ {\rm Myr}\ n_2^{-0.3},
\end{equation}
where $n_2 \equiv n / (100\ {\rm cm^{-3}})$,
can explain the H$_2$ abundances in the well-shielded regions
\begin{equation}\label{eq:xh2analytic}
2x_{\rm H_2} = 1 - e^{ -t_{\rm dyn}/t_{\rm H_2,form}  } = 1 - e^{ - 0.2 Z_d^{\prime} n_2^{0.7}  },
\end{equation}
which is over-plotted in the upper left panel of Fig.~\ref{fig:transitioncompareeqneq} as the thin dotted lines.
This simple analytic expression captures the average H$_2$ profile for the time-dependent model 
at all metallicities remarkably well,
except for the outermost part where the H$_2$ profile is truncated and overlaps with the profiles for the steady-state model,
which serve as an upper limits for the former.
In short,
the H/H$_2$ conversions are dictated mainly by the available time for H$_2$ formation,
which results in a slowly declining profile up to the surface of the sharp photodissociation front.

In comparison,
C/CO conversions occur at much higher densities with $n^*_{\rm CO} \approx 5 n^*_{\rm H_2}$.
Moreover,
unlike the H$_2$ profiles,
the CO profiles agree well with their steady-state counterparts,
which is not trivial as both the formation and destruction of CO are affected by the presence of H$_2$.
However,
in both models,
hydrogen is already fully molecular where the C/CO conversions occur, 
which implies the same CO formation rates.
The difference in the H$_2$ profiles in both models does lead to a difference in the
H$_2$ column densities which is very significant at low $Z^{\prime}$.
However,
as the C/CO conversions occur at much higher densities than H/H$_2$,
the effective H$_2$ column densities relevant for shielding are quite similar in both models.
Therefore,
it turns out that the time-dependent effect only shifts the C/CO conversions towards marginally higher density and column density.

Finally,
the time-dependent effect has the least impact on the C$^+$/C conversions.
This is expected as these are largely determined by the photoionization of C and recombination of C$^+$ (either radiative or on grains),
which has little to do with H$_2$ per se.
Interestingly,
the H/H$_2$ conversions occur (becomes half-molecular) 
at slightly higher column densities than C$^+$/C in the time-dependent case,
which is contrary to the steady-state model.

\subsection{Effective 1D PDR model}

\begin{figure}
	\centering
	\includegraphics[width=0.9\linewidth]{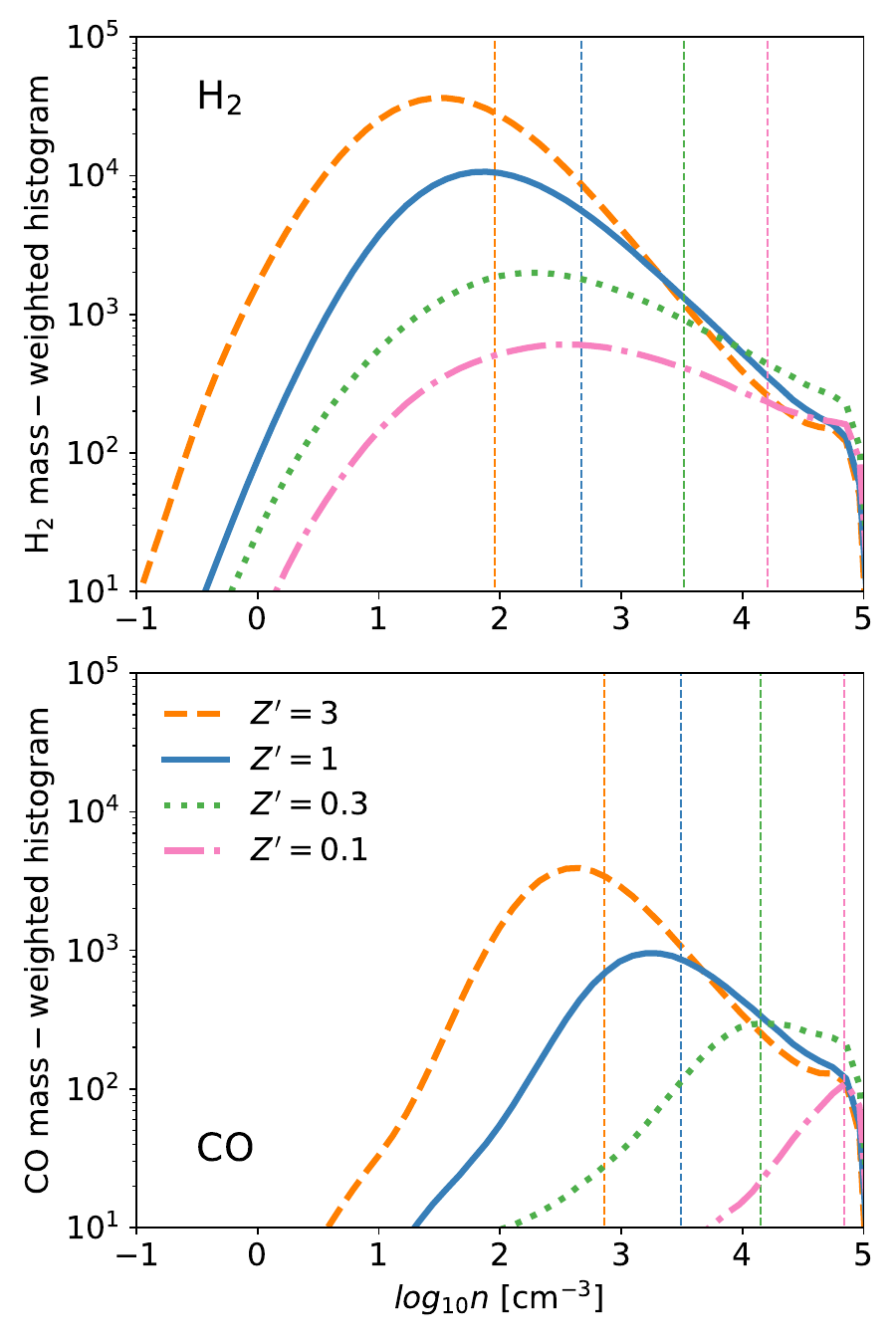}
	\caption{
		Time-averaged 
		density histograms weighted by the H$_2$ mass ($2m_{\rm g} x_{\rm H_2 }$, top panel) and normalized CO mass ($28 m_{\rm g} x_{\rm CO} / x_{\rm C,0}$, bottom panel) 
		at different metallicities.
		The vertical lines indicate the conversion densities as shown in Table \ref{tab:nNtrans}.
		The majority of H$_2$ is found at a density much lower than the H$_2$ conversion density (see Table \ref{tab:nNtrans}).
	}
	\label{fig:h2codensitypdf}
\end{figure}

\begin{figure}
	\centering
	\includegraphics[width=0.9\linewidth]{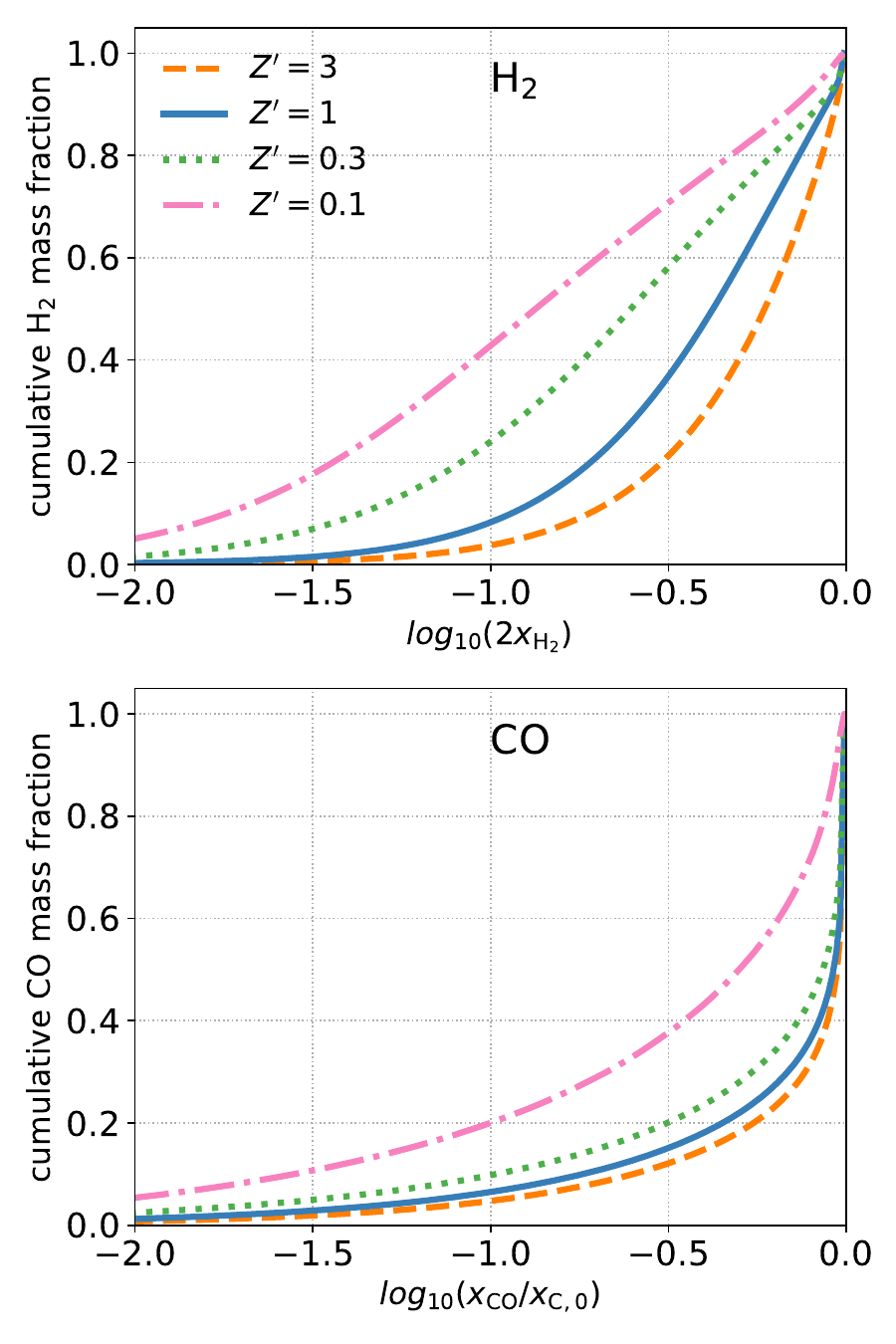}
	\caption{
		Time-averaged cumulative molecular mass fraction as a function of normalized chemical abundance at different metallicities.
		The top and bottom panels are for H$_2$ and CO, respectively.
		A significant fraction of the H$_2$ mass is found in low-H$_2$ abundance gas.
	}
	\label{fig:xh2xcocdfneq}
\end{figure}

\begin{figure}
	\centering
	\includegraphics[width=0.99\linewidth]{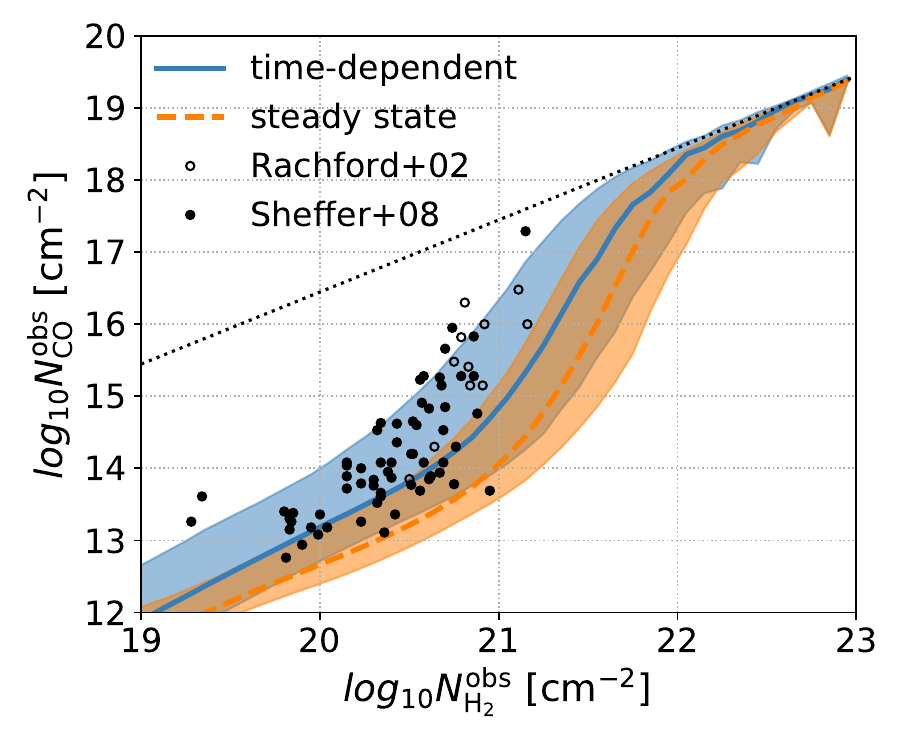}
	\caption{
		Time-averaged $N^{\rm obs}_{\rm H_2}$ - $N^{\rm obs}_{\rm CO}$ relationship in the $Z^{\prime} = 1$ run.
		The blue solid line shows the time-dependent H$_2$ model
		while the orange dashed line shows the steady-state H$_2$ model.
		The shaded area brackets the 16 and 84 percentiles in each $N^{\rm obs}_{\rm H_2}$-bin.
		The black circles are observational data from \citet{2002ApJ...577..221R} (empty) 
		and \citet{2008ApJ...687.1075S} (filled) for Galactic diffuse clouds.
		The dotted grey line indicates the upper limit for $N^{\rm obs}_{\rm CO}$ when all carbon is in the form of CO.
	}
	\label{fig:losnh2vsnconeqvseq}
\end{figure}

\begin{figure}
	\centering
	\includegraphics[trim=0cm 2.5cm 0cm 0cm, clip, width=0.99\linewidth]{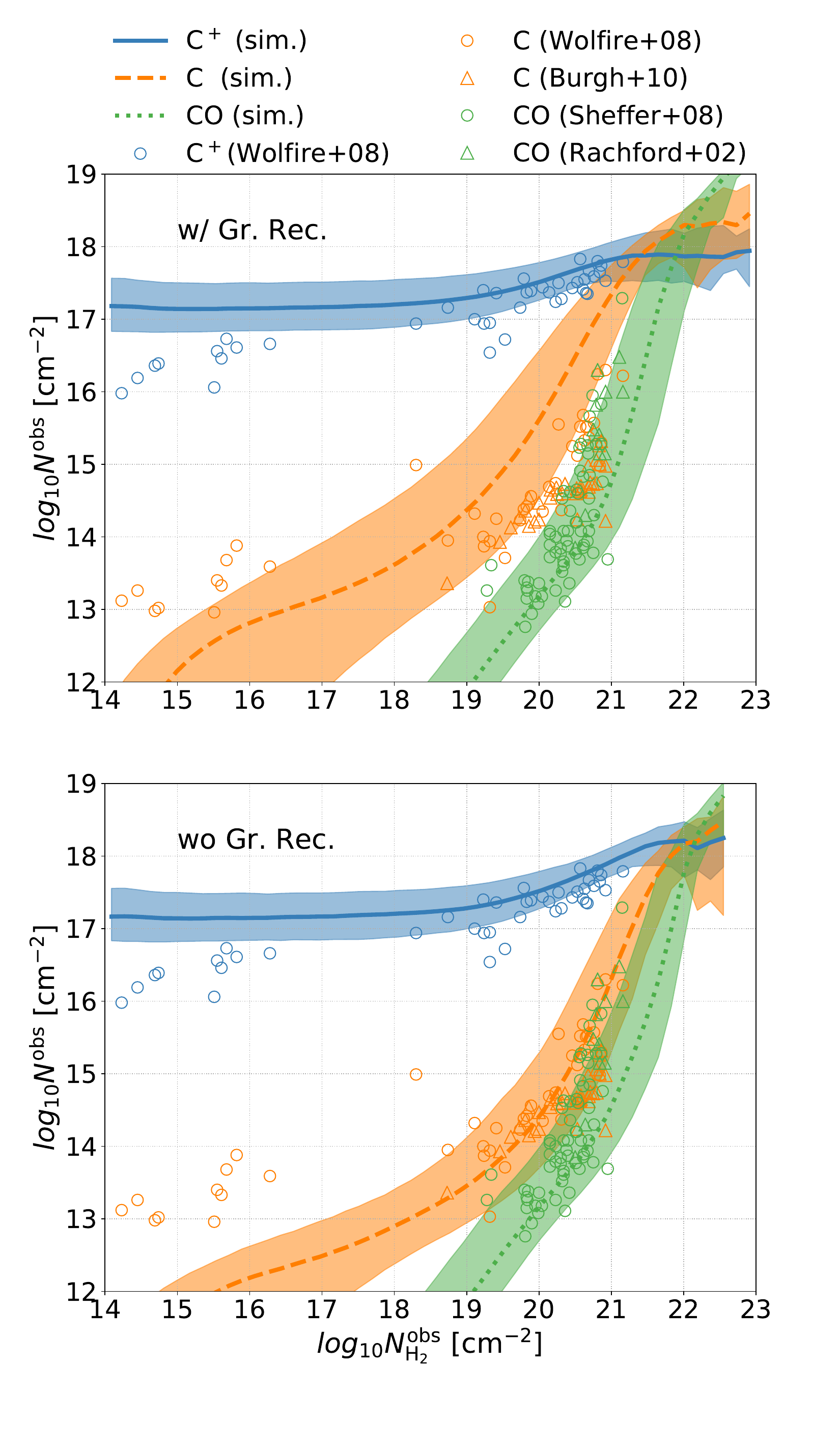}
	\caption{
			Time-averaged relationships of $N^{\rm obs}_{\rm H_2}$ vs. 
			$N^{\rm obs}_{\rm C^+}$ (solid blue), 
			$N^{\rm obs}_{\rm C}$ (dashed orange) 
			and $N^{\rm obs}_{\rm CO}$ (dotted green) in the $Z^{\prime} = 1$ run.
			The lines show the median values while the shaded areas bracket the 16 and 84 percentiles.			
			The top and bottom panels are models with and without recombination on grains, 
			respectively (both models are time-dependent).
			Observational data of the Galactic clouds are 
			taken from \citet{2008ApJ...680..384W} (blue circles for C$^+$ and orange circles for C),
			\citet{2010ApJ...708..334B} (orange triangles for C),			 
			\citet{2008ApJ...687.1075S} (green circles for CO) and \citet{2002ApJ...577..221R} (green triangles for CO).
			Switching off recombination on grains leads to better agreement with the observed $N^{\rm obs}_{\rm C}$ 
			except for the low-column density regime where both models under-produce the atomic carbon.
	}
	\label{fig:losnh2vsncpcico}
\end{figure}

The relationships in Fig.~\ref{fig:transitioncompareeqneq} are averaged over a large ensemble of clouds 
both spatially and temporally and therefore are representative of the abundance distributions of typical clouds 
(although smaller clouds may not have the depth to reach the highest $N^{\rm eff}$).
In fact,
we can construct an effective 1D PDR model that represents these distributions as follows.
Assuming the cloud follows a power-law density profile
$n(x) = B x^\beta$,
where $x = 0$ is the cloud center while the external radiation is illuminated from $\infty$.
Adopting a 1D slab geometry,
the column density integrated from $\infty$ to $x$ is
$N(x) \equiv - \int_{\infty}^{x} n(x^{\prime}) d x^{\prime} = -B x^{\beta+1} / (\beta+1)$.
Meanwhile,
the column density is correlated with $n$ (see Eq. \ref{eq:nvsN}) such that
$N(x) = A n(x)^\alpha = A B^\alpha x^{\alpha\beta}$.
Therefore,
it follows that
$\beta = 1 / (\alpha - 1)$ and $B = (A\alpha/(1-\alpha))^{1/(1-\alpha)}$.

We now can conduct 1D PDR calculations with this effective cloud profile, 
evolving up to $t_{\rm dyn}$ at each density bin to capture the time-dependent effect.
We adopt $I_{\rm UV} = 1$ and $\zeta_{\rm CR} = 10^{16}\ {\rm s^{-1}}$ as the time-averaged SFR is close to the solar-neighborhood value at all metallicities.
The temperature is fixed at 20 K as we find that the results are insensitive to moderate variations of temperature.
The results are compared with simulations in Fig.~\ref{fig:transitionneq},  
which shows the chemical abundances of 
H (black), H$_2$ (red, multiplied by 2), C$^+$ (blue), C (orange) and CO (green) in the time-dependent model 
as functions of $n$ (1st and 2nd columns) and $N^{\rm eff}$ (3rd and 4th columns) 
for $Z^{\prime} = $ 3, 1, 0.3 and 0.1 from top to bottom.
The solid lines show the medium abundances in each $n$- and $N^{\rm eff}$-bin while the shaded area brackets the 16 and 84 percentiles.
The dash-dotted lines show the effective 1D PDR model,
which agrees remarkably well with the median abundances in simulations.
There are some discrepancies in $N^{\rm eff}$ vs. $x_{\rm H}$ in the H$_2$-dominated regimes
despite the agreement in the corresponding $n$ vs. $x_{\rm H}$,
suggesting that the discrepancies originate from the $n - N^{\rm eff}$ relationship.
This is expected as $N^{\rm eff}$ represents, by definition, the amount of dust shielding for CO
and thus might not be a perfect proxy for H$_2$ self-shielding.

In Fig.~\ref{fig:cloudz},
we provide a
schematic illustration of 
the effective cloud and its chemical composition at increasing metallicity from right to left.
The top row shows the standard steady-state model \citep{2013ARA&A..51..207B}.
The transitions of H/H$_2$, C$^+$/C and C/CO 
occur at the boundaries of 
the white/purple, purple/red and red/magenta, respectively.
The transition surfaces are sharp and well-defined.
As the metallicity decreases,
the C$^+$/C and C/CO surfaces contract faster than the H/H$_2$ surface as H$_2$ self-shielding is very efficient.
In contrast,
the bottom row shows the time-dependent model,
where the H$_2$ abundance shows a shallow radial profile which declines gradually with increasing radius 
up to the sharp H$_2$ photodissociation front.
At low $Z^{\prime}$,
photodissociation has almost no effect on the H$_2$ profile 
as H$_2$ formation is mainly limited by the dynamical processes.
The C$^+$/C and C/CO transitions are largely unchanged.

\subsection{Density distributions of H$_2$ and CO}

Fig.~\ref{fig:h2codensitypdf}
shows the histograms of $n$
weighted by the H$_2$ mass ($2m_{\rm g} x_{\rm H_2 }$, top panel) 
and normalized CO mass ($28 m_{\rm g} x_{\rm CO} / x_{\rm C,0}$, bottom panel).
The vertical lines indicate the conversion densities as shown in Table \ref{tab:nNtrans}.

Most CO is found at $n \sim n^*_{\rm CO}$ (as indicated by the peak of the histogram),
while most H$_2$ is found at a density much lower than $n^*_{\rm H_2}$.
This is due to the difference in their profile shapes,
as the molecular mass-weighted density histogram is essentially 
a convolution of the total gas density histogram with the abundance profile.
Therefore,
the sharp CO profile means that most CO lives at $n \gtrsim n^*_{\rm CO}$,
while the shallow H$_2$ profile means that
gas with density below $n^*_{\rm H_2}$ can still make a significant contribution to the total H$_2$ mass budget,
shifting the peak towards a much lower density.
This can be more quantitatively demonstrated by the
cumulative molecular mass fraction as a function of the normalized molecular abundance shown in Fig.~\ref{fig:xh2xcocdfneq}.
The top and bottom panels are for H$_2$ and CO, respectively.
While most of the CO mass originates from gas with high $x_{\rm CO}$,
a significant fraction of the H$_2$ mass can be found in HI-dominated gas.
This is most prominent at $Z^{\prime} = 0.1$, 
where the shallow H$_2$ profile spans almost four orders of magnitude in $n$ 
before being truncated at the photodissociation surface,
and thus more than 40\% of H$_2$ is found in gas with $2x_{\rm H_2} < 0.1$.

\subsection{Comparison with the observed Galactic clouds}

The fact that H$_2$ is significantly affected by the time-dependent effect while CO is not 
leads to an interesting observational consequence.
Fig.~\ref{fig:losnh2vsnconeqvseq} shows 
the $N^{\rm obs}_{\rm H_2}$ - $N^{\rm obs}_{\rm CO}$ relationship in the $Z^{\prime} = 1$ run.
The blue solid line shows the time-dependent H$_2$ model
while the orange dashed line shows the steady-state H$_2$ model.
The shaded area brackets the 16 and 84 percentiles in each $N^{\rm obs}_{\rm H_2}$-bin.
The black circles are observational data from \citet{2002ApJ...577..221R} (empty) 
and \citet{2008ApJ...687.1075S} (filled) for Galactic diffuse clouds.
The dotted grey line indicates the upper limit for $N^{\rm obs}_{\rm CO}$ when all carbon is in the form of CO.
Our time-dependent H$_2$ model agrees well with observations especially in the low-column density regime,
which gives us confidence that our model is reasonably realistic.
In contrast,
since the steady-state H$_2$ model generally over-produces H$_2$ but not CO, 
the results are shifted rightward and away from the observed data.
This is an intriguing demonstration of how the time-dependent effect helps explain the observed $N^{\rm obs}_{\rm H_2}$ - $N^{\rm obs}_{\rm CO}$ relationship without resorting,
for example,
to non-thermal velocity distributions (boosted by the Alfv\'{e}n waves) 
for the reaction
C$^+$ + H$_2$ $\rightarrow$ CH$^+$ + H 
(e.g. \citealp{1996MNRAS.279L..41F, 2008ApJ...687.1075S, 2009A&A...503..323V}),
which we do not include.
That said,
it does not negate the need for this reaction to produce CH$^+$
given the widespread observations of CH$^+$ in the diffuse medium.
We note that \citet{2018ApJ...858...16G} have also reproduced the observed 
$N^{\rm obs}_{\rm H_2} - N^{\rm obs}_{\rm CO}$ relationship in their simulations.
However, they assumed steady-state chemistry which tends to over-produce H$_2$,
and it remains to be seen whether their results still hold if they account for the time-dependent effect.

Fig.~\ref{fig:losnh2vsncpcico} shows the time-averaged relationships of $N^{\rm obs}_{\rm H_2}$ vs. 
$N^{\rm obs}_{\rm C^+}$ (solid blue), 
$N^{\rm obs}_{\rm C}$ (dashed orange) 
and $N^{\rm obs}_{\rm CO}$ (dotted green) in the $Z^{\prime} = 1$ run.
The lines show the median values while the shaded areas bracket the 16 and 84 percentiles.			
The top and bottom panels are models with and without recombination on grains, 
respectively (both models are time-dependent).
Observational data of the Galactic clouds are 
taken from \citet{2008ApJ...680..384W} (blue circles for C$^+$ and orange circles for C),
\citet{2010ApJ...708..334B} (orange triangles for C),			 
\citet{2008ApJ...687.1075S} (green circles for CO) and \citet{2002ApJ...577..221R} (green triangles for CO).
Both models under-produces $N^{\rm obs}_{\rm C}$ in the low-column density regime where $10^{14} < N^{\rm obs}_{\rm H_2} / ({\rm cm^{-2}}) < 10^{16}$. 
Our fiducial model (top panel) largely reproduces the observed $N^{\rm obs}_{\rm C^+}$ and $N^{\rm obs}_{\rm CO}$,
but over-produces $N^{\rm obs}_{\rm C}$ by about a factor of ten in the regime where 
$N^{\rm obs}_{\rm H_2} > 10^{19} {\rm cm^{-2}}$. 
Switching off recombination on grains reduces $N^{\rm obs}_{\rm C}$ down to the observed values 
without hampering the agreement in $N^{\rm obs}_{\rm C^+}$ and $N^{\rm obs}_{\rm CO}$.
Despite the good agreement with observations,
we refrain from making it our default model as there is no physically justified reason to switch off recombination on grains.
Our results are consistent with \citet{2012MNRAS.421..116G} who showed that 
the C/CO ratio in their turbulent box simulations agrees with the observed values when recombination on grains is switched off 
but is overestimated otherwise.
On the other hand,
\citet{2017ApJ...843...38G}
found conflicting results in their PDR calculations
where the C/CO ratio is overestimated either with or without recombination on grains.
This is presumably because their CO abundances are significantly reduced without recombination on grains,
which is not the case in our simulations (see Sec. \ref{app:grRec}).

\subsection{Global properties and CO-dark H$_2$ fraction}

\begin{figure*}
	\centering
	\includegraphics[width=0.99\linewidth]{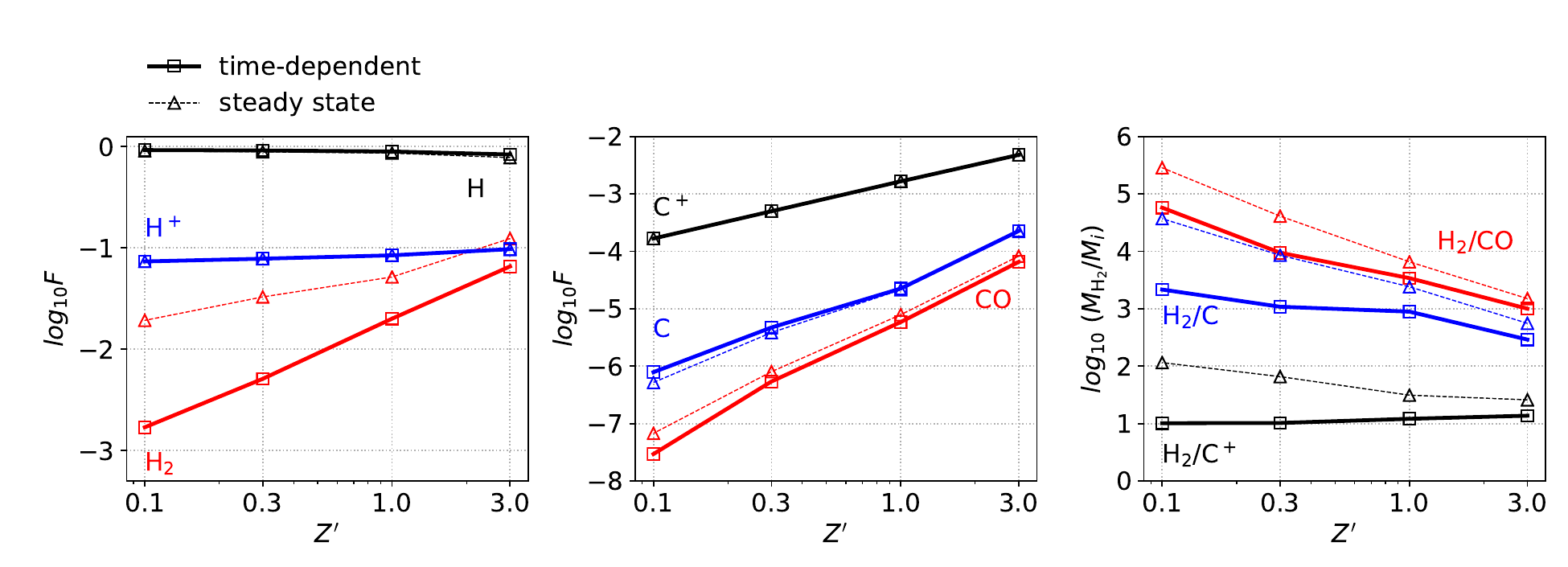}
	\caption{
		Time-averaged global chemical properties as a function of $Z^{\prime}$.
		Left:
		mass fractions of H$^+$, H and H$_2$ relative to the total hydrogen mass.
		Middle:
		mass fractions of C$^+$, C and CO relative to the total hydrogen mass.
		Right:
		mass ratio of H$_2$/C$^+$, H$_2$/C and H$_2$/CO.
		The time-dependent H$_2$ model is shown in filled symbols while the steady-state H$_2$ model is in empty symbols.
		The steady-state model significantly overestimates $F_{\rm H_2}$ but not $F_{\rm CO}$ at low $Z^{\prime}$.
	}
	\label{fig:ratioh2co}
\end{figure*}

We now present the time-averaged global properties of the simulations
which are summarized in Tables \ref{tab:sumstats} and \ref{tab:sumstatseq}.
$M_{\rm gas}$ is the total gas mass (including helium and metals)
and $F_{100}$ is the mass fraction with $n > 100\ {\rm cm^{-3}}$.
The total mass of species $i$ is denoted as $M_i$
where $i$ = H$^+$, H, H$_2$, C$^+$, C and CO,
while the mass fraction relative to the total hydrogen mass $M_{\rm H,0} = X_{\rm H} M_{\rm gas}$
is defined as $F_i = M_i / M_{\rm H,0}$.
$F^*_{\rm dark}{\rm (CO)}$ is the mass fraction of the CO-dark H$_2$ gas (see details below).
Fig.~\ref{fig:ratioh2co} shows $F_{\rm H^+}$, $F_{\rm H}$ and $F_{\rm H_2}$ in the left panel,
$F_{\rm C^+}$, $F_{\rm C}$ and $F_{\rm CO}$ in the middle panel,
and the mass ratios $M_{\rm H_2} / M_{\rm C^+}$, $M_{\rm H_2} / M_{\rm C}$ and $M_{\rm H_2} / M_{\rm CO}$
in the right panel as a function of $Z^{\prime}$, respectively.
The time-dependent H$_2$ model is shown in solid lines while the steady-state H$_2$ model is in dashed lines.

Both the SFR and $F_{100}$ are insensitive to $Z^{\prime}$,
as the density distribution varies only weakly with $Z^{\prime}$.
The global gas depletion time is $M_{\rm gas} / {\rm SFR} \approx 4$ Gyr,
slightly lower than the typical value of 2 Gyr in typical nearby spiral galaxies \citep{2008AJ....136.2846B,2008AJ....136.2782L}.
$F_{\rm H^+}$ is also insensitive to $Z^{\prime}$ as H$^+$ 
is mainly generated in the HII regions and the supernova remnants, 
which are tied to the SFR.
As H$_2$ only contributes to a small fraction of the total hydrogen mass,
$F_{\rm H} \approx 1 - F_{\rm H^+} - F_{\rm H_2}$ is also insensitive to $Z^{\prime}$ ($\sim$ 90\%).
In the steady-state H$_2$ model,
$F_{\rm H_2}$ scales sub-linearly with $Z^{\prime}$
because of the efficient H$_2$ self-shielding.
However,
in the time-dependent case,
$F_{\rm H_2}$ scales roughly linearly with $Z^{\prime}$ 
as H$_2$ is mainly limited by the dynamical time especially at low $Z^{\prime}$.
As such,
the steady-state H$_2$ model significantly overestimates $F_{\rm H_2}$,
by a factor of 1.8, 2.6, 6,3 and 11 for $Z^{\prime} = 3$, 1, 0.3 and 0.1, respectively.

The vast majority ($\sim$ 99\%) of carbon is in the form of C$^+$,
and thus $F_{\rm C^+}$ scales linearly with $Z^{\prime}$.
Both $F_{\rm C}$ and $F_{\rm CO}$ scale super-linearly with $Z^{\prime}$,
due to the effect of dust shielding in addition to the available carbon.
Contrary to $F_{\rm H_2}$,
$F_{\rm C}$ and $F_{\rm CO}$ are almost unaffected by the time-dependent effect
as the C/CO conversions are only weakly affected by the over-estimated H$_2$.
Therefore,
the ratios $M_{\rm H_2} / M_{\rm CO}$ and $M_{\rm H_2} / M_{\rm C}$ still anti-correlate with $Z^{\prime}$,
but not as much as what the steady-state H$_2$ model predicts.
Interestingly,
the ratios $M_{\rm H_2} / M_{\rm C^+}$ and $M_{\rm H_2} / M_{\rm C}$ both show a weaker dependence on $Z^{\prime}$
than $M_{\rm H_2} / M_{\rm CO}$.
This indicates that they could potentially be viable tracers for H$_2$ at low $Z^{\prime}$,
although a robust assessment requires information of the line intensities of these tracers,
which we leave for future work.

As $F_{\rm H_2}$ decreases with $Z^{\prime}$ while the SFR is insensitive to $Z^{\prime}$,
we confirm that the cold gas reservoir can be dominated by atomic hydrogen at low $Z^{\prime}$
as demonstrated by previous work \citep{2012ApJ...759....9K, 2012MNRAS.421....9G,2016MNRAS.458.3528H}.
Note that star formation occurs at $n\gtrsim 10^4 - 10^5\ {\rm cm^{-3}}$
which is above the H$_2$ conversion density.
This means that even if the cold gas reservoir is dominated by atomic hydrogen,
gas that undergoes runaway gravitational collapse will become molecular at some point before turning into stars.
Therefore,
our results that the cold gas reservoir is mostly atomic would still hold 
even if we were to adopt an H$_2$-based star formation recipe.
We stress that this is the case only because we resolve the gravitational collapse up to very high densities.
In cases where the H/H$_2$ conversion is unresolved (e.g. cosmological simulations),
the H$_2$-based star formation recipe leads to inaccurate ISM properties and thus should not be used.


\begin{figure*}
	\centering
	\includegraphics[trim=0 0 0 2cm, clip, width=0.99\linewidth]{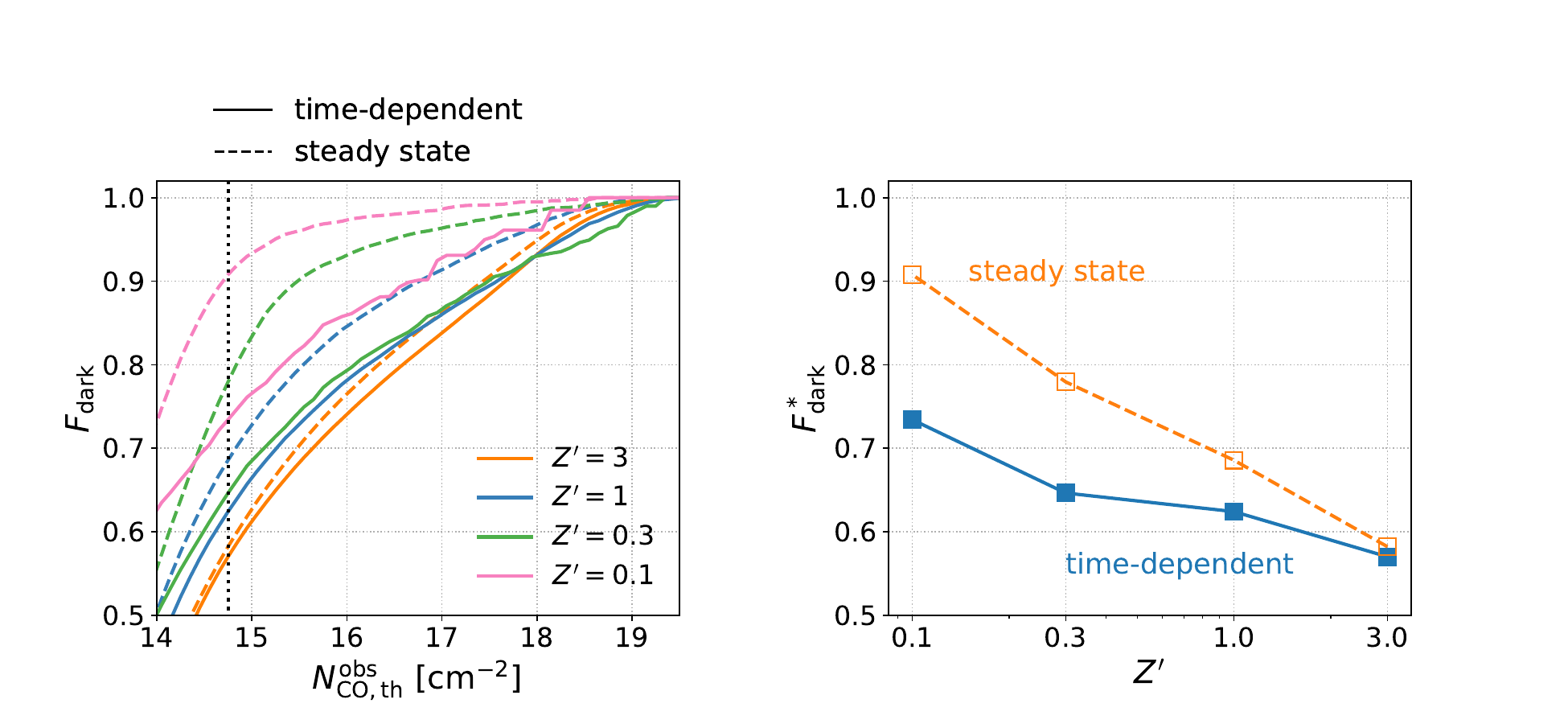}
	\caption{
		\textit{Left}: The CO-dark H$_2$ mass fraction as a function of the threshold observed CO column density $N^{\rm obs}_{\rm CO,th}$
		at different metallicities.
		The time-dependent H$_2$ model is shown in solid lines while the steady-state H$_2$ model is in dashed lines.
		The vertical black dotted line highlights $N^{\rm obs}_{\rm CO,th} = 6\times 10^{14} {\rm cm^{-3}}$.
		\textit{Right}: 
		The CO-dark H$_2$ mass fraction defined with
		$N^{\rm obs}_{\rm CO,th} = 6\times 10^{14} {\rm cm^{-3}}$
		as a function of $Z^{\prime}$.
		The steady-state model significantly over-predicts the CO-dark H$_2$ mass fraction at low $Z^{\prime}$.
	}
	\label{fig:codark}
\end{figure*}

\begin{deluxetable*}{ccclllllllc}
	\tablecaption{Time-averaged global properties. The chemical properties are from the time-dependent H$_2$ model.
		\label{tab:sumstats}}
	\tablewidth{0pt}
	\tablehead{
		\colhead{$Z^{\prime}$} &
		\colhead{$M_{\rm gas}$ }  &
		\colhead{${\rm SFR}$}  &
		\colhead{$F_{\rm 100}$}   &
		\colhead{$F_{\rm H^+}$}   &
		\colhead{$F_{\rm H  }$}   &
		\colhead{$F_{\rm H_2}$}   &
		\colhead{$F_{\rm C^+}$ }  &
		\colhead{$F_{\rm C  }$ }  &
		\colhead{$F_{\rm CO }$ }  &
		\colhead{$F_{\rm dark}$ } 
	}
	\decimalcolnumbers
	\startdata
	3   &$8.3\times 10^6$  &	 2.4$\times 10^{-3}$  &0.025 &  0.097	   &  0.84	  &  0.066   &   4.79$\times10^{-3}$   &  2.26$\times10^{-4}$  &	6.55$\times 10^{-5}$ &  0.57 	 \\
	1   &$8.5\times 10^6$  &	 2.6$\times 10^{-3}$  &0.023 &  0.085		 &  0.90	  &  0.020   &   1.65$\times10^{-3}$   &  2.25$\times10^{-5}$  &	5.87$\times 10^{-6}$ &  0.62 	 \\
	0.3 &$8.4\times 10^6$  &	 3.8$\times 10^{-3}$  &0.017 &  0.078	   &  0.92		&  0.0051  &   4.99$\times10^{-4}$   &  4.72$\times10^{-6}$  &	5.41$\times 10^{-7}$ &  0.65   \\
	0.1 &$8.6\times 10^6$  &	 2.5$\times 10^{-3}$  &0.013 &  0.074		 &  0.92		&  0.0017  &   1.67$\times10^{-4}$   &  7.87$\times10^{-7}$  &	2.95$\times 10^{-8}$ &  0.73   \\
	\enddata
	\tablecomments{
		(1) Normalized metallicity.
		(2) Total gas mass (including helium and metals) in units of ${\rm M_\odot}$.
		(3) Total star formation rate in units of ${\rm M_\odot\ yr^{-1}}$.
		(4) Mass fraction with $n > 100\ {\rm cm^{-3}}$.
		(5)-(10):
		Mass fraction (relative to total hydrogen mass) of H$^+$, H, H$_2$, C$^+$, C and CO.
		(11) Mass fraction of the CO-dark H$_2$ gas with $N^{\rm obs}_{\rm CO,th} = 6\times10^{14} {\rm cm^{-3}}$.
	}
\end{deluxetable*}

\begin{deluxetable*}{cllllllc}
	\tablecaption{Time-averaged global chemical properties in the steady-state H$_2$ model.
		\label{tab:sumstatseq}}
	\tablewidth{0pt}
	\tablehead{
		\colhead{$Z^{\prime}$} &
		\colhead{$F_{\rm H^+}$}   &
		\colhead{$F_{\rm H  }$}   &
		\colhead{$F_{\rm H_2}$}   &
		\colhead{$F_{\rm C^+}$ }  &
		\colhead{$F_{\rm C  }$ }  &
		\colhead{$F_{\rm CO }$ }  &
		\colhead{$F_{\rm dark}$ }
	}
	\decimalcolnumbers
	\startdata
	3   &  0.096	   &  0.78	  &  0.12    &   4.78$\times10^{-3}$   &  2.22$\times10^{-4}$   &  	8.25$\times 10^{-5}$ &  0.58  	\\
	1   &  0.085		 &  0.86	  &  0.052   &   1.66$\times10^{-3}$   &  2.14$\times10^{-5}$   &  	7.87$\times 10^{-6}$ &  0.69  	\\
	0.3 &  0.078	   &  0.89		&  0.033   &   5.00$\times10^{-4}$   &  3.84$\times10^{-6}$   &  	8.01$\times 10^{-7}$ &  0.80   \\
	0.1 &  0.074		 &  0.91		&  0.019   &   1.67$\times10^{-4}$   &  5.20$\times10^{-7}$   &  	6.76$\times 10^{-8}$ &  0.91   \\
	\enddata
	\tablecomments{
		(1) Normalized metallicity.
		(2)-(7):
		Mass fraction (relative to total hydrogen mass) of H$^+$, H, H$_2$, C$^+$, C and CO.
		(8) Mass fraction of the CO-dark H$_2$ gas with $N^{\rm obs}_{\rm CO,th} = 6\times10^{14} {\rm cm^{-3}}$.
	}
\end{deluxetable*}

The fact that the conversions for H/H$_2$ occur at lower volume and column densities than for C/CO 
leads to a natural outcome that CO does not trace H$_2$ perfectly,
which is readily visualized in Fig.~\ref{fig:loadmolmapsneq48}.
The H$_2$ gas that does not show a detectable CO emission is called ``CO-dark'',
which by definition depends on the detection limit.
However,
since we do not perform radiative transfer to obtain the CO emission in this work,
we instead use a threshold value of $N^{\rm obs}_{\rm CO}$ as a proxy of the detection limit.
The left panel in Fig.~\ref{fig:codark} shows the cumulative H$_2$ mass fraction as a function of $N^{\rm obs}_{\rm CO}$
at different metallicities.
Equivalently,
this is showing the mass fraction of the CO-dark H$_2$ gas, $F_{\rm dark}$,
as a function of the adopted threshold $N^{\rm obs}_{\rm CO,th}$.
The time-dependent H$_2$ model is shown in solid lines while the steady-state H$_2$ chemistry is in dashed lines.
The vertical black dotted line highlights our adopted threshold 
$N^{\rm obs}_{\rm CO,th} = 6\times 10^{14} {\rm cm^{-2}}$.
This value roughly corresponds to 
a CO(1-0) line intensity $W_{\rm CO}$ = 0.75 K km s$^{-1}$, 
which is the CO sensitivity limit for recent observations of nearby galaxies \citep{2020ApJ...897..122L},
assuming 
$T = $ 20 K, $n = 10^3$ cm$^{-3}$ and a linewidth of 3 km s$^{-1}$.

Our steady-state model suggests that
$F_{\rm dark}$ increases significantly as $Z^{\prime}$ decreases,
which is consistent with the standard PDR theories (e.g. \citealp{1997ApJ...483..200M, 1999ApJ...513..275B, 2010ApJ...716.1191W}).
However,
because of the time-dependent effect,
the actual H/H$_2$ conversions occur gradually 
and gas becomes half-molecular at a much higher density or column density.
Consequently,
$F_{\rm dark}$ becomes progressively lower than the steady-state prediction as $Z^{\prime}$ decreases.
To show the trend with $Z^{\prime}$ more explicitly, 
the right panel in Fig.~\ref{fig:codark} shows the CO-dark H$_2$ mass fraction defined with
$N^{\rm obs}_{\rm CO,th} = 6\times 10^{14} {\rm cm^{-3}}$ as a function of $Z^{\prime}$. 
The actual $F^{*}_{\rm dark}$ still increases inversely with $Z^{\prime}$ 
but not nearly as much as in the steady-state model,
as the latter significantly overestimates H$_2$ in the diffuse medium.


\subsection{Model limitations}

Our model does not account for spatial variations of the FUV radiation,
which can be much stronger than the background radiation field in the vicinity of young stars.
\citet{2017MNRAS.471.2151H} adopted a tree-based treatment assuming optically thin conditions 
and found that the H$_2$ fraction in a dwarf galaxy decreases only by a factor of two 
when switching from a constant FUV model to a variable FUV model,
as the majority of gas is still subject to a smoothly varied background radiation.
However, the optically thin assumption is expected to break down when the gas surface density or metallicity is high.
A full consideration requires radiative transfer which, 
if computationally feasible,
would not only capture the FUV spatial variations 
but also more naturally account for shielding without the need of the column density-based approximation.

For simplicity,
we have assumed that the dust-to-gas mass ratio (DGR) scales linearly with metallicity.
However,
observations have suggested a super-linear scaling at low metallicities \citep{2014A&A...563A..31R},
which means that the time-dependent effect should be even stronger than what we have shown at a given metallicity.
In addition,
the thermal equilibrium state will also be modified as 
photoelectric heating weakens faster than metal line cooling with decreasing metallicity.


\section{Discussion}\label{sec:discuss}

\subsection{Comparison with previous simulations}

In this section,
we compare our results with previous simulations in the literature.
As resolved multi-phase ISM simulations at low metallicities are rare,
we will focus on the solar-metallicity case which is most widely studied.

\citet{2014MNRAS.441.1628S} simulated a large patch of the Milky Way and
found $F_{\rm dark} = 46\%$
adopting $N^{\rm obs}_{\rm CO,th} = 10^{16}\ {\rm cm^{-2}}$.
If we adopt the same threshold,
we find $F_{\rm dark} = 78\%$ in the time-dependent case which is significantly higher. 
However,
their simulations did not include gravity and stellar feedback,
both of which are expected to change the dynamics of gas substantially.
In addition,
their simulation domain covers a galactocentric radius from 5 kpc to 10 kpc with a radial profile of gas surface density,
and so the results may not be directly comparable.
Finally,
their adopted chemistry network (NL97) tends to over-produce CO as shown in \citet{2012MNRAS.421..116G}.
More recently,
\citet{2020MNRAS.492.1594S} adopted a similar setup but included gravity and supernova feedback.
They found $F_{\rm H_2} = 21.4\%$ 
in their feedback-dominated model,
significantly higher than our $Z^{\prime} = 1$ run where $F_{\rm H_2} = 2.0\%$.
This again might be simply due to their larger simulation domain
(their ``zoom-in'' box is 3 kpc on a side and centered on the solar galactocentric radius),
which include regions with higher gas surface density.
The molecular fraction is expected to be rather sensitive to the total hydrogen surface density around 
$\Sigma_{\rm H,tot} = 10\ {\rm M_\odot pc^{-2}}$
as this is the transition between the atomic-dominated and molecular-dominated regimes \citep{2008AJ....136.2846B}.
In our case, 
$\Sigma_{\rm H,tot} \approx 5.8\ {\rm M_\odot pc^{-2}}$, 
so the ISM is expected to be dominated by atomic hydrogen.

\citet{2017MNRAS.472.4797S} performed
``zoom-in'' simulations of two molecular clouds taken from the SILCC simulations.
They found a CO-to-H$_2$ mass ratio of $1.8\times 10^{-4}$,
which is encouragingly similar to our value ($3.6\times 10^{-4}$ for $Z^{\prime} = 1$).
However,
we caution that their simulations are only run for a short amount of time (5 Myr) due to the nature of zoom-ins,
while in our case the results are averaged over 450 Myr with a time interval of 5 Myr 
and so the gas cycles are fairly well-sampled.
It is therefore unclear if the agreement is physical or coincidental.
As a follow-up work, 
\citet{2020MNRAS.492.1465S}
found $F_{\rm dark} = 15 - 65\%$ with a broad range of scatter.
Our time-dependent model finds $F_{\rm dark} = 62\%$,
which is within their scatter but is on the high end.
Again, the two results may not be directly comparable due to the differences in cloud sampling.
In addition,
they found the conversions of H/H$_2$ and C/CO occur at $A_V \sim 1$ and $A_V \sim 1.3$, respectively,
both of which are slightly lower than our time-dependent model 
($A_V = 1.5$ for H/H$_2$ and $A_V = 2.5$ for C/CO).

The fairest comparison we can make is with \citet{2018ApJ...858...16G},
who post-processed the TIGRESS simulations with a very similar setup to ours.
They reported time-averaged results over 60 Myr with a time interval of 5 Myr which roughly covers one gas cycle
and found $F_{\rm H_2} = 12 \%$, significantly higher than our 2\%.
However, 
as the TIGRESS simulations do not include time-dependent chemistry,
they had to evolve the H$_2$ abundance to steady state
and so their results should be compared to our steady-state H$_2$ model where $F_{\rm H_2} = 5.1 \%$.
In addition,
their column densities for shielding are integrated over the entire simulation box using a six-ray algorithm
and thus the shielding length is effectively 500 pc.
This could account for another factor of two difference in $F_{\rm H_2}$ as we show in Appendix \ref{app:Lsh}.
In fact, the steady-state H$_2$ abundance should be even more sensitive to $L_{\rm sh}$.
In terms of CO,
they found a total CO mass about four times higher than in our steady-state H$_2$ model,
which seems too much to be accounted for by the difference in $L_{\rm sh}$.
In fact,
it is more likely to be explained by the difference in the effective visual extinction where the C/CO conversions occur,
which in their case is around 1.3, slightly lower than our 2.2,
presumably due to the difference in the H$_2$ column densities.
As $A^{\rm eff}_{V} \propto n^{0.3}$ at $Z^{\prime} = 1$,
a small difference in $A^{\rm eff}_{V}$ implies a notable difference in the conversion density,
which is reflected in their CO density histogram which peaks at a much lower density $n\sim 300\ {\rm cm }^{-3}$.
For the CO-dark H$_2$ mass fraction,
they used an intensity-based threshold $W_{\rm CO}$ = 0.75 K km s$^{-1}$
and found $F_{\rm dark} = 56 \%$, 
while in their more recent model in \citet{2020ApJ...903..142G} (also with steady-state chemistry),
they found $F_{\rm dark} = 61 \%$. 
Our adopted threshold of $N^{\rm obs}_{\rm CO,th} = 6\times 10^{14}\ {\rm cm^{-2}}$ is designed to match their
threshold.
We find a slightly higher value of 69\% in our steady-state model.
Coincidentally, 
our time-dependent model shows $F_{\rm dark} = 62 \%$
which is in better agreement with their results.




\subsection{Observational implications}

Our results suggest that 
steady-state chemistry 
can significantly overestimate $F_{\rm dark}$ at low metallicities,
as the dynamical time is too short for H$_2$ to reach its steady-state abundance.
Recently,
\citet{2020A&A...643A.141M} investigated nearby dwarf galaxies 
using spectral synthesis models and concluded that $F_{\rm dark}$ ranges from 70\% to 100\%.
We caution that these numbers should be viewed as an upper limit as steady-state chemistry is used.
Admittedly,
our simulations do not cover the parameter space of different gas and stellar surface densities.
In low-surface density environments such as dwarf galaxies,
the SFR surface density is lower which may result in a less violent ISM and a longer dynamical time.
In such cases,
the time-dependent effect would be less pronounced.
We therefore expect 
that the actual $F_{\rm dark}$ in dwarf galaxies should be somewhere 
between our time-dependent and steady-state models 
as shown in Fig.~\ref{fig:codark},
which warrants future investigations.

We stress that the lower $F_{\rm dark}$ we find at low $Z^{\prime}$ is mainly due to the significantly under-abundant H$_2$.
CO emission, in the absolute sense, does not becomes brighter by the time-dependent effect.
If anything,
it should become slightly fainter.
CO traces H$_2$ better in the time-dependent model simply because there is little H$_2$ in the diffuse cold gas.
This raises the question
whether it is useful to obtain an accurate total H$_2$ mass
given that the correlation between H$_2$ and star formation breaks down at low $Z^{\prime}$.


\section{Summary} \label{sec:sum}

We conduct high-resolution 
(particle mass of $1\ {\rm M_\odot}$, effective spatial resolution $\sim$ 0.2 pc) hydrodynamical simulations 
to study the metallicity dependence of H$_2$ and CO in a multi-phase, self-regulating ISM,
covering a wide range of metallicities $0.1 < Z^{\prime} < 3$.
We adopt a hybrid technique 
where we use a simple chemistry network to follow H$_2$ and H$^+$ on-the-fly
and then post-process other species with an accurate chemistry network.
This allows us to simultaneously capture the time-dependent effect of H$_2$ 
and follow all the relevant CO formation and destruction channels.
Our post-processing chemistry network includes 31 species and 286 reactions 
which accurately reproduces the C$^+$/C/CO transitions as in classical PDR calculations 
(Fig.~\ref{fig:pdrf1test}).
Our chemistry code for post-processing is publicly available\footnote{\url{https://github.com/huchiayu/AstroChemistry.jl}} and is archived in Zenodo \citep{zenodo_astrochemisgtry}.
Our main findings can be summarized as follows.

\begin{enumerate}

\item
The ratios $M_{\rm H_2} / M_{\rm C^+}$ and $M_{\rm H_2} / M_{\rm C}$ both show a weaker dependence on $Z^{\prime}$
than $M_{\rm H_2} / M_{\rm CO}$,
which potentially indicates that C$^+$ and C could be viable alternative tracers for H$_2$ at low $Z^{\prime}$ in terms of mass budget.
At low $Z^{\prime}$,
the steady-state model significantly overestimates $M_{\rm H_2}$ but not $M_{\rm CO}$ (Fig.~\ref{fig:ratioh2co}).
As such,
the CO-dark H$_2$ mass fraction is significantly lower than what a steady-state model predicts at low $Z^{\prime}$
(Fig.~\ref{fig:codark}).
The global properties are summarized in Tables \ref{tab:sumstats} and \ref{tab:sumstatseq}.

\item 
The averaged SFR is insensitive to $Z^{\prime}$,
while the temporal fluctuation of SFR increases inversely with $Z^{\prime}$ (Fig.~\ref{fig:sfrtime}).
The median temperature as a function of density is insensitive to $Z^{\prime}$,
as the dominant cooling and heating processes both scale linearly with $Z^{\prime}$ and thus they cancel out.
At lower $Z^{\prime}$,
the median temperature in the cold gas is slightly higher (Figs. \ref{fig:pdmedian}).

\item 
The typical cloud size roughly follows the thermal Jeans length and is thus insensitive to $Z^{\prime}$.
This leads to a correlation between $n$ and $N^{\rm eff}$. 
As $Z^{\prime}$ increases,
dust extinction kicks in at a lower column density,
leading to a slightly flatter relationship
from $N^{\rm eff} \propto n^{0.39}$ at $Z^{\prime} = 0.1$
to $N^{\rm eff} \propto n^{0.3}$ at $Z^{\prime} = 3$.
(Fig.~\ref{fig:nhvsnh} and Eq. \ref{eq:nvsN}).

\item 
Using the correlation between $n$ and $N^{\rm eff}$,
we construct 1D effective density profiles of typical clouds
and conduct PDR calculations.
The results successfully capture the time-averaged chemical distributions in the actual simulations (Fig.~\ref{fig:transitionneq}).

\item
As $Z^{\prime}$ decreases,
H$_2$ becomes progressively under-abundant compared to its steady-state counterpart (which serves as an upper limit) as it is mainly limited by the dynamical time rather than by photodissociation.
The H$_2$ abundance follows a shallow profile 
which can be well-described by an analytic expression with a $Z^{\prime}$-indepedent dynamical time (Eq. \ref{eq:tdyn})
up to the photodissociation surface where it is sharply truncated.
In contrast,
the CO profile is sharp and controlled by photodissociation 
as CO forms rapidly and has reached steady state.
The under-abundant H$_2$ has little impact on the C$^+$/C/CO conversions (Figs. \ref{fig:transitioncompareeqneq} and \ref{fig:cloudz}).
The conversion densities, column densities and visual extinctions are summarized in Tables \ref{tab:nNtrans} and \ref{tab:nNtrans_eq}.


\item 
The shallow H$_2$ profile
leads to a significant contribution of gas with low H$_2$ abundance to the total H$_2$ reservoir.
In addition,
the density where most H$_2$ can be found is significantly lower than the H$_2$ conversion density
(Figs. \ref{fig:h2codensitypdf} and \ref{fig:xh2xcocdfneq}).

\item 
The time-dependent effect helps explain the observed relationship between 
$N^{\rm obs}_{\rm H_2}$ and $N^{\rm obs}_{\rm CO}$ for the Galactic clouds
without resorting to 
non-thermal velocity distributions 
for the reaction
C$^+$ + H$_2$ $\rightarrow$ CH$^+$ + H 
(Fig.~\ref{fig:losnh2vsnconeqvseq}).

\item
For a given $N^{\rm obs}_{\rm H_2}$,
our fiducial model reproduces the observed 
$N^{\rm obs}_{\rm C^+}$ and $N^{\rm obs}_{\rm CO}$ 
but significantly over-produces $N^{\rm obs}_{\rm C}$ in the Galactic clouds.
Only when we artificially switch off recombination on grains 
can we reproduce all the three quantities simultaneously (Fig.~\ref{fig:losnh2vsncpcico}).



\end{enumerate}


\section*{Acknowledgments}
We thank the anonymous referee for their constructive comments that improved our manuscript.
We thank Shmuel Bialy, Munan Gong and Daniel Seifried for fruitful discussions.
C.Y.H. is grateful for the constant support from Ting-Yi Wu and his SCDS support group.
C.Y.H. acknowledges support from the DFG via German-Israel Project Cooperation grant STE1869/2-1 GE625/17-1.
A.S. thanks the Center for Computational Astrophysics (CCA) of the Flatiron Institute,
and the Mathematics and Physical Science (MPS) division of the Simons Foundation for support.
All simulations were run on the Cobra and Draco supercomputers at the Max Planck Computing and Data Facility (MPCDF).

\bibliography{literatur}{}
\bibliographystyle{aasjournal}

\appendix

\section{Chemical timescales}\label{app:chemtime}

	In this section,
	we investigate the timescales for C$^+$, C and CO to reach steady state.
	
\begin{figure}
		\centering
		\includegraphics[width=0.99\linewidth]{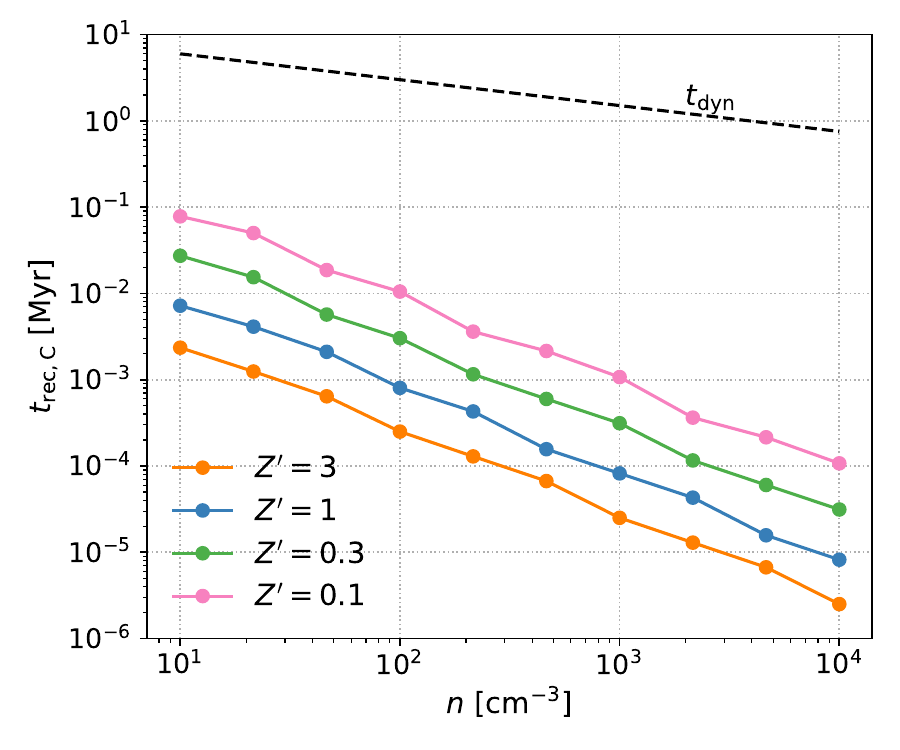}
		\caption{
				The C$^+$ recombination time $t_{\rm rec,C}$ as a function of $n$ at different metallicities.
				The dashed line shows the dynamical time $t_{\rm dyn}$ in our simulations.
				In all cases,
				$t_{\rm rec,C}$ is orders of magnitude shorter than $t_{\rm dyn}$,
				justifying our steady-state assumption.			
		}
		\label{fig:tcrec}
	\end{figure}

	For the C$^+$/C transition,
	the relevant timescale is C$^+$ recombination $t_{\rm rec,C}$.
	To measure $t_{\rm rec,C}$,
	we run the chemistry network in a one-zone setup where
	$T = 30$ K,
	$\zeta_{\rm CR} = 10^{-16} {\rm s^{-1}}$,
	$I_{\rm UV} = 0$
	and $n$ systematically varies from 10 to $10^4\ {\rm cm^{-3}}$.
	We adopt $I_{\rm UV} = 0$ such that C$^+$ will recombine even at low densities.
	Adopting a nonzero $I_{\rm UV}$ does not change $t_{\rm rec,C}$ significantly.
	All carbon is initially in the form of C$^+$ ($x_{\rm C^+} / x_{\rm C,0} = 1$)
	which gradually recombines as time evolves.
	We record $t_{\rm rec,C}$ as the time where $x_{\rm C^+} / x_{\rm C,0} < \exp(-1)$ first occurs.
	Fig.~\ref{fig:tcrec} shows $t_{\rm rec,C}$ as a function of $n$.
	The dashed line shows the dynamical time $t_{\rm dyn}$ in our simulations
	which is insensitive to $Z^{\prime}$ (see Sec. \ref{sec:transition}).
	$t_{\rm rec,C}$ scales roughly as $(nZ^{\prime})^{-1}$,
	as expected for two-body reactions.
	In all cases,
	$t_{\rm rec,C}$ is orders of magnitude shorter than $t_{\rm dyn}$,
	justifying our steady-state assumption.

\begin{figure*}
	\centering
	\includegraphics[width=0.99\linewidth]{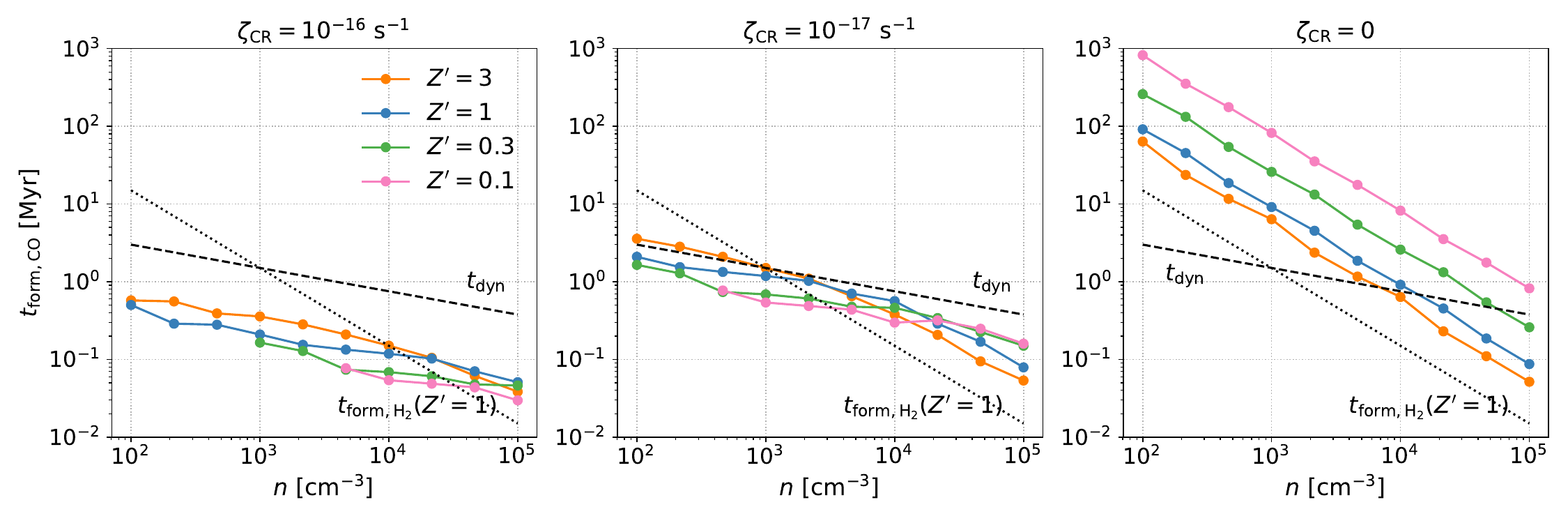}
	\caption{
			The CO formation time $t_{\rm form,CO}$ as a function of $n$ at different metallicities.
			The dashed line shows the dynamical time $t_{\rm dyn}$ in our simulations
			while the dotted line shows the CO formation time $t_{\rm form,H_2}$ at $Z^{\prime} = 1$.
			Hydrogen is initially 90\% in H$_2$ ($2x_{\rm H_2} = 0.9$).
			For $\zeta_{\rm CR} = 10^{-16} {\rm s^{-1}}$ (the averaged value in our simulations),
			$t_{\rm form,CO}$ is significantly lower than $t_{\rm dyn}$ at all densities and metallicities,
			justifying our steady-state assumption.
	}
	\label{fig:tcoform}
\end{figure*}

	For the C/CO transition,
	the relevant timescale is CO formation $t_{\rm form,CO}$.
	We are interested in the case where gas is already fully molecular ($2x_{\rm H_2} \lesssim 1$).
	The left panel in Fig.~\ref{fig:tcoform} shows $t_{\rm form,CO}$ as a function of $n$,
	with $T = 30$ K,
	$\zeta_{\rm CR} = 10^{-16} {\rm s^{-1}}$ and 
	$I_{\rm UV} = 0$.
	All carbon is initially in the form of C$^+$,
	while hydrogen is 90\% in H$_2$ ($2x_{\rm H_2} = 0.9$) and 10\% in atomic H.
	The dashed line shows the dynamical time $t_{\rm dyn}$ in our simulations
	while the dotted line is the H$_2$ formation time $t_{\rm form,H_2}$ at $Z^{\prime} = 1$ (see Eq. \ref{eq:tform}).
	We record $t_{\rm form,CO}$ as the time where $x_{\rm CO} / x_{\rm C,0} > 1-\exp(-1)$ 
	(or, equivalently, $x_{\rm C} / x_{\rm C,0} < \exp(-1)$) first occurs.
	CO forms mainly via two routes: the OH-channel and CH-channel.
	Both routes are facilitated by cosmic-ray ionization 
	which produces ions such as H$^+_3$ and initiates subsequent chemistry.
	As cosmic-ray ionization does not depend explicitly on either $n$ or $Z^{\prime}$,
	$t_{\rm form,CO}$ shows a much weaker dependence on both $n$ and $Z^{\prime}$,
	as opposed to $t_{\rm form,H_2}$ which scales as $(nZ^{\prime})^{-1}$.
	Instead,
	$t_{\rm form,CO}$ depends primarily on $\zeta_{\rm CR}$.
	For $\zeta_{\rm CR} = 10^{-16} {\rm s^{-1}}$ (the averaged value in our simulations),
	$t_{\rm form,CO}$ is significantly lower than $t_{\rm dyn}$ at all densities and metallicities,
	justifying our steady-state assumption.
	If we adopt $\zeta_{\rm CR} = 10^{-17} {\rm s^{-1}}$ (middle panel of Fig.~\ref{fig:tcoform}),
	$t_{\rm form,CO}$ increases by a factor of 3 to 5
	and becomes comparable to $t_{\rm dyn}$.
	In this case,
	the steady-state assumption is only marginally justified.
	Finally,
	if we adopt $\zeta_{\rm CR} = 0$ (right panel of Fig.~\ref{fig:tcoform}),
	$t_{\rm form,CO}$ becomes orders of magnitude longer especially at low $n$,
	and it scales roughly with $(nZ^{\prime})^{-1}$ as it is dictated by two-body reactions.
	Our $\zeta_{\rm CR} = 0$ case for $Z^{\prime} = 1$ agrees well with the timescale estimated by
	 \citet{2019MNRAS.484.1735J} ($t_{\rm form,CO} = 10^{3.7}  ( n/({\rm cm^{-3}}) ) ^{-1}$ Myr, their Eq. 38). 
	This makes sense as they adopted the chemistry network of NL97,
	which ignores the effect of cosmic ray ionization in CO formation,
	leading to a significantly overestimated $t_{\rm form,CO}$ that scales with $n^{-1}$.

\section{Model robustness}

\subsection{Shielding length}\label{app:Lsh}

\begin{figure*}
	\centering
	\includegraphics[trim=4cm 0cm 4cm 1cm, clip, width=0.95\linewidth]{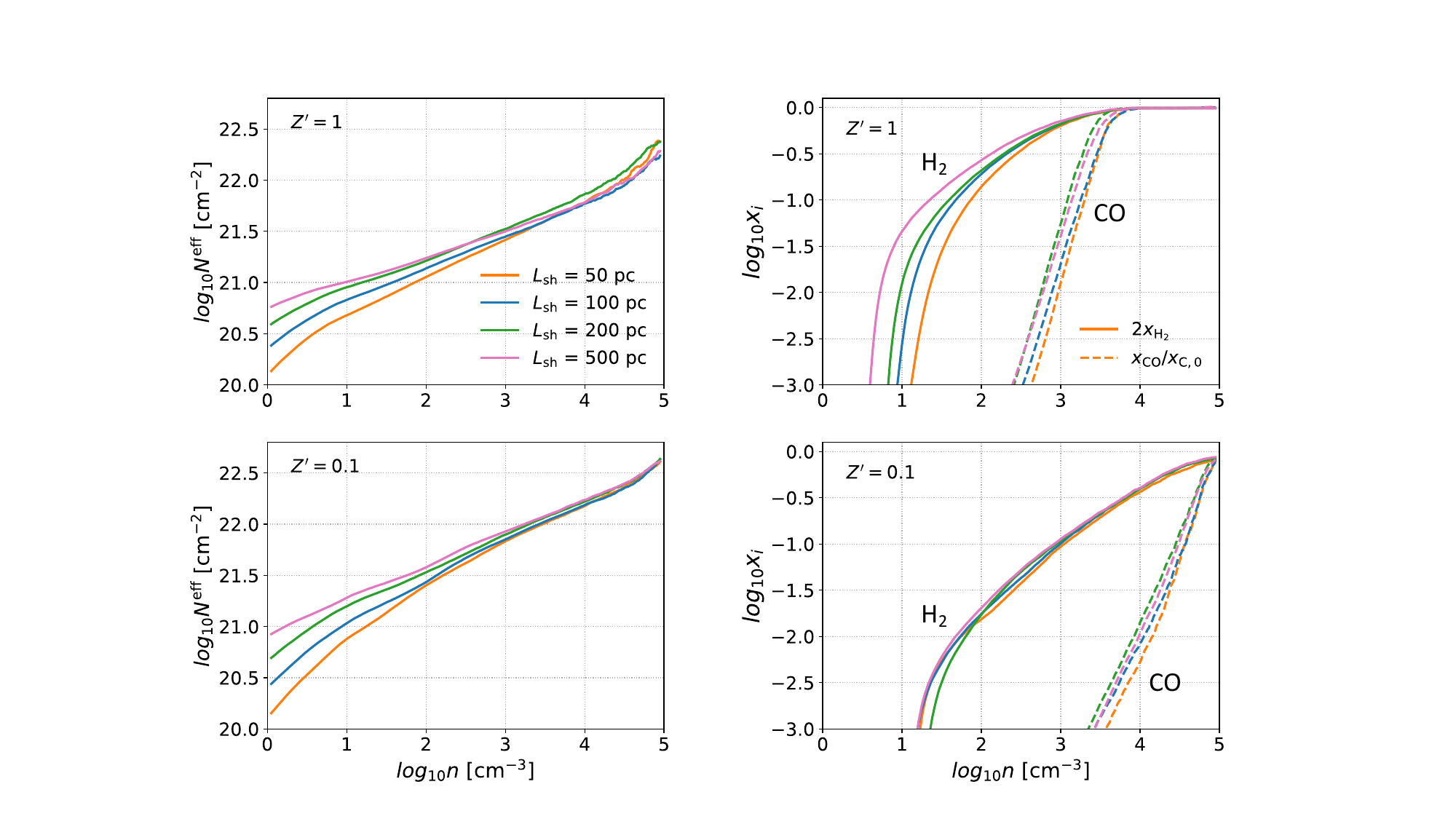}
	\caption{
		Effective column density (left) and normalized chemical abundances (right) as a function of density
		for $L_{\rm sh} = $ 50 (orange), 100 (blue), 200 (green) and 500 pc (pink)
		with $Z^{\prime} = 1$ (top) and $Z^{\prime} = 0.1$ (bottom).
		The solid and dashed lines show the normalized abundances of H$_2$ and CO, respectively.
		From $L_{\rm sh} = $ 50 pc to $L_{\rm sh} = $ 500 pc,
		$M_{\rm H_2}$ and $M_{\rm CO}$ increase, respectively,  
		by a factor of 2.4 and 1.6 at $Z^{\prime} = 1$,
		and 
		by a factor of 1.16 and 1.08 at $Z^{\prime} = 0.1$.
	}
	\label{fig:lshtransitioncompareeqneq}
\end{figure*}

In this section, 
we investigate how the choice of $L_{\rm sh}$ affects the molecular abundance.
Fig.~\ref{fig:lshtransitioncompareeqneq} shows
$N^{\rm eff}$ (left) and normalized chemical abundances (right) as a function of $n$
for $L_{\rm sh} = $ 50 (orange), 100 (blue), 200 (green) and 500 pc (pink)
with $Z^{\prime} = 1$ (top) and $Z^{\prime} = 0.1$ (bottom).
The solid and dashed lines show the normalized abundances of H$_2$ and CO, respectively.
For computational efficiency,
these simulations were run with $m_{\rm g} = 3\ {\rm M_\odot}$
(instead of $m_{\rm g} = 1\ {\rm M_\odot}$ as in our default runs).

In general, 
a larger $L_{\rm sh}$ leads to a higher $N^{\rm eff}$.
The largest difference is found in the diffuse medium,
within a factor of three at $n\sim 10\ {\rm cm^{-3}}$ between the two extreme cases.
In dense clouds,
the major contribution to shielding is from gas in the vicinity rather than from the diffuse volume-filling background,
and thus $N^{\rm eff}$ is insensitive to $L_{\rm sh}$ at high $n$.
As the H$_2$ profile is mainly controlled by the dynamical time,
the conversion density for H$_2$ (where $x_{\rm H} = x_{\rm H_2}$) is insensitive to $L_{\rm sh}$.
The effect of $L_{\rm sh}$ only appears near the photodissociation front in the $Z^{\prime} = 1$ run,
where a larger $L_{\rm sh}$ implies more shielding at a given density,
shifting the photodissociation front towards at a lower density.
At $Z^{\prime} = 0.1$,
the effect of $L_{\rm sh}$ becomes even weaker as the H$_2$ profile is mainly governed by the dynamical time.
In contrast,
the C/CO conversions are controlled by photodissociation 
which is shifted towards a lower density or column density as $L_{\rm sh}$ increases.
Fortunately,
this occurs at high enough column densities such that the corresponding density is similar, 
and thus $M_{\rm CO}$ is also insensitive to $L_{\rm sh}$.
From $L_{\rm sh} = $ 50 pc to $L_{\rm sh} = $ 500 pc,
$M_{\rm H_2}$ and $M_{\rm CO}$ increase, respectively,  
by a factor of 2.4 and 1.6 at $Z^{\prime} = 1$,
and 
by a factor of 1.16 and 1.08 at $Z^{\prime} = 0.1$.

\subsection{Recombination on grains}\label{app:grRec}

\begin{figure}
	\centering
	\includegraphics[width=0.9\linewidth]{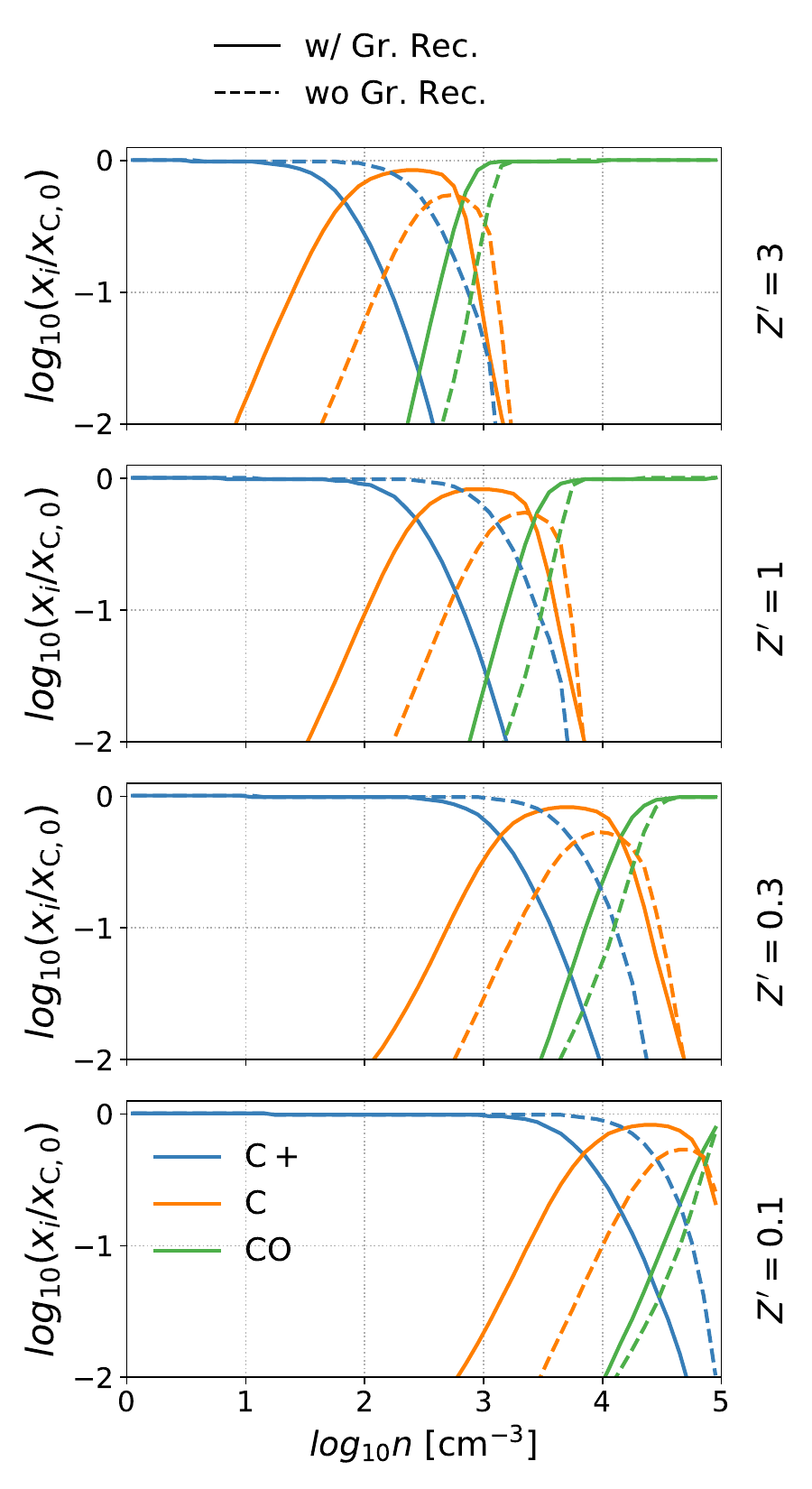}
	\caption{
		Normalized abundances of C$^+$ (blue), C (orange) and CO (green) as a function of density
		for $Z^{\prime} = $ 3, 1, 0.3 and 0.1 from top to bottom.
		The solid and dashed lines represent models with and without recombinations on grains.
		For $Z^{\prime} = $ 3, 1, 0.3 and 0.1,
		including recombination on grains 
		increases $M_{\rm C}$ by a factor of 
		3.9, 4.3, 5.9 and 4.5,
		while increases $M_{\rm CO}$ by a factor of 
		2, 1.6, 1.2 and 1.1, 
		respectively.
	}
	\label{fig:grrectransition}
\end{figure}

Recombination on grains can be more efficient than radiative recombination 
and becomes the dominant destruction process for positive ions in some circumstances
\citep{1978MNRAS.184..227W, 1987ApJ...320..803D, 1988ApJ...329..418L, 2001ApJ...563..842W}.
\citet{2017ApJ...843...38G} found that
including recombinations on grains is crucial in order to obtain the correct CO abundance in their PDR calculations.
Fig.~\ref{fig:grrectransition} shows
the normalized abundances of C$^+$ (blue), C (orange) and CO (green) as a function of density
for $Z^{\prime} = $ 3, 1, 0.3 and 0.1 from top to bottom.
The solid and dashed lines represent models with and without recombinations on grains.
Recombination on grains shifts the conversions of both C$^+$/C and C/CO to lower densities,
but the effect is significantly more pronounced in the former.
For $Z^{\prime} = $ 3, 1, 0.3 and 0.1,
including recombination on grains 
increases $M_{\rm C}$ by a factor of 
3.9, 4.3, 5.9 and 4.5,
while increases $M_{\rm CO}$ by a factor of 
2, 1.6, 1.2 and 1.1, 
respectively.
Contrary to \citet{2017ApJ...843...38G},
we find that $M_{\rm CO}$ is only marginally affected by recombination on grains 
as CO formation occurs at higher densities in our simulations.
Indeed,
we have confirmed that in one-zone PDR calculations, 
recombination on grains makes a increasingly larger difference on the CO abundance with decreasing density,
as also seen in \citet{2017ApJ...843...38G}.

\section{Numerical robustness}\label{app:num}

\subsection{Convergence study}\label{app:conv}

\begin{figure}
	\centering
	\includegraphics[width=0.9\linewidth]{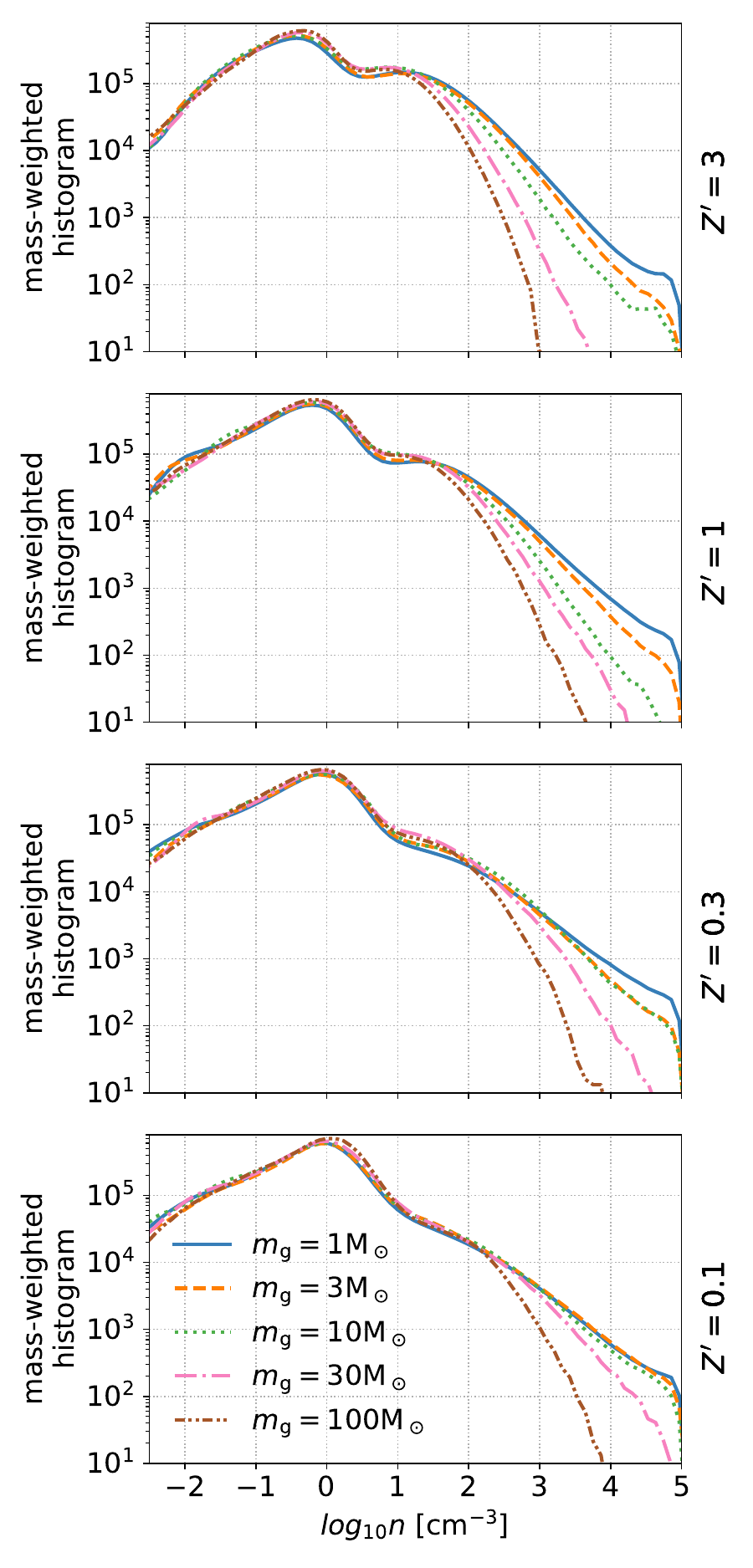}
	\caption{
		Time-averaged density histogram with five different mass resolutions 
		for $Z^{\prime} = $ 3, 1, 0.3 and 0.1 from top to bottom.
	 }
	\label{fig:convdenpdf}
\end{figure}

Fig.~\ref{fig:convdenpdf} shows the 
time-averaged density histogram 
with five different mass resolutions 
$m_{\rm g}$
for $Z^{\prime} = $ 3, 1, 0.3 and 0.1 from top to bottom.
The high density tail is gradually converged as $m_{\rm g}$ decreases.
However,
the convergence occurs faster at lower $Z^{\prime}$.
This is related to the difference in the star formation density threshold (see Fig.~\ref{fig:pdmedian}),
which is generally higher at lower $Z^{\prime}$ due to a higher temperature of clouds.
At high $Z^{\prime}$,
star formation occurs at lower densities which efficiently stops the gravitational collapse and removes the high-density gas.

\begin{figure}
	\centering
	\includegraphics[width=0.9\linewidth]{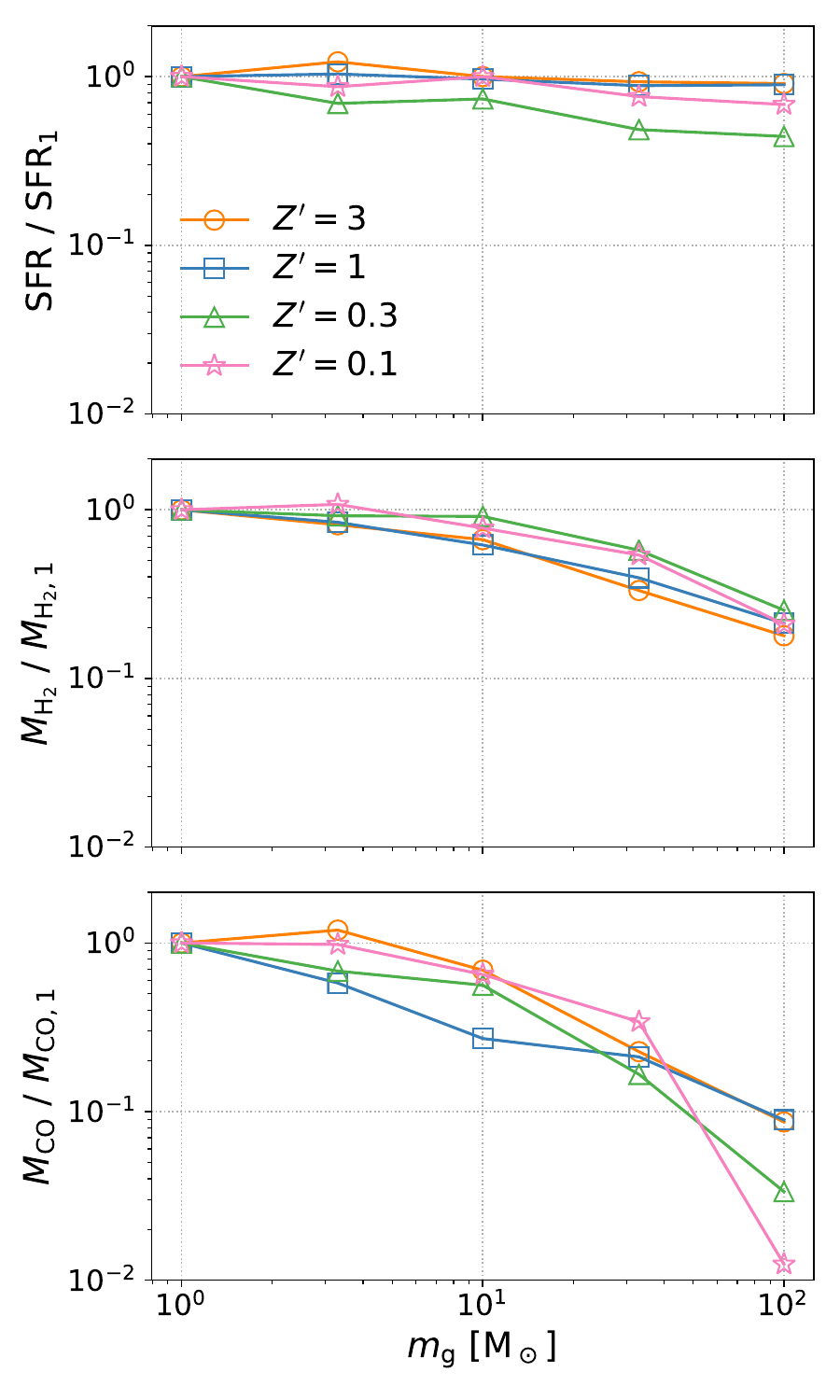}
	\caption{
		Time-averaged SFR (top), total H$_2$ mass (middle) and total CO mass (bottom) as a function of $m_{\rm g}$,
		normalized to the results in the highest resolution ($m_{\rm g}  = 1\ {\rm M_\odot}$).
	}
	\label{fig:convsfrh2co}
\end{figure}

In terms of the global properties,
Fig.~\ref{fig:convsfrh2co} shows the
time-averaged SFR (top), $M_{\rm H_2}$ (middle) and $M_{\rm CO}$ (bottom) as a function of $m_{\rm g}$,
normalized to the results in the highest resolution ($m_{\rm g}  = 1\ {\rm M_\odot}$).
Among the three quantities,
the SFR is the easiest to converge.
Even at $m_{\rm g}  = 1\ {\rm M_\odot}$,
the SFR is converged within a factor of 50\% and show a diverging trend with resolution.
On the other hand,
both $M_{\rm H_2}$  and $M_{\rm CO}$ show a clear trend of diverging (underestimation) as $m_{\rm g}$ increases,
with $M_{\rm CO}$ diverging faster.
This is expected as CO forms at higher densities than H$_2$ which requires a higher resolution to resolve. 
Interestingly,
the $Z^{\prime} = 0.1$ run performs as well as the $Z^{\prime} = 3$ run in terms of convergence,
with both $M_{\rm H_2}$ and $M_{\rm CO}$ converge within 10\% at $m_{\rm g}  = 3\ {\rm M_\odot}$.
This is because while both H$_2$ and CO form at higher densities at lower $Z^{\prime}$,
its density histogram is also converged up to a higher density
due to a higher SFR density threshold.
In contrast,
the $Z^{\prime} = 1$ and $Z^{\prime} = 0.3$ runs have $M_{\rm H_2}$ and $M_{\rm CO}$ converged within, respectively,  
around 20\% and 50\% at $m_{\rm g}  = 3\ {\rm M_\odot}$,
which is slightly worse than the other two runs.
This demonstrates that
there is no universal resolution requirement for the convergence of molecular abundances,
which occurs once the density histogram converges up to the conversion densities for H/H$_2$ and C/CO.

\subsection{Instantaneous star formation threshold}\label{app:instantSF}

\begin{figure}
	\centering
	\includegraphics[width=0.9\linewidth]{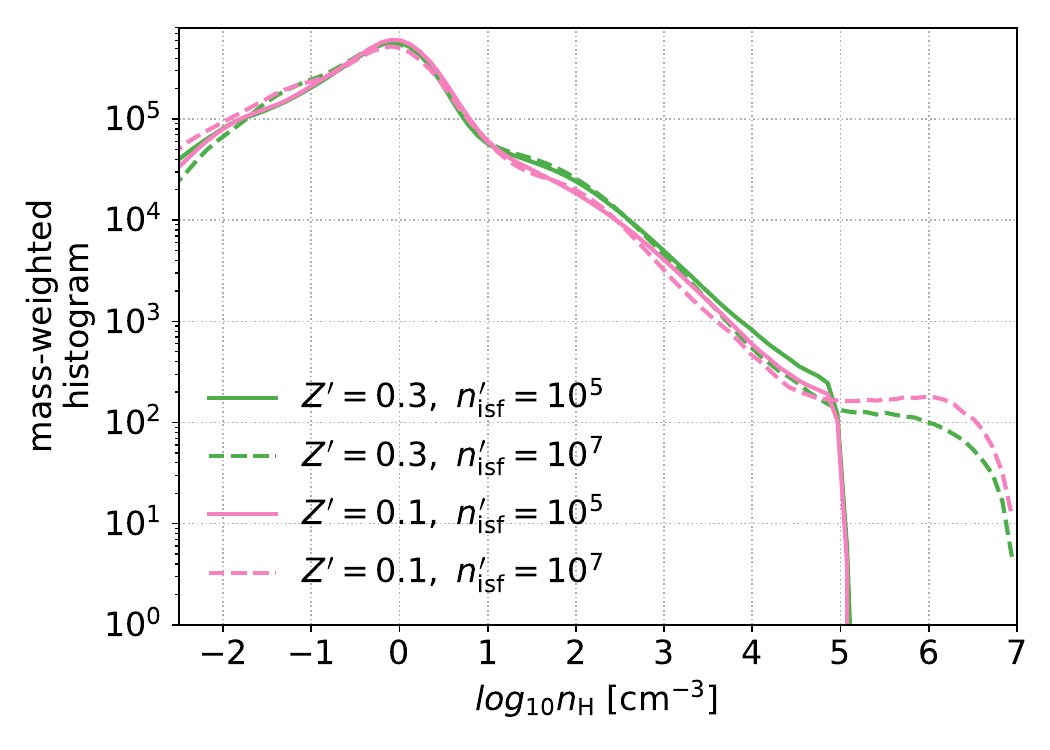}
	\caption{
		Density histogram at $Z^{\prime} = 0.3$ (green) and $Z^{\prime} = 0.1$ (pink).
		The solid and dashed lines represent $n_{\rm isf} = 10^5\ {\rm cm^{-3}}$ and $n_{\rm isf} = 10^7\ {\rm cm^{-3}}$, respectively.
	}
	\label{fig:instantsfdenpdf}
\end{figure}

In this section,
we demonstrate the effect of instantaneous star formation density threshold $n_{\rm isf}$ (see Sec. \ref{sec:SFrecipe})
by running two additional simulations with $n_{\rm isf} = 10^7\ {\rm cm^{-3}}$.
Fig.~\ref{fig:instantsfdenpdf} shows the 
density histogram at $Z^{\prime} = 0.3$ (green) and $Z^{\prime} = 0.1$ (pink).
The solid and dashed lines represent $n_{\rm isf} = 10^5\ {\rm cm^{-3}}$ and $n_{\rm isf} = 10^7\ {\rm cm^{-3}}$, respectively.

With $n_{\rm isf} = 10^7\ {\rm cm^{-3}}$,
the gas collapses way beyond the Jeans mass threshold,
which corresponds to $n\sim 3\times 10^4\ {\rm cm^{-3}}$ (see Fig.~\ref{fig:pdmedian}),
and piles up at high densities.
As such,
the histogram becomes flat above $10^5\  {\rm cm^{-3}}$.
This leads to $M_{\rm CO}$ increasing by a factor of 1.2 and 4.0 in the $Z^{\prime} = 0.3$ and $Z^{\prime} = 0.1$ runs,
respectively.
The effect is less significant at higher $Z^{\prime}$ as the C/CO conversions occur at lower densities 
where the density distribution is still unaffected.
Likewise,
as H$_2$ forms at much lower densities,
$M_{\rm H_2}$ is increased by a factor of 1.05 for $Z^{\prime} = 0.3$
and is even decreased by a factor of 0.96 for $Z^{\prime} = 0.1$.
Therefore,
although our choice of $n_{\rm isf}$ is numerics-motivated and somewhat arbitrary,
it only has a significant effect on $M_{\rm CO}$ at $Z^{\prime} = 0.1$.
Moreover,
we argue that the flat density distribution with $n_{\rm isf} = 10^7\ {\rm cm^{-3}}$ 
is unrealistic as self-gravitating clouds are expected to follow a declining power-law distribution.

The stochastic star formation recipe we adopt has been the standard approach in cosmological simulations.
Adding an instantaneous density threshold in the stochastic method
has been proposed in \citet{2019ApJ...879L..18L} with $4 {\rm M_\odot}$-resolution.
Moreover,
the ``sink-particle'' method, 
which is commonly used in cloud-scale simulations,
is also an instantaneous star formation model.
It therefore appears that removing gravitationally unresolved gas instantaneously 
(rather than waiting for a few free-fall times)
is numerically beneficial as one goes from cosmological scales to ISM scales.

\subsection{Chemistry iteration}\label{app:chem_iter}

\begin{figure}
	\centering
	\includegraphics[trim=9cm 0cm 9cm 0cm, clip, width=0.99\linewidth]{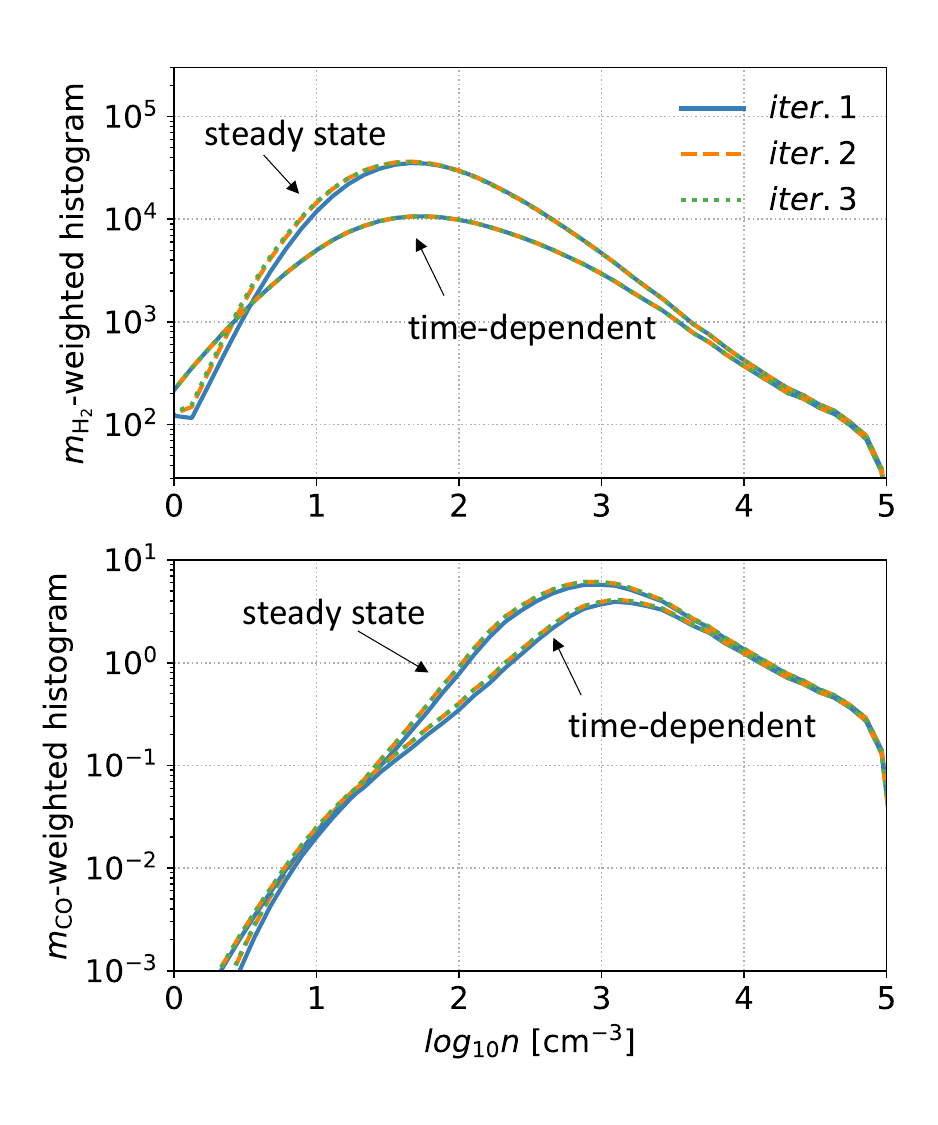}
	\caption{
		Time-averaged density histograms of H$_2$ (top) and CO (bottom) for the $Z^{\prime} = 1$ case for each iteration (iter. 1, 2 and 3) when solving the post-processing chemistry network.
		Both $x_{\rm H_2}$ and $x_{\rm CO}$ converge rapidly.
	}
	\label{fig:iteration}
\end{figure}

As discussed in Sec. \ref{sec:shieldcolumn},
solving the chemistry network in post-processing requires iterations as shielding from gas depends on the nonlocal quantities 
$x_{\rm H_2}$ and $x_{\rm CO}$.
In Fig.~\ref{fig:iteration},
we show the time-averaged density histograms of H$_2$ and CO for the $Z^{\prime} = 1$ case for each iteration.
For the time-dependent case,
$x_{\rm H_2}$ is treated as given and therefore remains fixed during the iterations.
$x_{\rm CO}$ converges rapidly ($M_{\rm CO}/{\rm M_\odot}$ = 33.7, 35.4 and 35.4 from the first to third iteration)
as dust shielding and H$_2$ shielding for CO are already known.
The difference in $x_{\rm CO}$ between the first and the final iterations reflects the effect of CO self-shielding,
which only increases $M_{\rm CO}$ by 5\%.
For the steady-state case,
both $x_{\rm H_2}$ and $x_{\rm CO}$ require iterations.
We use the time-dependent $x_{\rm H_2}$ as the initial guess in the first iteration which facilitates the convergence.
From the first to third iteration,
$M_{\rm H_2}/(10^5{\rm M_\odot})$ = 2.98, 3.15, 3.14 while $M_{\rm CO}/{\rm M_\odot}$ = 44.7, 47.5 and 47.5, respectively.

\end{document}